\pdfoutput=1
\documentclass[sigconf]{acmart}

\copyrightyear{2022}
\acmYear{2022}
\setcopyright{acmlicensed}\acmConference[ASE '22]{37th IEEE/ACM International Conference on Automated Software Engineering}{October 10--14, 2022}{Rochester, MI, USA}
\acmBooktitle{37th IEEE/ACM International Conference on Automated Software Engineering (ASE '22), October 10--14, 2022, Rochester, MI, USA}
\acmPrice{15.00}
\acmDOI{10.1145/3551349.3556958}
\acmISBN{978-1-4503-9475-8/22/10}

\usepackage{xspace}
\usepackage{xcolor}
\usepackage{graphicx}
\usepackage{amsmath}
\usepackage{enumitem}
\usepackage{semantic} %

\usepackage{hyperref}

\usepackage{booktabs}
\usepackage{multirow}
\usepackage{makecell}
\usepackage{ragged2e}

\usepackage{caption}
\usepackage{subcaption}
\usepackage{wrapfig}

\usepackage{pifont}
\usepackage{listings}

\usepackage{algorithm}
\usepackage[noend]{algpseudocode}

\usepackage{tikz}
\usepackage{forest} %
\usetikzlibrary{arrows.meta}
\usetikzlibrary{backgrounds}
\usetikzlibrary{calc}
\usetikzlibrary{decorations.pathreplacing}
\usetikzlibrary{fit}
\usetikzlibrary{positioning}
\usetikzlibrary{shapes.geometric}

\newcommand{\XSpace}[1]{}
\newcommand{\XComment}[1]{}

\newcommand{\DefMacro}[2]{\expandafter\newcommand\csname rmk-#1\endcsname{#2}}
\newcommand{\UseMacro}[1]{\csname rmk-#1\endcsname}

\newcommand{\MyPara}[1]{\vspace{1pt}\noindent\textbf{#1}.}

\newcommand{\InputWithSpace}[1]{\bgroup\def\arraystretch{1.2}\input{#1}\egroup}
\newcommand{\Code}[1]{{\ifmmode{\mathtt{#1}}\else$\mathtt{#1}$\fi}}
\newcommand{\CodeIn}[1]{\texttt{\small #1}}

\newcommand{\Capitalize}[1]{\expandafter\MakeUppercase #1}
\newcommand{\Num}[1]{#1}    %

\newcolumntype{R}[1]{>{\RaggedLeft\arraybackslash}p{#1}}
\newcolumntype{L}[1]{>{\RaggedRight\arraybackslash}p{#1}}

\definecolor{gray}{RGB}{211,211,211}
\newcommand{\jbasicstyle}{\small\sffamily} %

\newcommand{\jnumberstyle}{\scriptsize}

\lstdefinelanguage{pseudo}
{
morekeywords={},
keywordstyle=\bfseries,
lineskip=-0.1em,
numbers=left, %
numberstyle=\jnumberstyle,
numbersep=4pt,
basicstyle=\jbasicstyle,
breaklines=true,
breakautoindent=true,
tabsize=2,
columns=fullflexible,
morecomment=*[l][\textsl]{//},
mathescape=true,
xleftmargin=10pt,
}

\lstdefinelanguage{todo-comment}
{
morekeywords={},
keywordstyle=\bfseries,
lineskip=-0.1em,
numbers=none,
basicstyle=\jbasicstyle,
breaklines=true,
breakautoindent=true,
tabsize=2,
columns=fullflexible,
morecomment=*[l][\textsl]{//},
mathescape=true,
xleftmargin=-10pt,
}

\lstdefinelanguage{bug}
{
language=java,
numbers=none,
basicstyle=\scriptsize\ttfamily,
numberstyle=\scriptsize,
breaklines=true,
columns=fullflexible,
showstringspaces=false,
escapeinside={(*@}{@*)}
}

\lstdefinelanguage{sketchy}
{
language=java,
numbers=left,
basicstyle=\scriptsize\ttfamily,
numberstyle=\scriptsize,
breaklines=true,
columns=fullflexible,
xleftmargin=16pt,
showstringspaces=false,
emph={@Entry}, emphstyle=\color{blue},
escapeinside={(*@}{@*)}
}

\lstdefinelanguage{sketchy-display}
{
language=sketchy,
numbers=none,
basicstyle=\footnotesize\ttfamily,
xleftmargin=0pt,
emph={@Entry,intId,intVal,arithmetic,relation,logic,SUB,ADD,MUL,DIV,LE,GE,LT,GT,EQ,NE,AND,OR}, emphstyle=\color{blue},
}

\lstdefinelanguage{sketchy-appendix}
{
language=java,
numbers=left,
basicstyle=\scriptsize\ttfamily,
numberstyle=\scriptsize,
breaklines=true,
columns=fullflexible,
xleftmargin=16pt,
showstringspaces=false,
emph={@Entry}, emphstyle=\color{blue},
emph={intId,intVal,arithmetic,relation,logic,@Entry,SUB,ADD,MUL,DIV,LE,GE,LT,GT,EQ,NE,AND,OR}, emphstyle=\color{blue},
emph={[2]eval}, emphstyle={[2]\color{purple}\bfseries},
escapeinside={(*@}{@*)}
}

\lstdefinelanguage{syntax}
{
numbers=none,
basicstyle=\footnotesize\ttfamily,
breaklines=true,
columns=fullflexible,
xleftmargin=16pt,
showstringspaces=false,
}

\lstdefinelanguage{semantics}
{
morekeywords={if,else,where}, keywordstyle=\bfseries,
numbers=none,
basicstyle=\footnotesize\ttfamily,
breaklines=true,
columns=fullflexible,
xleftmargin=16pt,
showstringspaces=false,
}

\lstdefinelanguage{api-list}
{
language=java,
numbers=none,
basicstyle=\footnotesize\ttfamily,
breaklines=true,
columns=fullflexible,
showstringspaces=false,
emph={boolVal,intVal,boolId,intId,arithmetic,relation,logic,alt,refId,exprStmt,ifStmt,whileStmt,tryStmt,assign}, emphstyle=\color{blue},
}

\lstdefinelanguage{java-pretty}
{
language=java,
numbers=left,
basicstyle=\footnotesize\ttfamily,
numberstyle=\footnotesize,
breaklines=true,
columns=fullflexible,
xleftmargin=16pt,
showstringspaces=false,
}

\lstdefinelanguage{java-display}
{
language=java-pretty,
numbers=none
}

\algnewcommand\Break{\textbf{break}}%

\algdef{SE}[DOWHILE]{Do}{EndDoWhile}{\algorithmicdo}[1]{\algorithmicwhile\ #1}

\algdef{SL}[INPUT]{Input}{0}{\textbf{Input: }}
\algdef{SL}[OUTPUT]{Output}{0}{\textbf{Output: }}

\makeatletter
\algrenewcommand\alglinenumber[1]{\footnotesize #1:} %
\algrenewcommand\ALG@beginalgorithmic{\footnotesize}
\algrenewcommand\algorithmiccomment[2][\footnotesize]{{#1\hfill\(\triangleright\) #2}} %
\makeatother

\newcommand{\Title}{Compiler Testing using Template Java Programs}
\newcommand{\JTool}{\textsc{JAttack}\xspace}

\newcommand{\API}{API\xspace}
\newcommand{\APIs}{APIs\xspace}
\newcommand{\JIT}{JIT\xspace}
\newcommand{\COne}{C1\xspace}
\newcommand{\CTwo}{C2\xspace}

\newcommand{\JVM}{JVM\xspace}

\newcommand{\DSL}{DSL\xspace}
\newcommand{\eAST}{eAST\xspace}
\newcommand{\eASTs}{eASTs\xspace}
\newcommand{\sketch}{template\xspace}
\newcommand{\Sketch}{Template\xspace}
\newcommand{\sketches}{templates\xspace}
\newcommand{\Sketches}{Templates\xspace}
\newcommand{\sketching}{templating\xspace}
\newcommand{\entryannot}{\CodeIn{@Entry}\xspace}
\newcommand{\hole}{hole\xspace}
\newcommand{\holes}{holes\xspace}

\newcommand{\candidate}{candidate\xspace}
\newcommand{\candidates}{candidates\xspace}
\newcommand{\Candidates}{Candidates\xspace}
\newcommand{\node}{node\xspace}
\newcommand{\nodes}{nodes\xspace}

\newcommand{\sketchprogram}{\sketch program\xspace}
\newcommand{\sketchprograms}{\sketch programs\xspace}
\newcommand{\Sketchprogram}{\Sketch program\xspace}

\newcommand{\generatedprogram}{generated program\xspace}
\newcommand{\generatedprograms}{generated programs\xspace}

\newcommand{\sketchmethod}{\sketch entry method\xspace}
\newcommand{\sketchmethods}{\sketch entry methods\xspace}

\newcommand{\generatedmethod}{generated entry method\xspace}

\newcommand{\Staticss}{Static\xspace}
\newcommand{\hotfilling}{hot filling\xspace}
\newcommand{\Hotfilling}{Hot filling\xspace}
\newcommand{\HotFilling}{Hot Filling\xspace}
\newcommand{\stopearly}{early stop\xspace}
\newcommand{\Stopearly}{Early stop\xspace}
\newcommand{\StopEarly}{Early Stop\xspace}
\newcommand{\solveraid}{eager pruning\xspace}
\newcommand{\Solveraid}{Eager pruning\xspace}

\newcommand{\staticgen}{\Staticss}

\newcommand{\nonopt}{Non-opt.\xspace}
\newcommand{\stopearlyopt}{\StopEarly}
\newcommand{\hotfillingopt}{\HotFilling}
\newcommand{\solveraidopt}{Eager Pruning}
\newcommand{\fullopt}{Full Opt.\xspace}

\newcommand{\fullitrs}{Full Itrs.\xspace}
\newcommand{\fullitrsmaster}{Full Itrs. w/ Master\xspace}
\newcommand{\fullitrsco}{Full Itrs. w/ Compile Only\xspace}
\newcommand{\fullitrsmasterco}{Full Itrs. w/ Master \& Compile Only\xspace}
\newcommand{\oneitr}{One Itr.\xspace}
\newcommand{\oneitrmaster}{One Itr. w/ Master\xspace}
\newcommand{\oneitrco}{One Itr. w/ Compile Only\xspace}
\newcommand{\oneitrmasterco}{One Itr. w/ Master \& Compile Only\xspace}

\newcommand{\OracleJDK}{Oracle JDK\xspace}
\newcommand{\OpenJDK}{OpenJDK\xspace}
\newcommand{\OpenJNine}{OpenJ9\xspace}

\DefMacro{oraclejdkdefault}{Default\xspace}
\DefMacro{oraclejdkgraal}{Graal\xspace}
\DefMacro{oraclejdklevel4}{Level 4\xspace}
\DefMacro{oraclejdklevel3}{Level 3\xspace}
\DefMacro{oraclejdklevel2}{Level 2\xspace}
\DefMacro{oraclejdklevel1}{Level 1\xspace}
\DefMacro{oraclejdklevel0}{Level 0\xspace}
\DefMacro{oraclejdkc1}{C1\xspace}
\DefMacro{oraclejdkc2}{C2\xspace}
\DefMacro{openjdkhotspotdefault}{\makecell{OpenJDK w/ \\ HotSpot Default\xspace}}
\DefMacro{openjdkhotspotgraal}{\makecell{OpenJDK w/ \\ HotSpot Graal\xspace}}
\DefMacro{openjdkhotspotlevel4}{\makecell{OpenJDK w/ \\ HotSpot Level 4\xspace}}
\DefMacro{openjdkhotspotlevel3}{\makecell{OpenJDK w/ \\ HotSpot Level 3\xspace}}
\DefMacro{openjdkhotspotlevel2}{\makecell{OpenJDK w/ \\ HotSpot Level 2\xspace}}
\DefMacro{openjdkhotspotlevel1}{\makecell{OpenJDK w/ \\ HotSpot Level 1\xspace}}
\DefMacro{openjdkhotspotlevel0}{\makecell{OpenJDK w/ \\ HotSpot Level 0\xspace}}
\DefMacro{openjdkhotspotc1}{\makecell{OpenJDK w/ \\ HotSpot C1\xspace}}
\DefMacro{openjdkhotspotc2}{\makecell{OpenJDK w/ \\ HotSpot C2\xspace}}
\DefMacro{openjdkopenj9default}{\makecell{OpenJDK w/ \\ OpenJ9 Default\xspace}}

\newcommand{\csmith}{Csmith\xspace}

\newcommand{\javafuzzer}{Java* Fuzzer\xspace}

\newcommand{\astgen}{ASTGen\xspace}
\newcommand{\udita}{UDITA\xspace}

\newcommand{\edsketch}{EdSketch\xspace}

\newcommand{\aTool}{jattack}
\newcommand{\aAttack}{Boom}
\newcommand{\aHole}[1]{\ensuremath{[\![}#1\ensuremath{]\!]}}
\newcommand{\aConfig}[1]{\ensuremath{\langle} #1 \ensuremath{\rangle}}

\let\oldding\ding%
\renewcommand{\ding}[2][1]{\scalebox{#1}{\oldding{#2}}}%
\newcommand{\cicrlenumber}[1]{\ding[1.1]{\number\numexpr181 + #1}}

\newcommand{\annotation}[2]{\cicrlenumber{#1}\label{#2}}
\newcommand{\highlightApi}[1]{{\color{blue}#1}}
\newcommand{\highlightEval}[1]{{\color{purple}\textbf{#1}}}
\colorlet{colorhole1}{red!20}
\colorlet{colorhole2}{green!20}
\colorlet{colorhole3}{yellow!20}
\colorlet{colorhole4}{teal!20}
\colorlet{colorhole5}{brown!20}

\DefMacro{table-jittack-javafuzzer-results-comparison-row0-tool}{\JTool}
\DefMacro{table-jittack-javafuzzer-results-comparison-row0-gen}{50609}
\DefMacro{table-jittack-javafuzzer-results-comparison-row0-timeout}{1243}
\DefMacro{table-jittack-javafuzzer-results-comparison-row0-fail}{137}
\DefMacro{table-jittack-javafuzzer-results-comparison-row0-coverage-c1}{84.3}
\DefMacro{table-jittack-javafuzzer-results-comparison-row0-coverage-c2}{80.3}
\DefMacro{table-jittack-javafuzzer-results-comparison-row1-tool}{\javafuzzer}
\DefMacro{table-jittack-javafuzzer-results-comparison-row1-gen}{15931}
\DefMacro{table-jittack-javafuzzer-results-comparison-row1-timeout}{2336}
\DefMacro{table-jittack-javafuzzer-results-comparison-row1-fail}{0}
\DefMacro{table-jittack-javafuzzer-results-comparison-row1-coverage-c1}{80.6}
\DefMacro{table-jittack-javafuzzer-results-comparison-row1-coverage-c2}{67.5}

\DefMacro{table-jittack-javafuzzer-results-comparison-float-env}{table}
\DefMacro{table-jittack-javafuzzer-results-comparison-fontsize}{footnotesize}
\DefMacro{table-jittack-javafuzzer-results-comparison-col-gen}{\#Generated}
\DefMacro{table-jittack-javafuzzer-results-comparison-col-timeout}{\#Timeout}
\DefMacro{table-jittack-javafuzzer-results-comparison-col-fail}{\#Failures}
\DefMacro{table-jittack-javafuzzer-results-comparison-col-coverage}{Coverage(\%)}
\DefMacro{table-jittack-javafuzzer-results-comparison-col-coverage-c1}{\COne}
\DefMacro{table-jittack-javafuzzer-results-comparison-col-coverage-c2}{\CTwo}
\DefMacro{table-jittack-javafuzzer-results-comparison-caption}{Comparing results of \JTool and \javafuzzer.\label{tab:comparing-results-of-jitattack-and-javafuzzer}\vspace{-5pt}}

\DefMacro{table-extraction-short-version-total-numProjects}{77}
\DefMacro{table-extraction-short-version-total-numFiles}{15{,}325}
\DefMacro{table-extraction-short-version-total-numMethods}{16{,}309}
\DefMacro{table-extraction-short-version-total-numTemplates}{5{,}419}
\DefMacro{table-extraction-short-version-total-numGen}{50{,}609}
\DefMacro{table-extraction-short-version-total-numFail}{137}

\DefMacro{CaptionTableExtractionResultsShortVersion}{\Sketchprogram
extraction from existing Java projects.
\UseMacro{TableExtractionResultsShortVersionHeadProjects}
is
projects; \UseMacro{TableExtractionResultsShortVersionHeadTemplates}
is \sketches; \UseMacro{TableExtractionResultsShortVersionHeadGen}
is \generatedprograms.
\label{tab:extraction-results-short-version}}
\DefMacro{TableExtractionResultsShortVersionHeadProjects}{Proj.}
\DefMacro{TableExtractionResultsShortVersionHeadNumProjects}{\# \UseMacro{TableExtractionResultsShortVersionHeadProjects}}
\DefMacro{TableExtractionResultsShortVersionHeadFiles}{Classes}
\DefMacro{TableExtractionResultsShortVersionHeadNumFiles}{\# \UseMacro{TableExtractionResultsShortVersionHeadFiles}}
\DefMacro{TableExtractionResultsShortVersionHeadMethods}{Methods}
\DefMacro{TableExtractionResultsShortVersionHeadNumMethods}{\# \UseMacro{TableExtractionResultsShortVersionHeadMethods}}
\DefMacro{TableExtractionResultsShortVersionHeadTemplates}{Tmpl.}
\DefMacro{TableExtractionResultsShortVersionHeadNumTemplates}{\# \UseMacro{TableExtractionResultsShortVersionHeadTemplates}}
\DefMacro{TableExtractionResultsShortVersionHeadGen}{Gen.}
\DefMacro{TableExtractionResultsShortVersionHeadNumGen}{\# \UseMacro{TableExtractionResultsShortVersionHeadGen}}
\DefMacro{TableExtractionResultsShortVersionHeadFail}{Failures}
\DefMacro{TableExtractionResultsShortVersionHeadNumFail}{\# \UseMacro{TableExtractionResultsShortVersionHeadFail}}

\DefMacro{table-extraction-checkstyle-name}{checkstyle}
\DefMacro{table-extraction-checkstyle-numTemplates}{194}
\DefMacro{table-extraction-checkstyle-numGen}{1{,}219}
\DefMacro{table-extraction-checkstyle-numFail}{23}
\DefMacro{table-extraction-codec-name}{commons-codec}
\DefMacro{table-extraction-codec-numTemplates}{145}
\DefMacro{table-extraction-codec-numGen}{1{,}339}
\DefMacro{table-extraction-codec-numFail}{45}
\DefMacro{table-extraction-compress-name}{commons-compress}
\DefMacro{table-extraction-compress-numTemplates}{140}
\DefMacro{table-extraction-compress-numGen}{1{,}350}
\DefMacro{table-extraction-compress-numFail}{0}
\DefMacro{table-extraction-configuration-name}{commons-configuration}
\DefMacro{table-extraction-configuration-numTemplates}{8}
\DefMacro{table-extraction-configuration-numGen}{58}
\DefMacro{table-extraction-configuration-numFail}{0}
\DefMacro{table-extraction-gson-name}{gson}
\DefMacro{table-extraction-gson-numTemplates}{11}
\DefMacro{table-extraction-gson-numGen}{98}
\DefMacro{table-extraction-gson-numFail}{0}
\DefMacro{table-extraction-jfreechart-name}{jfreechart}
\DefMacro{table-extraction-jfreechart-numTemplates}{158}
\DefMacro{table-extraction-jfreechart-numGen}{1{,}482}
\DefMacro{table-extraction-jfreechart-numFail}{1}
\DefMacro{table-extraction-jxpath-name}{commons-jxpath}
\DefMacro{table-extraction-jxpath-numTemplates}{37}
\DefMacro{table-extraction-jxpath-numGen}{355}
\DefMacro{table-extraction-jxpath-numFail}{0}
\DefMacro{table-extraction-lang-name}{commons-lang}
\DefMacro{table-extraction-lang-numTemplates}{733}
\DefMacro{table-extraction-lang-numGen}{7{,}036}
\DefMacro{table-extraction-lang-numFail}{0}
\DefMacro{table-extraction-libgdx-name}{libgdx}
\DefMacro{table-extraction-libgdx-numTemplates}{359}
\DefMacro{table-extraction-libgdx-numGen}{3{,}466}
\DefMacro{table-extraction-libgdx-numFail}{0}
\DefMacro{table-extraction-math-name}{commons-math}
\DefMacro{table-extraction-math-numTemplates}{564}
\DefMacro{table-extraction-math-numGen}{5{,}484}
\DefMacro{table-extraction-math-numFail}{9}
\DefMacro{table-extraction-openfire-name}{Openfire}
\DefMacro{table-extraction-openfire-numTemplates}{137}
\DefMacro{table-extraction-openfire-numGen}{1{,}030}
\DefMacro{table-extraction-openfire-numFail}{0}
\DefMacro{table-extraction-vectorz-name}{vectorz}
\DefMacro{table-extraction-vectorz-numTemplates}{398}
\DefMacro{table-extraction-vectorz-numGen}{3{,}622}
\DefMacro{table-extraction-vectorz-numFail}{0}
\DefMacro{table-extraction-zxing-name}{zxing}
\DefMacro{table-extraction-zxing-numTemplates}{240}
\DefMacro{table-extraction-zxing-numGen}{2{,}090}
\DefMacro{table-extraction-zxing-numFail}{2}
\DefMacro{table-extraction-bootique-name}{bootique}
\DefMacro{table-extraction-bootique-numTemplates}{6}
\DefMacro{table-extraction-bootique-numGen}{60}
\DefMacro{table-extraction-bootique-numFail}{0}
\DefMacro{table-extraction-dockermavenplugin-name}{docker-maven-plugin}
\DefMacro{table-extraction-dockermavenplugin-numTemplates}{88}
\DefMacro{table-extraction-dockermavenplugin-numGen}{861}
\DefMacro{table-extraction-dockermavenplugin-numFail}{0}
\DefMacro{table-extraction-easyrandom-name}{easy-random}
\DefMacro{table-extraction-easyrandom-numTemplates}{5}
\DefMacro{table-extraction-easyrandom-numGen}{50}
\DefMacro{table-extraction-easyrandom-numFail}{0}
\DefMacro{table-extraction-findsecbugs-name}{find-sec-bugs}
\DefMacro{table-extraction-findsecbugs-numTemplates}{207}
\DefMacro{table-extraction-findsecbugs-numGen}{2{,}070}
\DefMacro{table-extraction-findsecbugs-numFail}{0}
\DefMacro{table-extraction-gameserver-name}{game-server}
\DefMacro{table-extraction-gameserver-numTemplates}{112}
\DefMacro{table-extraction-gameserver-numGen}{1{,}075}
\DefMacro{table-extraction-gameserver-numFail}{2}
\DefMacro{table-extraction-graviteegateway-name}{gravitee-gateway}
\DefMacro{table-extraction-graviteegateway-numTemplates}{1}
\DefMacro{table-extraction-graviteegateway-numGen}{10}
\DefMacro{table-extraction-graviteegateway-numFail}{0}
\DefMacro{table-extraction-jupiter-name}{Jupiter}
\DefMacro{table-extraction-jupiter-numTemplates}{61}
\DefMacro{table-extraction-jupiter-numGen}{455}
\DefMacro{table-extraction-jupiter-numFail}{0}
\DefMacro{table-extraction-mango-name}{mango}
\DefMacro{table-extraction-mango-numTemplates}{38}
\DefMacro{table-extraction-mango-numGen}{347}
\DefMacro{table-extraction-mango-numFail}{0}
\DefMacro{table-extraction-mysqlperfanalyzer-name}{mysql\_perf\_analyzer}
\DefMacro{table-extraction-mysqlperfanalyzer-numTemplates}{1}
\DefMacro{table-extraction-mysqlperfanalyzer-numGen}{10}
\DefMacro{table-extraction-mysqlperfanalyzer-numFail}{0}
\DefMacro{table-extraction-springcloudconfig-name}{spring-cloud-config}
\DefMacro{table-extraction-springcloudconfig-numTemplates}{3}
\DefMacro{table-extraction-springcloudconfig-numGen}{16}
\DefMacro{table-extraction-springcloudconfig-numFail}{0}
\DefMacro{table-extraction-streamex-name}{streamex}
\DefMacro{table-extraction-streamex-numTemplates}{6}
\DefMacro{table-extraction-streamex-numGen}{60}
\DefMacro{table-extraction-streamex-numFail}{0}
\DefMacro{table-extraction-ta4j-name}{ta4j}
\DefMacro{table-extraction-ta4j-numTemplates}{22}
\DefMacro{table-extraction-ta4j-numGen}{220}
\DefMacro{table-extraction-ta4j-numFail}{0}
\DefMacro{table-extraction-whatsmars-name}{whatsmars}
\DefMacro{table-extraction-whatsmars-numTemplates}{58}
\DefMacro{table-extraction-whatsmars-numGen}{575}
\DefMacro{table-extraction-whatsmars-numFail}{0}
\DefMacro{table-extraction-aviatorscript-name}{aviatorscript}
\DefMacro{table-extraction-aviatorscript-numTemplates}{39}
\DefMacro{table-extraction-aviatorscript-numGen}{354}
\DefMacro{table-extraction-aviatorscript-numFail}{0}
\DefMacro{table-extraction-bytecodeviewer-name}{bytecode-viewer}
\DefMacro{table-extraction-bytecodeviewer-numTemplates}{52}
\DefMacro{table-extraction-bytecodeviewer-numGen}{442}
\DefMacro{table-extraction-bytecodeviewer-numFail}{0}
\DefMacro{table-extraction-chroniclequeue-name}{Chronicle-Queue}
\DefMacro{table-extraction-chroniclequeue-numTemplates}{37}
\DefMacro{table-extraction-chroniclequeue-numGen}{362}
\DefMacro{table-extraction-chroniclequeue-numFail}{0}
\DefMacro{table-extraction-dockerjava-name}{docker-java}
\DefMacro{table-extraction-dockerjava-numTemplates}{4}
\DefMacro{table-extraction-dockerjava-numGen}{40}
\DefMacro{table-extraction-dockerjava-numFail}{0}
\DefMacro{table-extraction-flyway-name}{flyway}
\DefMacro{table-extraction-flyway-numTemplates}{47}
\DefMacro{table-extraction-flyway-numGen}{428}
\DefMacro{table-extraction-flyway-numFail}{0}
\DefMacro{table-extraction-javacpp-name}{javacpp}
\DefMacro{table-extraction-javacpp-numTemplates}{10}
\DefMacro{table-extraction-javacpp-numGen}{100}
\DefMacro{table-extraction-javacpp-numFail}{0}
\DefMacro{table-extraction-jsonschema2pojo-name}{jsonschema2pojo}
\DefMacro{table-extraction-jsonschema2pojo-numTemplates}{1}
\DefMacro{table-extraction-jsonschema2pojo-numGen}{10}
\DefMacro{table-extraction-jsonschema2pojo-numFail}{0}
\DefMacro{table-extraction-jsoup-name}{jsoup}
\DefMacro{table-extraction-jsoup-numTemplates}{79}
\DefMacro{table-extraction-jsoup-numGen}{790}
\DefMacro{table-extraction-jsoup-numFail}{0}
\DefMacro{table-extraction-jsqlparser-name}{JSqlParser}
\DefMacro{table-extraction-jsqlparser-numTemplates}{12}
\DefMacro{table-extraction-jsqlparser-numGen}{120}
\DefMacro{table-extraction-jsqlparser-numFail}{0}
\DefMacro{table-extraction-metrics-name}{metrics}
\DefMacro{table-extraction-metrics-numTemplates}{0}
\DefMacro{table-extraction-metrics-numGen}{0}
\DefMacro{table-extraction-metrics-numFail}{0}
\DefMacro{table-extraction-recaf-name}{Recaf}
\DefMacro{table-extraction-recaf-numTemplates}{187}
\DefMacro{table-extraction-recaf-numGen}{1{,}816}
\DefMacro{table-extraction-recaf-numFail}{2}
\DefMacro{table-extraction-ripme-name}{ripme}
\DefMacro{table-extraction-ripme-numTemplates}{29}
\DefMacro{table-extraction-ripme-numGen}{238}
\DefMacro{table-extraction-ripme-numFail}{38}
\DefMacro{table-extraction-springcloudgateway-name}{spring-cloud-gateway}
\DefMacro{table-extraction-springcloudgateway-numTemplates}{16}
\DefMacro{table-extraction-springcloudgateway-numGen}{160}
\DefMacro{table-extraction-springcloudgateway-numFail}{0}
\DefMacro{table-extraction-springdatajpa-name}{spring-data-jpa}
\DefMacro{table-extraction-springdatajpa-numTemplates}{5}
\DefMacro{table-extraction-springdatajpa-numGen}{44}
\DefMacro{table-extraction-springdatajpa-numFail}{0}
\DefMacro{table-extraction-wepush-name}{WePush}
\DefMacro{table-extraction-wepush-numTemplates}{26}
\DefMacro{table-extraction-wepush-numGen}{251}
\DefMacro{table-extraction-wepush-numFail}{0}
\DefMacro{table-extraction-bytebuddy-name}{byte-buddy}
\DefMacro{table-extraction-bytebuddy-numTemplates}{36}
\DefMacro{table-extraction-bytebuddy-numGen}{359}
\DefMacro{table-extraction-bytebuddy-numFail}{0}
\DefMacro{table-extraction-commonsbcel-name}{commons-bcel}
\DefMacro{table-extraction-commonsbcel-numTemplates}{57}
\DefMacro{table-extraction-commonsbcel-numGen}{570}
\DefMacro{table-extraction-commonsbcel-numFail}{0}
\DefMacro{table-extraction-digester-name}{commons-digester}
\DefMacro{table-extraction-digester-numTemplates}{2}
\DefMacro{table-extraction-digester-numGen}{20}
\DefMacro{table-extraction-digester-numFail}{0}
\DefMacro{table-extraction-dnsjava-name}{dnsjava}
\DefMacro{table-extraction-dnsjava-numTemplates}{85}
\DefMacro{table-extraction-dnsjava-numGen}{831}
\DefMacro{table-extraction-dnsjava-numFail}{0}
\DefMacro{table-extraction-easybatch-name}{easy-batch}
\DefMacro{table-extraction-easybatch-numTemplates}{4}
\DefMacro{table-extraction-easybatch-numGen}{40}
\DefMacro{table-extraction-easybatch-numFail}{0}
\DefMacro{table-extraction-equalsverifier-name}{equalsverifier}
\DefMacro{table-extraction-equalsverifier-numTemplates}{4}
\DefMacro{table-extraction-equalsverifier-numGen}{10}
\DefMacro{table-extraction-equalsverifier-numFail}{0}
\DefMacro{table-extraction-graphqlspqr-name}{graphql-spqr}
\DefMacro{table-extraction-graphqlspqr-numTemplates}{7}
\DefMacro{table-extraction-graphqlspqr-numGen}{70}
\DefMacro{table-extraction-graphqlspqr-numFail}{0}
\DefMacro{table-extraction-jamesmime4j-name}{james-mime4j}
\DefMacro{table-extraction-jamesmime4j-numTemplates}{21}
\DefMacro{table-extraction-jamesmime4j-numGen}{204}
\DefMacro{table-extraction-jamesmime4j-numFail}{14}
\DefMacro{table-extraction-javassist-name}{javassist}
\DefMacro{table-extraction-javassist-numTemplates}{224}
\DefMacro{table-extraction-javassist-numGen}{2{,}227}
\DefMacro{table-extraction-javassist-numFail}{0}
\DefMacro{table-extraction-jsqlinjection-name}{jsql-injection}
\DefMacro{table-extraction-jsqlinjection-numTemplates}{8}
\DefMacro{table-extraction-jsqlinjection-numGen}{80}
\DefMacro{table-extraction-jsqlinjection-numFail}{0}
\DefMacro{table-extraction-karafcellar-name}{karaf-cellar}
\DefMacro{table-extraction-karafcellar-numTemplates}{0}
\DefMacro{table-extraction-karafcellar-numGen}{0}
\DefMacro{table-extraction-karafcellar-numFail}{0}
\DefMacro{table-extraction-lambda-name}{lambda}
\DefMacro{table-extraction-lambda-numTemplates}{2}
\DefMacro{table-extraction-lambda-numGen}{20}
\DefMacro{table-extraction-lambda-numFail}{0}
\DefMacro{table-extraction-mug-name}{mug}
\DefMacro{table-extraction-mug-numTemplates}{9}
\DefMacro{table-extraction-mug-numGen}{90}
\DefMacro{table-extraction-mug-numFail}{0}
\DefMacro{table-extraction-ormlitecore-name}{ormlite-core}
\DefMacro{table-extraction-ormlitecore-numTemplates}{8}
\DefMacro{table-extraction-ormlitecore-numGen}{64}
\DefMacro{table-extraction-ormlitecore-numFail}{0}
\DefMacro{table-extraction-pebble-name}{pebble}
\DefMacro{table-extraction-pebble-numTemplates}{10}
\DefMacro{table-extraction-pebble-numGen}{100}
\DefMacro{table-extraction-pebble-numFail}{0}
\DefMacro{table-extraction-simplejavamail-name}{simple-java-mail}
\DefMacro{table-extraction-simplejavamail-numTemplates}{3}
\DefMacro{table-extraction-simplejavamail-numGen}{13}
\DefMacro{table-extraction-simplejavamail-numFail}{0}
\DefMacro{table-extraction-springcloudaws-name}{spring-cloud-aws}
\DefMacro{table-extraction-springcloudaws-numTemplates}{1}
\DefMacro{table-extraction-springcloudaws-numGen}{10}
\DefMacro{table-extraction-springcloudaws-numFail}{0}
\DefMacro{table-extraction-springcloudcommons-name}{spring-cloud-commons}
\DefMacro{table-extraction-springcloudcommons-numTemplates}{1}
\DefMacro{table-extraction-springcloudcommons-numGen}{10}
\DefMacro{table-extraction-springcloudcommons-numFail}{0}
\DefMacro{table-extraction-springdataneo4j-name}{spring-data-neo4j}
\DefMacro{table-extraction-springdataneo4j-numTemplates}{7}
\DefMacro{table-extraction-springdataneo4j-numGen}{70}
\DefMacro{table-extraction-springdataneo4j-numFail}{0}
\DefMacro{table-extraction-springdatarest-name}{spring-data-rest}
\DefMacro{table-extraction-springdatarest-numTemplates}{1}
\DefMacro{table-extraction-springdatarest-numGen}{10}
\DefMacro{table-extraction-springdatarest-numFail}{0}
\DefMacro{table-extraction-typescriptgenerator-name}{typescript-generator}
\DefMacro{table-extraction-typescriptgenerator-numTemplates}{4}
\DefMacro{table-extraction-typescriptgenerator-numGen}{24}
\DefMacro{table-extraction-typescriptgenerator-numFail}{0}
\DefMacro{table-extraction-uimauimafit-name}{uima-uimafit}
\DefMacro{table-extraction-uimauimafit-numTemplates}{61}
\DefMacro{table-extraction-uimauimafit-numGen}{603}
\DefMacro{table-extraction-uimauimafit-numFail}{0}
\DefMacro{table-extraction-carina-name}{carina}
\DefMacro{table-extraction-carina-numTemplates}{41}
\DefMacro{table-extraction-carina-numGen}{306}
\DefMacro{table-extraction-carina-numFail}{0}
\DefMacro{table-extraction-fluo-name}{fluo}
\DefMacro{table-extraction-fluo-numTemplates}{22}
\DefMacro{table-extraction-fluo-numGen}{212}
\DefMacro{table-extraction-fluo-numFail}{0}
\DefMacro{table-extraction-jexl-name}{commons-jexl}
\DefMacro{table-extraction-jexl-numTemplates}{36}
\DefMacro{table-extraction-jexl-numGen}{360}
\DefMacro{table-extraction-jexl-numFail}{0}
\DefMacro{table-extraction-mavenassemblyplugin-name}{maven-assembly-plugin}
\DefMacro{table-extraction-mavenassemblyplugin-numTemplates}{5}
\DefMacro{table-extraction-mavenassemblyplugin-numGen}{50}
\DefMacro{table-extraction-mavenassemblyplugin-numFail}{0}
\DefMacro{table-extraction-mavenindexer-name}{maven-indexer}
\DefMacro{table-extraction-mavenindexer-numTemplates}{7}
\DefMacro{table-extraction-mavenindexer-numGen}{52}
\DefMacro{table-extraction-mavenindexer-numFail}{0}
\DefMacro{table-extraction-mavenwagon-name}{maven-wagon}
\DefMacro{table-extraction-mavenwagon-numTemplates}{4}
\DefMacro{table-extraction-mavenwagon-numGen}{40}
\DefMacro{table-extraction-mavenwagon-numFail}{0}
\DefMacro{table-extraction-numbers-name}{commons-numbers}
\DefMacro{table-extraction-numbers-numTemplates}{42}
\DefMacro{table-extraction-numbers-numGen}{415}
\DefMacro{table-extraction-numbers-numFail}{0}
\DefMacro{table-extraction-ognl-name}{commons-ognl}
\DefMacro{table-extraction-ognl-numTemplates}{2}
\DefMacro{table-extraction-ognl-numGen}{4}
\DefMacro{table-extraction-ognl-numFail}{0}
\DefMacro{table-extraction-onenio-name}{one-nio}
\DefMacro{table-extraction-onenio-numTemplates}{298}
\DefMacro{table-extraction-onenio-numGen}{2{,}899}
\DefMacro{table-extraction-onenio-numFail}{0}
\DefMacro{table-extraction-phoenicis-name}{phoenicis}
\DefMacro{table-extraction-phoenicis-numTemplates}{1}
\DefMacro{table-extraction-phoenicis-numGen}{10}
\DefMacro{table-extraction-phoenicis-numFail}{0}
\DefMacro{table-extraction-text-name}{commons-text}
\DefMacro{table-extraction-text-numTemplates}{80}
\DefMacro{table-extraction-text-numGen}{798}
\DefMacro{table-extraction-text-numFail}{1}
\DefMacro{table-extraction-turbine-name}{turbine}
\DefMacro{table-extraction-turbine-numTemplates}{21}
\DefMacro{table-extraction-turbine-numGen}{204}
\DefMacro{table-extraction-turbine-numFail}{0}
\DefMacro{table-extraction-validator-name}{commons-validator}
\DefMacro{table-extraction-validator-numTemplates}{22}
\DefMacro{table-extraction-validator-numGen}{195}
\DefMacro{table-extraction-validator-numFail}{0}
\DefMacro{table-extraction-velocitytools-name}{velocity-tools}
\DefMacro{table-extraction-velocitytools-numTemplates}{7}
\DefMacro{table-extraction-velocitytools-numGen}{46}
\DefMacro{table-extraction-velocitytools-numFail}{0}
\DefMacro{table-extraction-total-numTemplates}{5{,}419}
\DefMacro{table-extraction-total-numGen}{50{,}609}
\DefMacro{table-extraction-total-numFail}{137}

\DefMacro{FontsizeTableExtractionResults}{tiny}
\DefMacro{FloatEnvTableExtractionResults}{table*}
\DefMacro{CaptionTableExtractionResults}{Results of \sketchprogram
extraction from existing Java projects, including the number of
\sketches, the total number of \generatedprograms and the number of
\generatedprograms that failed \JIT-testing per project.\label{tab:extraction-results}}
\DefMacro{TableExtractionResultsHeadProject}{Project}
\DefMacro{TableExtractionResultsHeadNumTemplates}{\# Templates}
\DefMacro{TableExtractionResultsHeadNumGen}{\# Generated}
\DefMacro{TableExtractionResultsHeadNumFail}{\# Failures}
\DefMacro{TableExtractionResultsCellSum}{$\Sigma$}

\DefMacro{table-timing-gen-SkJDK8243670-name}{B1}
\DefMacro{table-timing-gen-SkJDK8243670-genTime-non_opt}{05:00:36}
\DefMacro{table-timing-gen-SkJDK8243670-genTime-stop_early_opt}{00:14}
\DefMacro{table-timing-gen-SkJDK8243670-reduction-stop_early_opt}{99.92}
\DefMacro{table-timing-gen-SkJDK8243670-genTime-hot_filling_opt}{01:31}
\DefMacro{table-timing-gen-SkJDK8243670-reduction-hot_filling_opt}{99.50}
\DefMacro{table-timing-gen-SkJDK8243670-genTime-solver_aid_opt}{02:55:07}
\DefMacro{table-timing-gen-SkJDK8243670-reduction-solver_aid_opt}{41.74}
\DefMacro{table-timing-gen-SkJDK8243670-genTime-full_opt}{00:56}
\DefMacro{table-timing-gen-SkJDK8243670-reduction-full_opt}{99.69}
\DefMacro{table-timing-gen-SkJDK8248552-name}{B2}
\DefMacro{table-timing-gen-SkJDK8248552-genTime-non_opt}{12:51:25}
\DefMacro{table-timing-gen-SkJDK8248552-genTime-stop_early_opt}{00:14}
\DefMacro{table-timing-gen-SkJDK8248552-reduction-stop_early_opt}{99.97}
\DefMacro{table-timing-gen-SkJDK8248552-genTime-hot_filling_opt}{04:14}
\DefMacro{table-timing-gen-SkJDK8248552-reduction-hot_filling_opt}{99.45}
\DefMacro{table-timing-gen-SkJDK8248552-genTime-solver_aid_opt}{12:53:50}
\DefMacro{table-timing-gen-SkJDK8248552-reduction-solver_aid_opt}{-0.31}
\DefMacro{table-timing-gen-SkJDK8248552-genTime-full_opt}{01:03}
\DefMacro{table-timing-gen-SkJDK8248552-reduction-full_opt}{99.86}
\DefMacro{table-timing-gen-SkJDK8251535-name}{B3}
\DefMacro{table-timing-gen-SkJDK8251535-genTime-non_opt}{01:19}
\DefMacro{table-timing-gen-SkJDK8251535-genTime-stop_early_opt}{00:12}
\DefMacro{table-timing-gen-SkJDK8251535-reduction-stop_early_opt}{84.33}
\DefMacro{table-timing-gen-SkJDK8251535-genTime-hot_filling_opt}{00:33}
\DefMacro{table-timing-gen-SkJDK8251535-reduction-hot_filling_opt}{58.57}
\DefMacro{table-timing-gen-SkJDK8251535-genTime-solver_aid_opt}{02:01}
\DefMacro{table-timing-gen-SkJDK8251535-reduction-solver_aid_opt}{-53.59}
\DefMacro{table-timing-gen-SkJDK8251535-genTime-full_opt}{00:13}
\DefMacro{table-timing-gen-SkJDK8251535-reduction-full_opt}{84.08}
\DefMacro{table-timing-gen-SkJDK8255763-name}{B4}
\DefMacro{table-timing-gen-SkJDK8255763-genTime-non_opt}{01:43}
\DefMacro{table-timing-gen-SkJDK8255763-genTime-stop_early_opt}{01:41}
\DefMacro{table-timing-gen-SkJDK8255763-reduction-stop_early_opt}{2.37}
\DefMacro{table-timing-gen-SkJDK8255763-genTime-hot_filling_opt}{01:41}
\DefMacro{table-timing-gen-SkJDK8255763-reduction-hot_filling_opt}{2.03}
\DefMacro{table-timing-gen-SkJDK8255763-genTime-solver_aid_opt}{01:41}
\DefMacro{table-timing-gen-SkJDK8255763-reduction-solver_aid_opt}{2.45}
\DefMacro{table-timing-gen-SkJDK8255763-genTime-full_opt}{01:43}
\DefMacro{table-timing-gen-SkJDK8255763-reduction-full_opt}{0.29}
\DefMacro{table-timing-gen-Sk1-name}{M1}
\DefMacro{table-timing-gen-Sk1-genTime-non_opt}{11:32}
\DefMacro{table-timing-gen-Sk1-genTime-stop_early_opt}{00:10}
\DefMacro{table-timing-gen-Sk1-reduction-stop_early_opt}{98.60}
\DefMacro{table-timing-gen-Sk1-genTime-hot_filling_opt}{00:42}
\DefMacro{table-timing-gen-Sk1-reduction-hot_filling_opt}{93.96}
\DefMacro{table-timing-gen-Sk1-genTime-solver_aid_opt}{10:17}
\DefMacro{table-timing-gen-Sk1-reduction-solver_aid_opt}{10.82}
\DefMacro{table-timing-gen-Sk1-genTime-full_opt}{00:33}
\DefMacro{table-timing-gen-Sk1-reduction-full_opt}{95.22}
\DefMacro{table-timing-gen-Sk2-name}{M2}
\DefMacro{table-timing-gen-Sk2-genTime-non_opt}{12:57}
\DefMacro{table-timing-gen-Sk2-genTime-stop_early_opt}{00:12}
\DefMacro{table-timing-gen-Sk2-reduction-stop_early_opt}{98.49}
\DefMacro{table-timing-gen-Sk2-genTime-hot_filling_opt}{00:44}
\DefMacro{table-timing-gen-Sk2-reduction-hot_filling_opt}{94.29}
\DefMacro{table-timing-gen-Sk2-genTime-solver_aid_opt}{14:40}
\DefMacro{table-timing-gen-Sk2-reduction-solver_aid_opt}{-13.22}
\DefMacro{table-timing-gen-Sk2-genTime-full_opt}{00:34}
\DefMacro{table-timing-gen-Sk2-reduction-full_opt}{95.59}
\DefMacro{table-timing-gen-Sk3-name}{M3}
\DefMacro{table-timing-gen-Sk3-genTime-non_opt}{11:00}
\DefMacro{table-timing-gen-Sk3-genTime-stop_early_opt}{02:19}
\DefMacro{table-timing-gen-Sk3-reduction-stop_early_opt}{78.90}
\DefMacro{table-timing-gen-Sk3-genTime-hot_filling_opt}{01:05}
\DefMacro{table-timing-gen-Sk3-reduction-hot_filling_opt}{90.14}
\DefMacro{table-timing-gen-Sk3-genTime-solver_aid_opt}{13:04}
\DefMacro{table-timing-gen-Sk3-reduction-solver_aid_opt}{-18.88}
\DefMacro{table-timing-gen-Sk3-genTime-full_opt}{00:56}
\DefMacro{table-timing-gen-Sk3-reduction-full_opt}{91.49}
\DefMacro{table-timing-gen-Sk4-name}{M4}
\DefMacro{table-timing-gen-Sk4-genTime-non_opt}{19:44}
\DefMacro{table-timing-gen-Sk4-genTime-stop_early_opt}{04:31}
\DefMacro{table-timing-gen-Sk4-reduction-stop_early_opt}{77.12}
\DefMacro{table-timing-gen-Sk4-genTime-hot_filling_opt}{00:28}
\DefMacro{table-timing-gen-Sk4-reduction-hot_filling_opt}{97.60}
\DefMacro{table-timing-gen-Sk4-genTime-solver_aid_opt}{20:35}
\DefMacro{table-timing-gen-Sk4-reduction-solver_aid_opt}{-4.26}
\DefMacro{table-timing-gen-Sk4-genTime-full_opt}{00:31}
\DefMacro{table-timing-gen-Sk4-reduction-full_opt}{97.39}
\DefMacro{table-timing-gen-Sk5-name}{M5}
\DefMacro{table-timing-gen-Sk5-genTime-non_opt}{05:14:14}
\DefMacro{table-timing-gen-Sk5-genTime-stop_early_opt}{46:01}
\DefMacro{table-timing-gen-Sk5-reduction-stop_early_opt}{85.36}
\DefMacro{table-timing-gen-Sk5-genTime-hot_filling_opt}{01:04}
\DefMacro{table-timing-gen-Sk5-reduction-hot_filling_opt}{99.66}
\DefMacro{table-timing-gen-Sk5-genTime-solver_aid_opt}{04:20:07}
\DefMacro{table-timing-gen-Sk5-reduction-solver_aid_opt}{17.22}
\DefMacro{table-timing-gen-Sk5-genTime-full_opt}{00:46}
\DefMacro{table-timing-gen-Sk5-reduction-full_opt}{99.76}
\DefMacro{table-timing-gen-Sk6-name}{M6}
\DefMacro{table-timing-gen-Sk6-genTime-non_opt}{03:05}
\DefMacro{table-timing-gen-Sk6-genTime-stop_early_opt}{00:09}
\DefMacro{table-timing-gen-Sk6-reduction-stop_early_opt}{95.08}
\DefMacro{table-timing-gen-Sk6-genTime-hot_filling_opt}{00:45}
\DefMacro{table-timing-gen-Sk6-reduction-hot_filling_opt}{75.79}
\DefMacro{table-timing-gen-Sk6-genTime-solver_aid_opt}{04:21}
\DefMacro{table-timing-gen-Sk6-reduction-solver_aid_opt}{-41.40}
\DefMacro{table-timing-gen-Sk6-genTime-full_opt}{00:38}
\DefMacro{table-timing-gen-Sk6-reduction-full_opt}{79.43}
\DefMacro{table-timing-gen-Sk7-name}{M7}
\DefMacro{table-timing-gen-Sk7-genTime-non_opt}{05:24}
\DefMacro{table-timing-gen-Sk7-genTime-stop_early_opt}{05:31}
\DefMacro{table-timing-gen-Sk7-reduction-stop_early_opt}{-2.40}
\DefMacro{table-timing-gen-Sk7-genTime-hot_filling_opt}{00:38}
\DefMacro{table-timing-gen-Sk7-reduction-hot_filling_opt}{88.19}
\DefMacro{table-timing-gen-Sk7-genTime-solver_aid_opt}{05:41}
\DefMacro{table-timing-gen-Sk7-reduction-solver_aid_opt}{-5.39}
\DefMacro{table-timing-gen-Sk7-genTime-full_opt}{00:49}
\DefMacro{table-timing-gen-Sk7-reduction-full_opt}{84.74}
\DefMacro{table-timing-gen-Sk8-name}{M8}
\DefMacro{table-timing-gen-Sk8-genTime-non_opt}{19:19}
\DefMacro{table-timing-gen-Sk8-genTime-stop_early_opt}{12:26}
\DefMacro{table-timing-gen-Sk8-reduction-stop_early_opt}{35.62}
\DefMacro{table-timing-gen-Sk8-genTime-hot_filling_opt}{01:08}
\DefMacro{table-timing-gen-Sk8-reduction-hot_filling_opt}{94.14}
\DefMacro{table-timing-gen-Sk8-genTime-solver_aid_opt}{21:10}
\DefMacro{table-timing-gen-Sk8-reduction-solver_aid_opt}{-9.62}
\DefMacro{table-timing-gen-Sk8-genTime-full_opt}{01:16}
\DefMacro{table-timing-gen-Sk8-reduction-full_opt}{93.43}
\DefMacro{table-timing-gen-Sk9-name}{M9}
\DefMacro{table-timing-gen-Sk9-genTime-non_opt}{02:25}
\DefMacro{table-timing-gen-Sk9-genTime-stop_early_opt}{00:14}
\DefMacro{table-timing-gen-Sk9-reduction-stop_early_opt}{90.42}
\DefMacro{table-timing-gen-Sk9-genTime-hot_filling_opt}{00:28}
\DefMacro{table-timing-gen-Sk9-reduction-hot_filling_opt}{80.59}
\DefMacro{table-timing-gen-Sk9-genTime-solver_aid_opt}{02:49}
\DefMacro{table-timing-gen-Sk9-reduction-solver_aid_opt}{-16.18}
\DefMacro{table-timing-gen-Sk9-genTime-full_opt}{00:20}
\DefMacro{table-timing-gen-Sk9-reduction-full_opt}{85.98}
\DefMacro{table-timing-gen-Sk10-name}{M10}
\DefMacro{table-timing-gen-Sk10-genTime-non_opt}{09:07}
\DefMacro{table-timing-gen-Sk10-genTime-stop_early_opt}{02:55}
\DefMacro{table-timing-gen-Sk10-reduction-stop_early_opt}{67.96}
\DefMacro{table-timing-gen-Sk10-genTime-hot_filling_opt}{00:41}
\DefMacro{table-timing-gen-Sk10-reduction-hot_filling_opt}{92.50}
\DefMacro{table-timing-gen-Sk10-genTime-solver_aid_opt}{10:32}
\DefMacro{table-timing-gen-Sk10-reduction-solver_aid_opt}{-15.63}
\DefMacro{table-timing-gen-Sk10-genTime-full_opt}{00:39}
\DefMacro{table-timing-gen-Sk10-reduction-full_opt}{92.90}
\DefMacro{table-timing-gen-Sk11-name}{M11}
\DefMacro{table-timing-gen-Sk11-genTime-non_opt}{10:58}
\DefMacro{table-timing-gen-Sk11-genTime-stop_early_opt}{00:10}
\DefMacro{table-timing-gen-Sk11-reduction-stop_early_opt}{98.48}
\DefMacro{table-timing-gen-Sk11-genTime-hot_filling_opt}{02:15}
\DefMacro{table-timing-gen-Sk11-reduction-hot_filling_opt}{79.44}
\DefMacro{table-timing-gen-Sk11-genTime-solver_aid_opt}{08:35}
\DefMacro{table-timing-gen-Sk11-reduction-solver_aid_opt}{21.79}
\DefMacro{table-timing-gen-Sk11-genTime-full_opt}{01:02}
\DefMacro{table-timing-gen-Sk11-reduction-full_opt}{90.62}
\DefMacro{table-timing-gen-Sk12-name}{M12}
\DefMacro{table-timing-gen-Sk12-genTime-non_opt}{04:23}
\DefMacro{table-timing-gen-Sk12-genTime-stop_early_opt}{04:40}
\DefMacro{table-timing-gen-Sk12-reduction-stop_early_opt}{-6.38}
\DefMacro{table-timing-gen-Sk12-genTime-hot_filling_opt}{00:36}
\DefMacro{table-timing-gen-Sk12-reduction-hot_filling_opt}{86.51}
\DefMacro{table-timing-gen-Sk12-genTime-solver_aid_opt}{05:24}
\DefMacro{table-timing-gen-Sk12-reduction-solver_aid_opt}{-22.88}
\DefMacro{table-timing-gen-Sk12-genTime-full_opt}{00:43}
\DefMacro{table-timing-gen-Sk12-reduction-full_opt}{83.64}
\DefMacro{table-timing-gen-Sk55-name}{M13}
\DefMacro{table-timing-gen-Sk55-genTime-non_opt}{11:35:19}
\DefMacro{table-timing-gen-Sk55-genTime-stop_early_opt}{11:26:37}
\DefMacro{table-timing-gen-Sk55-reduction-stop_early_opt}{1.25}
\DefMacro{table-timing-gen-Sk55-genTime-hot_filling_opt}{01:06}
\DefMacro{table-timing-gen-Sk55-reduction-hot_filling_opt}{99.84}
\DefMacro{table-timing-gen-Sk55-genTime-solver_aid_opt}{00:57}
\DefMacro{table-timing-gen-Sk55-reduction-solver_aid_opt}{99.86}
\DefMacro{table-timing-gen-Sk55-genTime-full_opt}{00:51}
\DefMacro{table-timing-gen-Sk55-reduction-full_opt}{99.88}
\DefMacro{table-timing-gen-SkOptLockCoarsening-name}{M14}
\DefMacro{table-timing-gen-SkOptLockCoarsening-genTime-non_opt}{11:40:46}
\DefMacro{table-timing-gen-SkOptLockCoarsening-genTime-stop_early_opt}{05:43:54}
\DefMacro{table-timing-gen-SkOptLockCoarsening-reduction-stop_early_opt}{50.93}
\DefMacro{table-timing-gen-SkOptLockCoarsening-genTime-hot_filling_opt}{01:38}
\DefMacro{table-timing-gen-SkOptLockCoarsening-reduction-hot_filling_opt}{99.77}
\DefMacro{table-timing-gen-SkOptLockCoarsening-genTime-solver_aid_opt}{11:58:25}
\DefMacro{table-timing-gen-SkOptLockCoarsening-reduction-solver_aid_opt}{-2.52}
\DefMacro{table-timing-gen-SkOptLockCoarsening-genTime-full_opt}{01:17}
\DefMacro{table-timing-gen-SkOptLockCoarsening-reduction-full_opt}{99.82}
\DefMacro{table-timing-gen-SkOptLoopPeeling-name}{M15}
\DefMacro{table-timing-gen-SkOptLoopPeeling-genTime-non_opt}{03:55:35}
\DefMacro{table-timing-gen-SkOptLoopPeeling-genTime-stop_early_opt}{00:09}
\DefMacro{table-timing-gen-SkOptLoopPeeling-reduction-stop_early_opt}{99.94}
\DefMacro{table-timing-gen-SkOptLoopPeeling-genTime-hot_filling_opt}{00:45}
\DefMacro{table-timing-gen-SkOptLoopPeeling-reduction-hot_filling_opt}{99.68}
\DefMacro{table-timing-gen-SkOptLoopPeeling-genTime-solver_aid_opt}{02:49:16}
\DefMacro{table-timing-gen-SkOptLoopPeeling-reduction-solver_aid_opt}{28.15}
\DefMacro{table-timing-gen-SkOptLoopPeeling-genTime-full_opt}{00:14}
\DefMacro{table-timing-gen-SkOptLoopPeeling-reduction-full_opt}{99.90}
\DefMacro{table-timing-gen-SkOptLoopUnrolling-name}{M16}
\DefMacro{table-timing-gen-SkOptLoopUnrolling-genTime-non_opt}{07:38:42}
\DefMacro{table-timing-gen-SkOptLoopUnrolling-genTime-stop_early_opt}{07:47:57}
\DefMacro{table-timing-gen-SkOptLoopUnrolling-reduction-stop_early_opt}{-2.02}
\DefMacro{table-timing-gen-SkOptLoopUnrolling-genTime-hot_filling_opt}{01:02}
\DefMacro{table-timing-gen-SkOptLoopUnrolling-reduction-hot_filling_opt}{99.77}
\DefMacro{table-timing-gen-SkOptLoopUnrolling-genTime-solver_aid_opt}{07:57:36}
\DefMacro{table-timing-gen-SkOptLoopUnrolling-reduction-solver_aid_opt}{-4.12}
\DefMacro{table-timing-gen-SkOptLoopUnrolling-genTime-full_opt}{01:09}
\DefMacro{table-timing-gen-SkOptLoopUnrolling-reduction-full_opt}{99.75}
\DefMacro{table-timing-gen-SkOptLoopUnswitching-name}{M17}
\DefMacro{table-timing-gen-SkOptLoopUnswitching-genTime-non_opt}{10:38:24}
\DefMacro{table-timing-gen-SkOptLoopUnswitching-genTime-stop_early_opt}{11:28:05}
\DefMacro{table-timing-gen-SkOptLoopUnswitching-reduction-stop_early_opt}{-7.78}
\DefMacro{table-timing-gen-SkOptLoopUnswitching-genTime-hot_filling_opt}{03:12}
\DefMacro{table-timing-gen-SkOptLoopUnswitching-reduction-hot_filling_opt}{99.50}
\DefMacro{table-timing-gen-SkOptLoopUnswitching-genTime-solver_aid_opt}{10:51:56}
\DefMacro{table-timing-gen-SkOptLoopUnswitching-reduction-solver_aid_opt}{-2.12}
\DefMacro{table-timing-gen-SkOptLoopUnswitching-genTime-full_opt}{03:32}
\DefMacro{table-timing-gen-SkOptLoopUnswitching-reduction-full_opt}{99.45}
\DefMacro{table-timing-gen-SkOptRangeCheckElimination-name}{M18}
\DefMacro{table-timing-gen-SkOptRangeCheckElimination-genTime-non_opt}{04:57}
\DefMacro{table-timing-gen-SkOptRangeCheckElimination-genTime-stop_early_opt}{00:09}
\DefMacro{table-timing-gen-SkOptRangeCheckElimination-reduction-stop_early_opt}{96.81}
\DefMacro{table-timing-gen-SkOptRangeCheckElimination-genTime-hot_filling_opt}{00:41}
\DefMacro{table-timing-gen-SkOptRangeCheckElimination-reduction-hot_filling_opt}{86.07}
\DefMacro{table-timing-gen-SkOptRangeCheckElimination-genTime-solver_aid_opt}{05:43}
\DefMacro{table-timing-gen-SkOptRangeCheckElimination-reduction-solver_aid_opt}{-15.45}
\DefMacro{table-timing-gen-SkOptRangeCheckElimination-genTime-full_opt}{00:32}
\DefMacro{table-timing-gen-SkOptRangeCheckElimination-reduction-full_opt}{89.08}
\DefMacro{table-timing-gen-SkOptScalarReplacement-name}{M19}
\DefMacro{table-timing-gen-SkOptScalarReplacement-genTime-non_opt}{05:47}
\DefMacro{table-timing-gen-SkOptScalarReplacement-genTime-stop_early_opt}{05:53}
\DefMacro{table-timing-gen-SkOptScalarReplacement-reduction-stop_early_opt}{-1.82}
\DefMacro{table-timing-gen-SkOptScalarReplacement-genTime-hot_filling_opt}{01:01}
\DefMacro{table-timing-gen-SkOptScalarReplacement-reduction-hot_filling_opt}{82.34}
\DefMacro{table-timing-gen-SkOptScalarReplacement-genTime-solver_aid_opt}{03:51}
\DefMacro{table-timing-gen-SkOptScalarReplacement-reduction-solver_aid_opt}{33.41}
\DefMacro{table-timing-gen-SkOptScalarReplacement-genTime-full_opt}{01:02}
\DefMacro{table-timing-gen-SkOptScalarReplacement-reduction-full_opt}{82.16}
\DefMacro{table-timing-gen-genTime-non_opt-sum}{2:22:38:42}
\DefMacro{table-timing-gen-reduction-non_opt-sum}{0.00}
\DefMacro{table-timing-gen-genTime-stop_early_opt-sum}{1:13:54:24}
\DefMacro{table-timing-gen-reduction-stop_early_opt-sum}{46.34}
\DefMacro{table-timing-gen-genTime-hot_filling_opt-sum}{27:59}
\DefMacro{table-timing-gen-reduction-hot_filling_opt-sum}{99.34}
\DefMacro{table-timing-gen-genTime-solver_aid_opt-sum}{2:07:57:38}
\DefMacro{table-timing-gen-reduction-solver_aid_opt-sum}{20.79}
\DefMacro{table-timing-gen-genTime-full_opt-sum}{21:20}
\DefMacro{table-timing-gen-reduction-full_opt-sum}{99.50}

\DefMacro{FontsizeTableTimingGen}{scriptsize}
\DefMacro{FloatEnvTableTimingGen}{table}
\DefMacro{CaptionTableTimingGen}{Time ((dd:)hh:mm:ss) and
relative reduction to \nonopt (\%) to
generate \NumGeneratedProgramsPerSketch programs in various
configurations. \UseMacro{TableTimingGenHeadSketch}
is \Sketches; \UseMacro{TableTimingGenHeadReduction} is the
relative reduction. \label{tab:timing-gen}\vspace{-5pt}}
\DefMacro{TableTimingGenHeadSketch}{Tmpl.}
\DefMacro{TableTimingGenHeadNonOpt}{\nonopt}
\DefMacro{TableTimingGenHeadStopEarlyOpt}{\stopearlyopt}
\DefMacro{TableTimingGenHeadHotFillingOpt}{\hotfillingopt}
\DefMacro{TableTimingGenHeadSolverAidOpt}{\solveraidopt}
\DefMacro{TableTimingGenHeadFullOpt}{\fullopt}
\DefMacro{TableTimingGenHeadStatic}{\staticgen}
\DefMacro{TableTimingGenHeadTime}{Time}
\DefMacro{TableTimingGenHeadReduction}{Rdct.}
\DefMacro{TableTimingGenCellSum}{$\Sigma$}
\DefMacro{TableTimingGenCellNA}{/}

\DefMacro{table-timing-static-gen-SkJDK8243670-name}{B1}
\DefMacro{table-timing-static-gen-SkJDK8243670-genTime-full_opt}{00:56}
\DefMacro{table-timing-static-gen-SkJDK8243670-genTime-static}{00:14}
\DefMacro{table-timing-static-gen-SkJDK8248552-name}{B2}
\DefMacro{table-timing-static-gen-SkJDK8248552-genTime-full_opt}{01:03}
\DefMacro{table-timing-static-gen-SkJDK8248552-genTime-static}{00:13}
\DefMacro{table-timing-static-gen-SkJDK8251535-name}{B3}
\DefMacro{table-timing-static-gen-SkJDK8251535-genTime-full_opt}{00:13}
\DefMacro{table-timing-static-gen-SkJDK8251535-genTime-static}{00:10}
\DefMacro{table-timing-static-gen-SkJDK8255763-name}{B4}
\DefMacro{table-timing-static-gen-SkJDK8255763-genTime-full_opt}{01:43}
\DefMacro{table-timing-static-gen-SkJDK8255763-genTime-static}{00:10}
\DefMacro{table-timing-static-gen-Sk1-name}{M1}
\DefMacro{table-timing-static-gen-Sk1-genTime-full_opt}{00:33}
\DefMacro{table-timing-static-gen-Sk1-genTime-static}{00:09}
\DefMacro{table-timing-static-gen-Sk2-name}{M2}
\DefMacro{table-timing-static-gen-Sk2-genTime-full_opt}{00:34}
\DefMacro{table-timing-static-gen-Sk2-genTime-static}{00:10}
\DefMacro{table-timing-static-gen-Sk3-name}{M3}
\DefMacro{table-timing-static-gen-Sk3-genTime-full_opt}{00:56}
\DefMacro{table-timing-static-gen-Sk3-genTime-static}{00:09}
\DefMacro{table-timing-static-gen-Sk4-name}{M4}
\DefMacro{table-timing-static-gen-Sk4-genTime-full_opt}{00:31}
\DefMacro{table-timing-static-gen-Sk4-genTime-static}{00:09}
\DefMacro{table-timing-static-gen-Sk5-name}{M5}
\DefMacro{table-timing-static-gen-Sk5-genTime-full_opt}{00:46}
\DefMacro{table-timing-static-gen-Sk5-genTime-static}{00:09}
\DefMacro{table-timing-static-gen-Sk6-name}{M6}
\DefMacro{table-timing-static-gen-Sk6-genTime-full_opt}{00:38}
\DefMacro{table-timing-static-gen-Sk6-genTime-static}{\UseMacro{TableTimingStaticGenCellNA}}
\DefMacro{table-timing-static-gen-Sk7-name}{M7}
\DefMacro{table-timing-static-gen-Sk7-genTime-full_opt}{00:49}
\DefMacro{table-timing-static-gen-Sk7-genTime-static}{00:09}
\DefMacro{table-timing-static-gen-Sk8-name}{M8}
\DefMacro{table-timing-static-gen-Sk8-genTime-full_opt}{01:16}
\DefMacro{table-timing-static-gen-Sk8-genTime-static}{00:09}
\DefMacro{table-timing-static-gen-Sk9-name}{M9}
\DefMacro{table-timing-static-gen-Sk9-genTime-full_opt}{00:20}
\DefMacro{table-timing-static-gen-Sk9-genTime-static}{00:09}
\DefMacro{table-timing-static-gen-Sk10-name}{M10}
\DefMacro{table-timing-static-gen-Sk10-genTime-full_opt}{00:39}
\DefMacro{table-timing-static-gen-Sk10-genTime-static}{00:09}
\DefMacro{table-timing-static-gen-Sk11-name}{M11}
\DefMacro{table-timing-static-gen-Sk11-genTime-full_opt}{01:02}
\DefMacro{table-timing-static-gen-Sk11-genTime-static}{00:10}
\DefMacro{table-timing-static-gen-Sk12-name}{M12}
\DefMacro{table-timing-static-gen-Sk12-genTime-full_opt}{00:43}
\DefMacro{table-timing-static-gen-Sk12-genTime-static}{00:09}
\DefMacro{table-timing-static-gen-Sk55-name}{M13}
\DefMacro{table-timing-static-gen-Sk55-genTime-full_opt}{00:51}
\DefMacro{table-timing-static-gen-Sk55-genTime-static}{\UseMacro{TableTimingStaticGenCellNA}}
\DefMacro{table-timing-static-gen-SkOptLockCoarsening-name}{M14}
\DefMacro{table-timing-static-gen-SkOptLockCoarsening-genTime-full_opt}{01:17}
\DefMacro{table-timing-static-gen-SkOptLockCoarsening-genTime-static}{00:08}
\DefMacro{table-timing-static-gen-SkOptLoopPeeling-name}{M15}
\DefMacro{table-timing-static-gen-SkOptLoopPeeling-genTime-full_opt}{00:14}
\DefMacro{table-timing-static-gen-SkOptLoopPeeling-genTime-static}{00:09}
\DefMacro{table-timing-static-gen-SkOptLoopUnrolling-name}{M16}
\DefMacro{table-timing-static-gen-SkOptLoopUnrolling-genTime-full_opt}{01:09}
\DefMacro{table-timing-static-gen-SkOptLoopUnrolling-genTime-static}{00:09}
\DefMacro{table-timing-static-gen-SkOptLoopUnswitching-name}{M17}
\DefMacro{table-timing-static-gen-SkOptLoopUnswitching-genTime-full_opt}{03:32}
\DefMacro{table-timing-static-gen-SkOptLoopUnswitching-genTime-static}{00:09}
\DefMacro{table-timing-static-gen-SkOptRangeCheckElimination-name}{M18}
\DefMacro{table-timing-static-gen-SkOptRangeCheckElimination-genTime-full_opt}{00:32}
\DefMacro{table-timing-static-gen-SkOptRangeCheckElimination-genTime-static}{\UseMacro{TableTimingStaticGenCellNA}}
\DefMacro{table-timing-static-gen-SkOptScalarReplacement-name}{M19}
\DefMacro{table-timing-static-gen-SkOptScalarReplacement-genTime-full_opt}{01:02}
\DefMacro{table-timing-static-gen-SkOptScalarReplacement-genTime-static}{00:09}
\DefMacro{table-timing-static-gen-genTime-full_opt-sum}{21:20}
\DefMacro{table-timing-static-gen-genTime-static-sum}{03:12}

\DefMacro{FontsizeTableTimingStaticGen}{small}
\DefMacro{FloatEnvTableTimingStaticGen}{table}
\DefMacro{CaptionTableTimingStaticGen}{Time (mm:ss) to
generate \NumGeneratedProgramsPerSketch programs with
execution-based and static approach respectively. Some \sketches
involve \holes constructed from runtime values of variables and
thus static approach cannot work (See an example in Section
``Execution-Based vs Static Generation''%
, which is marked with ``\UseMacro{TableTimingStaticGenCellNA}'' in
the table. \label{tab:timing-static-gen}}
\DefMacro{TableTimingStaticGenHeadSketch}{\Sketches}
\DefMacro{TableTimingStaticGenHeadFullOpt}{Execution-based (\fullopt)}
\DefMacro{TableTimingStaticGenHeadStatic}{\staticgen}
\DefMacro{TableTimingStaticGenCellSum}{$\Sigma$}
\DefMacro{TableTimingStaticGenCellNA}{/}

\DefMacro{TableFontsizeSketchyExecTimeRandom}{small}
\DefMacro{FloatEnvTableSketchyExecTimeRandom}{table*}
\DefMacro{TableCaptionSketchyExecTimeRandom}{Time ((hh:)mm:ss) to
execute \NumGeneratedProgramsPerSketch programs in various
configurations. \UseMacro{TableHeaderSketchyExecTimeSketchesRandom}
is \sketches; \UseMacro{TableHeaderSketchyExecTimeFullItrsRandom}
is \fullitrs; \UseMacro{TableHeaderSketchyExecTimeFullItrsMasterRandom}
is \fullitrsmaster; \UseMacro{TableHeaderSketchyExecTimeOneItrCoRandom}
is \oneitrco; \UseMacro{TableHeaderSketchyExecTimeOneItrMasterCoRandom}
is \oneitrmasterco; \UseMacro{TableHeaderSketchyExecTimeOracleLevel4Random}
is \UseMacro{oraclejdklevel4}; \UseMacro{TableHeaderSketchyExecTimeOracleLevel1Random}
is \UseMacro{oraclejdklevel1}.
\label{tab:exec-time-sketchy:a-single-generator:random}}
\DefMacro{TableHeaderSketchyExecTimeSketchesRandom}{Tmpl.}
\DefMacro{TableHeaderSketchyExecTimeFullItrsRandom}{F.I.}
\DefMacro{TableHeaderSketchyExecTimeFullItrsMasterRandom}{F.I.M.}
\DefMacro{TableHeaderSketchyExecTimeOneItrCoRandom}{O.I.C.}
\DefMacro{TableHeaderSketchyExecTimeOneItrMasterCoRandom}{O.I.M.C.}
\DefMacro{TableHeaderSketchyExecTimeOracleLevel4Random}{L4}
\DefMacro{TableHeaderSketchyExecTimeOracleLevel1Random}{L1}
\DefMacro{TableHeaderSketchyExecTimeOracleC2Random}{C2}

\DefMacro{TableFontsizeSketchyExecTimeStatic}{small}
\DefMacro{FloatEnvTableSketchyExecTimeStatic}{table*}
\DefMacro{TableCaptionSketchyExecTimeStatic}{Time ((hh:)mm:ss) to
execute \NumGeneratedProgramsPerSketch programs generated in static approach in various
configurations. \UseMacro{TableHeaderSketchyExecTimeSketchesStatic}
is \sketches; \UseMacro{TableHeaderSketchyExecTimeFullItrsStatic}
is \fullitrs; \UseMacro{TableHeaderSketchyExecTimeFullItrsMasterStatic}
is \fullitrsmaster; \UseMacro{TableHeaderSketchyExecTimeOneItrCoStatic}
is \oneitrco; \UseMacro{TableHeaderSketchyExecTimeOneItrMasterCoStatic}
is \oneitrmasterco; \UseMacro{TableHeaderSketchyExecTimeOracleLevel4Static}
is \UseMacro{oraclejdklevel4}; \UseMacro{TableHeaderSketchyExecTimeOracleLevel1Static}
is \UseMacro{oraclejdklevel1}.
\label{tab:exec-time-sketchy:a-single-generator:static}}
\DefMacro{TableHeaderSketchyExecTimeSketchesStatic}{Tmpl.}
\DefMacro{TableHeaderSketchyExecTimeFullItrsStatic}{F.I.}
\DefMacro{TableHeaderSketchyExecTimeFullItrsMasterStatic}{F.I.M.}
\DefMacro{TableHeaderSketchyExecTimeOneItrCoStatic}{O.I.C.}
\DefMacro{TableHeaderSketchyExecTimeOneItrMasterCoStatic}{O.I.M.C.}
\DefMacro{TableHeaderSketchyExecTimeOracleLevel4Static}{L4}
\DefMacro{TableHeaderSketchyExecTimeOracleLevel1Static}{L1}
\DefMacro{TableHeaderSketchyExecTimeOracleC2Static}{C2}

\DefMacro{table-tracking-holes-SkJDK8243670-name}{B1}
\DefMacro{table-tracking-holes-SkJDK8243670-dynamic-reached}{10100}
\DefMacro{table-tracking-holes-SkJDK8243670-dynamic-filled}{10100}
\DefMacro{table-tracking-holes-SkJDK8243670-dynamic-percentage}{100.00}
\DefMacro{table-tracking-holes-SkJDK8243670-static-reached}{7610}
\DefMacro{table-tracking-holes-SkJDK8243670-static-filled}{11000}
\DefMacro{table-tracking-holes-SkJDK8243670-static-percentage}{69.18}
\DefMacro{table-tracking-holes-SkJDK8248552-name}{B2}
\DefMacro{table-tracking-holes-SkJDK8248552-dynamic-reached}{4108}
\DefMacro{table-tracking-holes-SkJDK8248552-dynamic-filled}{4108}
\DefMacro{table-tracking-holes-SkJDK8248552-dynamic-percentage}{100.00}
\DefMacro{table-tracking-holes-SkJDK8248552-static-reached}{3793}
\DefMacro{table-tracking-holes-SkJDK8248552-static-filled}{5904}
\DefMacro{table-tracking-holes-SkJDK8248552-static-percentage}{64.24}
\DefMacro{table-tracking-holes-SkJDK8251535-name}{B3}
\DefMacro{table-tracking-holes-SkJDK8251535-dynamic-reached}{2729}
\DefMacro{table-tracking-holes-SkJDK8251535-dynamic-filled}{2729}
\DefMacro{table-tracking-holes-SkJDK8251535-dynamic-percentage}{100.00}
\DefMacro{table-tracking-holes-SkJDK8251535-static-reached}{2750}
\DefMacro{table-tracking-holes-SkJDK8251535-static-filled}{3000}
\DefMacro{table-tracking-holes-SkJDK8251535-static-percentage}{91.67}
\DefMacro{table-tracking-holes-SkJDK8255763-name}{B4}
\DefMacro{table-tracking-holes-SkJDK8255763-dynamic-reached}{6000}
\DefMacro{table-tracking-holes-SkJDK8255763-dynamic-filled}{6000}
\DefMacro{table-tracking-holes-SkJDK8255763-dynamic-percentage}{100.00}
\DefMacro{table-tracking-holes-SkJDK8255763-static-reached}{6000}
\DefMacro{table-tracking-holes-SkJDK8255763-static-filled}{6000}
\DefMacro{table-tracking-holes-SkJDK8255763-static-percentage}{100.00}
\DefMacro{table-tracking-holes-Sk1-name}{M1}
\DefMacro{table-tracking-holes-Sk1-dynamic-reached}{6497}
\DefMacro{table-tracking-holes-Sk1-dynamic-filled}{6497}
\DefMacro{table-tracking-holes-Sk1-dynamic-percentage}{100.00}
\DefMacro{table-tracking-holes-Sk1-static-reached}{6473}
\DefMacro{table-tracking-holes-Sk1-static-filled}{8000}
\DefMacro{table-tracking-holes-Sk1-static-percentage}{80.91}
\DefMacro{table-tracking-holes-Sk2-name}{M2}
\DefMacro{table-tracking-holes-Sk2-dynamic-reached}{6497}
\DefMacro{table-tracking-holes-Sk2-dynamic-filled}{6497}
\DefMacro{table-tracking-holes-Sk2-dynamic-percentage}{100.00}
\DefMacro{table-tracking-holes-Sk2-static-reached}{6440}
\DefMacro{table-tracking-holes-Sk2-static-filled}{8000}
\DefMacro{table-tracking-holes-Sk2-static-percentage}{80.50}
\DefMacro{table-tracking-holes-Sk3-name}{M3}
\DefMacro{table-tracking-holes-Sk3-dynamic-reached}{3000}
\DefMacro{table-tracking-holes-Sk3-dynamic-filled}{3000}
\DefMacro{table-tracking-holes-Sk3-dynamic-percentage}{100.00}
\DefMacro{table-tracking-holes-Sk3-static-reached}{3000}
\DefMacro{table-tracking-holes-Sk3-static-filled}{8000}
\DefMacro{table-tracking-holes-Sk3-static-percentage}{37.50}
\DefMacro{table-tracking-holes-Sk4-name}{M4}
\DefMacro{table-tracking-holes-Sk4-dynamic-reached}{6793}
\DefMacro{table-tracking-holes-Sk4-dynamic-filled}{6793}
\DefMacro{table-tracking-holes-Sk4-dynamic-percentage}{100.00}
\DefMacro{table-tracking-holes-Sk4-static-reached}{3199}
\DefMacro{table-tracking-holes-Sk4-static-filled}{7000}
\DefMacro{table-tracking-holes-Sk4-static-percentage}{45.70}
\DefMacro{table-tracking-holes-Sk5-name}{M5}
\DefMacro{table-tracking-holes-Sk5-dynamic-reached}{3592}
\DefMacro{table-tracking-holes-Sk5-dynamic-filled}{3592}
\DefMacro{table-tracking-holes-Sk5-dynamic-percentage}{100.00}
\DefMacro{table-tracking-holes-Sk5-static-reached}{3486}
\DefMacro{table-tracking-holes-Sk5-static-filled}{6000}
\DefMacro{table-tracking-holes-Sk5-static-percentage}{58.10}
\DefMacro{table-tracking-holes-Sk6-name}{M6}
\DefMacro{table-tracking-holes-Sk6-dynamic-reached}{2516}
\DefMacro{table-tracking-holes-Sk6-dynamic-filled}{2516}
\DefMacro{table-tracking-holes-Sk6-dynamic-percentage}{100.00}
\DefMacro{table-tracking-holes-Sk6-static-reached}{\UseMacro{TableTrackingHolesCellNA}}
\DefMacro{table-tracking-holes-Sk6-static-filled}{\UseMacro{TableTrackingHolesCellNA}}
\DefMacro{table-tracking-holes-Sk6-static-percentage}{\UseMacro{TableTrackingHolesCellNA}}
\DefMacro{table-tracking-holes-Sk7-name}{M7}
\DefMacro{table-tracking-holes-Sk7-dynamic-reached}{6736}
\DefMacro{table-tracking-holes-Sk7-dynamic-filled}{6736}
\DefMacro{table-tracking-holes-Sk7-dynamic-percentage}{100.00}
\DefMacro{table-tracking-holes-Sk7-static-reached}{6736}
\DefMacro{table-tracking-holes-Sk7-static-filled}{7000}
\DefMacro{table-tracking-holes-Sk7-static-percentage}{96.23}
\DefMacro{table-tracking-holes-Sk8-name}{M8}
\DefMacro{table-tracking-holes-Sk8-dynamic-reached}{4360}
\DefMacro{table-tracking-holes-Sk8-dynamic-filled}{4360}
\DefMacro{table-tracking-holes-Sk8-dynamic-percentage}{100.00}
\DefMacro{table-tracking-holes-Sk8-static-reached}{4338}
\DefMacro{table-tracking-holes-Sk8-static-filled}{5000}
\DefMacro{table-tracking-holes-Sk8-static-percentage}{86.76}
\DefMacro{table-tracking-holes-Sk9-name}{M9}
\DefMacro{table-tracking-holes-Sk9-dynamic-reached}{6931}
\DefMacro{table-tracking-holes-Sk9-dynamic-filled}{6931}
\DefMacro{table-tracking-holes-Sk9-dynamic-percentage}{100.00}
\DefMacro{table-tracking-holes-Sk9-static-reached}{6929}
\DefMacro{table-tracking-holes-Sk9-static-filled}{7000}
\DefMacro{table-tracking-holes-Sk9-static-percentage}{98.99}
\DefMacro{table-tracking-holes-Sk10-name}{M10}
\DefMacro{table-tracking-holes-Sk10-dynamic-reached}{5546}
\DefMacro{table-tracking-holes-Sk10-dynamic-filled}{5546}
\DefMacro{table-tracking-holes-Sk10-dynamic-percentage}{100.00}
\DefMacro{table-tracking-holes-Sk10-static-reached}{5577}
\DefMacro{table-tracking-holes-Sk10-static-filled}{6000}
\DefMacro{table-tracking-holes-Sk10-static-percentage}{92.95}
\DefMacro{table-tracking-holes-Sk11-name}{M11}
\DefMacro{table-tracking-holes-Sk11-dynamic-reached}{5400}
\DefMacro{table-tracking-holes-Sk11-dynamic-filled}{5400}
\DefMacro{table-tracking-holes-Sk11-dynamic-percentage}{100.00}
\DefMacro{table-tracking-holes-Sk11-static-reached}{5392}
\DefMacro{table-tracking-holes-Sk11-static-filled}{6000}
\DefMacro{table-tracking-holes-Sk11-static-percentage}{89.87}
\DefMacro{table-tracking-holes-Sk12-name}{M12}
\DefMacro{table-tracking-holes-Sk12-dynamic-reached}{5502}
\DefMacro{table-tracking-holes-Sk12-dynamic-filled}{5502}
\DefMacro{table-tracking-holes-Sk12-dynamic-percentage}{100.00}
\DefMacro{table-tracking-holes-Sk12-static-reached}{5466}
\DefMacro{table-tracking-holes-Sk12-static-filled}{6000}
\DefMacro{table-tracking-holes-Sk12-static-percentage}{91.10}
\DefMacro{table-tracking-holes-Sk55-name}{M13}
\DefMacro{table-tracking-holes-Sk55-dynamic-reached}{71501}
\DefMacro{table-tracking-holes-Sk55-dynamic-filled}{71501}
\DefMacro{table-tracking-holes-Sk55-dynamic-percentage}{100.00}
\DefMacro{table-tracking-holes-Sk55-static-reached}{\UseMacro{TableTrackingHolesCellNA}}
\DefMacro{table-tracking-holes-Sk55-static-filled}{\UseMacro{TableTrackingHolesCellNA}}
\DefMacro{table-tracking-holes-Sk55-static-percentage}{\UseMacro{TableTrackingHolesCellNA}}
\DefMacro{table-tracking-holes-SkOptLockCoarsening-name}{M14}
\DefMacro{table-tracking-holes-SkOptLockCoarsening-dynamic-reached}{3325}
\DefMacro{table-tracking-holes-SkOptLockCoarsening-dynamic-filled}{3325}
\DefMacro{table-tracking-holes-SkOptLockCoarsening-dynamic-percentage}{100.00}
\DefMacro{table-tracking-holes-SkOptLockCoarsening-static-reached}{3349}
\DefMacro{table-tracking-holes-SkOptLockCoarsening-static-filled}{4000}
\DefMacro{table-tracking-holes-SkOptLockCoarsening-static-percentage}{83.72}
\DefMacro{table-tracking-holes-SkOptLoopPeeling-name}{M15}
\DefMacro{table-tracking-holes-SkOptLoopPeeling-dynamic-reached}{4519}
\DefMacro{table-tracking-holes-SkOptLoopPeeling-dynamic-filled}{4519}
\DefMacro{table-tracking-holes-SkOptLoopPeeling-dynamic-percentage}{100.00}
\DefMacro{table-tracking-holes-SkOptLoopPeeling-static-reached}{4525}
\DefMacro{table-tracking-holes-SkOptLoopPeeling-static-filled}{5000}
\DefMacro{table-tracking-holes-SkOptLoopPeeling-static-percentage}{90.50}
\DefMacro{table-tracking-holes-SkOptLoopUnrolling-name}{M16}
\DefMacro{table-tracking-holes-SkOptLoopUnrolling-dynamic-reached}{2373}
\DefMacro{table-tracking-holes-SkOptLoopUnrolling-dynamic-filled}{2373}
\DefMacro{table-tracking-holes-SkOptLoopUnrolling-dynamic-percentage}{100.00}
\DefMacro{table-tracking-holes-SkOptLoopUnrolling-static-reached}{2338}
\DefMacro{table-tracking-holes-SkOptLoopUnrolling-static-filled}{3000}
\DefMacro{table-tracking-holes-SkOptLoopUnrolling-static-percentage}{77.93}
\DefMacro{table-tracking-holes-SkOptLoopUnswitching-name}{M17}
\DefMacro{table-tracking-holes-SkOptLoopUnswitching-dynamic-reached}{3000}
\DefMacro{table-tracking-holes-SkOptLoopUnswitching-dynamic-filled}{3000}
\DefMacro{table-tracking-holes-SkOptLoopUnswitching-dynamic-percentage}{100.00}
\DefMacro{table-tracking-holes-SkOptLoopUnswitching-static-reached}{3000}
\DefMacro{table-tracking-holes-SkOptLoopUnswitching-static-filled}{4000}
\DefMacro{table-tracking-holes-SkOptLoopUnswitching-static-percentage}{75.00}
\DefMacro{table-tracking-holes-SkOptRangeCheckElimination-name}{M18}
\DefMacro{table-tracking-holes-SkOptRangeCheckElimination-dynamic-reached}{2495}
\DefMacro{table-tracking-holes-SkOptRangeCheckElimination-dynamic-filled}{2495}
\DefMacro{table-tracking-holes-SkOptRangeCheckElimination-dynamic-percentage}{100.00}
\DefMacro{table-tracking-holes-SkOptRangeCheckElimination-static-reached}{\UseMacro{TableTrackingHolesCellNA}}
\DefMacro{table-tracking-holes-SkOptRangeCheckElimination-static-filled}{\UseMacro{TableTrackingHolesCellNA}}
\DefMacro{table-tracking-holes-SkOptRangeCheckElimination-static-percentage}{\UseMacro{TableTrackingHolesCellNA}}
\DefMacro{table-tracking-holes-SkOptScalarReplacement-name}{M19}
\DefMacro{table-tracking-holes-SkOptScalarReplacement-dynamic-reached}{4462}
\DefMacro{table-tracking-holes-SkOptScalarReplacement-dynamic-filled}{4462}
\DefMacro{table-tracking-holes-SkOptScalarReplacement-dynamic-percentage}{100.00}
\DefMacro{table-tracking-holes-SkOptScalarReplacement-static-reached}{4438}
\DefMacro{table-tracking-holes-SkOptScalarReplacement-static-filled}{5000}
\DefMacro{table-tracking-holes-SkOptScalarReplacement-static-percentage}{88.76}
\DefMacro{table-tracking-holes-dynamic-reached-sum}{177982}
\DefMacro{table-tracking-holes-dynamic-filled-sum}{177982}
\DefMacro{table-tracking-holes-dynamic-percentage-sum}{100.00}
\DefMacro{table-tracking-holes-static-reached-sum}{94839}
\DefMacro{table-tracking-holes-static-filled-sum}{120904}
\DefMacro{table-tracking-holes-static-percentage-sum}{78.44}

\DefMacro{FontsizeTableTrackingHoles}{small}
\DefMacro{FloatEnvTableTrackingHoles}{table*}
\DefMacro{CaptionTableTrackingHoles}{Number of holes reached and filled in generated programs from execution-based and static approach respectively. \UseMacro{TableTrackingHolesHeadSketch} is \sketches; \UseMacro{TableTrackingHolesHeadPercentage} is reachability of \holes, which is defined as the number of reached \holes over filled \holes. \label{tab:tracking-holes}}
\DefMacro{TableTrackingHolesHeadSketch}{Tmpl.}
\DefMacro{TableTrackingHolesHeadDynamic}{Execution-based}
\DefMacro{TableTrackingHolesHeadStatic}{Static}
\DefMacro{TableTrackingHolesHeadReached}{\# Reached}
\DefMacro{TableTrackingHolesHeadFilled}{\# Filled}
\DefMacro{TableTrackingHolesHeadPercentage}{Rchb. (\%)}
\DefMacro{TableTrackingHolesCellSum}{$\Sigma$}
\DefMacro{TableTrackingHolesCellNA}{/}

\DefMacro{sketchy-differential-testing-a-single-generator-sketchy_rand-SkJDK8243670-name}{B1}
\DefMacro{sketchy-differential-testing-a-single-generator-sketchy_rand-SkJDK8243670-fullItrs-fail}{0}
\DefMacro{sketchy-differential-testing-a-single-generator-sketchy_rand-SkJDK8243670-fullItrsMaster-fail}{0}
\DefMacro{sketchy-differential-testing-a-single-generator-sketchy_rand-SkJDK8243670-fullItrsCo-fail}{0}
\DefMacro{sketchy-differential-testing-a-single-generator-sketchy_rand-SkJDK8243670-fullItrsMasterCo-fail}{0}
\DefMacro{sketchy-differential-testing-a-single-generator-sketchy_rand-SkJDK8243670-oneItr-fail}{0}
\DefMacro{sketchy-differential-testing-a-single-generator-sketchy_rand-SkJDK8243670-oneItrMaster-fail}{0}
\DefMacro{sketchy-differential-testing-a-single-generator-sketchy_rand-SkJDK8243670-oneItrCo-fail}{0}
\DefMacro{sketchy-differential-testing-a-single-generator-sketchy_rand-SkJDK8243670-oneItrMasterCo-fail}{0}
\DefMacro{sketchy-differential-testing-a-single-generator-sketchy_rand-SkJDK8248552-name}{B2}
\DefMacro{sketchy-differential-testing-a-single-generator-sketchy_rand-SkJDK8248552-fullItrs-fail}{6}
\DefMacro{sketchy-differential-testing-a-single-generator-sketchy_rand-SkJDK8248552-fullItrsMaster-fail}{\UseMacro{TableCellSketchyDifferentialTestingCrashRandom}}
\DefMacro{sketchy-differential-testing-a-single-generator-sketchy_rand-SkJDK8248552-fullItrsCo-fail}{6}
\DefMacro{sketchy-differential-testing-a-single-generator-sketchy_rand-SkJDK8248552-fullItrsMasterCo-fail}{\UseMacro{TableCellSketchyDifferentialTestingCrashRandom}}
\DefMacro{sketchy-differential-testing-a-single-generator-sketchy_rand-SkJDK8248552-oneItr-fail}{9}
\DefMacro{sketchy-differential-testing-a-single-generator-sketchy_rand-SkJDK8248552-oneItrMaster-fail}{\UseMacro{TableCellSketchyDifferentialTestingCrashRandom}}
\DefMacro{sketchy-differential-testing-a-single-generator-sketchy_rand-SkJDK8248552-oneItrCo-fail}{9}
\DefMacro{sketchy-differential-testing-a-single-generator-sketchy_rand-SkJDK8248552-oneItrMasterCo-fail}{\UseMacro{TableCellSketchyDifferentialTestingCrashRandom}}
\DefMacro{sketchy-differential-testing-a-single-generator-sketchy_rand-SkJDK8251535-name}{B3}
\DefMacro{sketchy-differential-testing-a-single-generator-sketchy_rand-SkJDK8251535-fullItrs-fail}{0}
\DefMacro{sketchy-differential-testing-a-single-generator-sketchy_rand-SkJDK8251535-fullItrsMaster-fail}{0}
\DefMacro{sketchy-differential-testing-a-single-generator-sketchy_rand-SkJDK8251535-fullItrsCo-fail}{0}
\DefMacro{sketchy-differential-testing-a-single-generator-sketchy_rand-SkJDK8251535-fullItrsMasterCo-fail}{0}
\DefMacro{sketchy-differential-testing-a-single-generator-sketchy_rand-SkJDK8251535-oneItr-fail}{0}
\DefMacro{sketchy-differential-testing-a-single-generator-sketchy_rand-SkJDK8251535-oneItrMaster-fail}{0}
\DefMacro{sketchy-differential-testing-a-single-generator-sketchy_rand-SkJDK8251535-oneItrCo-fail}{0}
\DefMacro{sketchy-differential-testing-a-single-generator-sketchy_rand-SkJDK8251535-oneItrMasterCo-fail}{0}
\DefMacro{sketchy-differential-testing-a-single-generator-sketchy_rand-SkJDK8255763-name}{B4}
\DefMacro{sketchy-differential-testing-a-single-generator-sketchy_rand-SkJDK8255763-fullItrs-fail}{0}
\DefMacro{sketchy-differential-testing-a-single-generator-sketchy_rand-SkJDK8255763-fullItrsMaster-fail}{0}
\DefMacro{sketchy-differential-testing-a-single-generator-sketchy_rand-SkJDK8255763-fullItrsCo-fail}{0}
\DefMacro{sketchy-differential-testing-a-single-generator-sketchy_rand-SkJDK8255763-fullItrsMasterCo-fail}{0}
\DefMacro{sketchy-differential-testing-a-single-generator-sketchy_rand-SkJDK8255763-oneItr-fail}{56}
\DefMacro{sketchy-differential-testing-a-single-generator-sketchy_rand-SkJDK8255763-oneItrMaster-fail}{56}
\DefMacro{sketchy-differential-testing-a-single-generator-sketchy_rand-SkJDK8255763-oneItrCo-fail}{56}
\DefMacro{sketchy-differential-testing-a-single-generator-sketchy_rand-SkJDK8255763-oneItrMasterCo-fail}{56}
\DefMacro{sketchy-differential-testing-a-single-generator-sketchy_rand-Sk1-name}{M1}
\DefMacro{sketchy-differential-testing-a-single-generator-sketchy_rand-Sk1-fullItrs-fail}{0}
\DefMacro{sketchy-differential-testing-a-single-generator-sketchy_rand-Sk1-fullItrsMaster-fail}{0}
\DefMacro{sketchy-differential-testing-a-single-generator-sketchy_rand-Sk1-fullItrsCo-fail}{0}
\DefMacro{sketchy-differential-testing-a-single-generator-sketchy_rand-Sk1-fullItrsMasterCo-fail}{0}
\DefMacro{sketchy-differential-testing-a-single-generator-sketchy_rand-Sk1-oneItr-fail}{0}
\DefMacro{sketchy-differential-testing-a-single-generator-sketchy_rand-Sk1-oneItrMaster-fail}{0}
\DefMacro{sketchy-differential-testing-a-single-generator-sketchy_rand-Sk1-oneItrCo-fail}{0}
\DefMacro{sketchy-differential-testing-a-single-generator-sketchy_rand-Sk1-oneItrMasterCo-fail}{0}
\DefMacro{sketchy-differential-testing-a-single-generator-sketchy_rand-Sk2-name}{M2}
\DefMacro{sketchy-differential-testing-a-single-generator-sketchy_rand-Sk2-fullItrs-fail}{0}
\DefMacro{sketchy-differential-testing-a-single-generator-sketchy_rand-Sk2-fullItrsMaster-fail}{0}
\DefMacro{sketchy-differential-testing-a-single-generator-sketchy_rand-Sk2-fullItrsCo-fail}{0}
\DefMacro{sketchy-differential-testing-a-single-generator-sketchy_rand-Sk2-fullItrsMasterCo-fail}{0}
\DefMacro{sketchy-differential-testing-a-single-generator-sketchy_rand-Sk2-oneItr-fail}{0}
\DefMacro{sketchy-differential-testing-a-single-generator-sketchy_rand-Sk2-oneItrMaster-fail}{0}
\DefMacro{sketchy-differential-testing-a-single-generator-sketchy_rand-Sk2-oneItrCo-fail}{0}
\DefMacro{sketchy-differential-testing-a-single-generator-sketchy_rand-Sk2-oneItrMasterCo-fail}{0}
\DefMacro{sketchy-differential-testing-a-single-generator-sketchy_rand-Sk3-name}{M3}
\DefMacro{sketchy-differential-testing-a-single-generator-sketchy_rand-Sk3-fullItrs-fail}{0}
\DefMacro{sketchy-differential-testing-a-single-generator-sketchy_rand-Sk3-fullItrsMaster-fail}{0}
\DefMacro{sketchy-differential-testing-a-single-generator-sketchy_rand-Sk3-fullItrsCo-fail}{0}
\DefMacro{sketchy-differential-testing-a-single-generator-sketchy_rand-Sk3-fullItrsMasterCo-fail}{0}
\DefMacro{sketchy-differential-testing-a-single-generator-sketchy_rand-Sk3-oneItr-fail}{0}
\DefMacro{sketchy-differential-testing-a-single-generator-sketchy_rand-Sk3-oneItrMaster-fail}{0}
\DefMacro{sketchy-differential-testing-a-single-generator-sketchy_rand-Sk3-oneItrCo-fail}{0}
\DefMacro{sketchy-differential-testing-a-single-generator-sketchy_rand-Sk3-oneItrMasterCo-fail}{0}
\DefMacro{sketchy-differential-testing-a-single-generator-sketchy_rand-Sk4-name}{M4}
\DefMacro{sketchy-differential-testing-a-single-generator-sketchy_rand-Sk4-fullItrs-fail}{1}
\DefMacro{sketchy-differential-testing-a-single-generator-sketchy_rand-Sk4-fullItrsMaster-fail}{\UseMacro{TableCellSketchyDifferentialTestingCrashRandom}}
\DefMacro{sketchy-differential-testing-a-single-generator-sketchy_rand-Sk4-fullItrsCo-fail}{1}
\DefMacro{sketchy-differential-testing-a-single-generator-sketchy_rand-Sk4-fullItrsMasterCo-fail}{\UseMacro{TableCellSketchyDifferentialTestingCrashRandom}}
\DefMacro{sketchy-differential-testing-a-single-generator-sketchy_rand-Sk4-oneItr-fail}{0}
\DefMacro{sketchy-differential-testing-a-single-generator-sketchy_rand-Sk4-oneItrMaster-fail}{0}
\DefMacro{sketchy-differential-testing-a-single-generator-sketchy_rand-Sk4-oneItrCo-fail}{0}
\DefMacro{sketchy-differential-testing-a-single-generator-sketchy_rand-Sk4-oneItrMasterCo-fail}{0}
\DefMacro{sketchy-differential-testing-a-single-generator-sketchy_rand-Sk5-name}{M5}
\DefMacro{sketchy-differential-testing-a-single-generator-sketchy_rand-Sk5-fullItrs-fail}{0}
\DefMacro{sketchy-differential-testing-a-single-generator-sketchy_rand-Sk5-fullItrsMaster-fail}{0}
\DefMacro{sketchy-differential-testing-a-single-generator-sketchy_rand-Sk5-fullItrsCo-fail}{0}
\DefMacro{sketchy-differential-testing-a-single-generator-sketchy_rand-Sk5-fullItrsMasterCo-fail}{0}
\DefMacro{sketchy-differential-testing-a-single-generator-sketchy_rand-Sk5-oneItr-fail}{0}
\DefMacro{sketchy-differential-testing-a-single-generator-sketchy_rand-Sk5-oneItrMaster-fail}{0}
\DefMacro{sketchy-differential-testing-a-single-generator-sketchy_rand-Sk5-oneItrCo-fail}{0}
\DefMacro{sketchy-differential-testing-a-single-generator-sketchy_rand-Sk5-oneItrMasterCo-fail}{0}
\DefMacro{sketchy-differential-testing-a-single-generator-sketchy_rand-Sk6-name}{M6}
\DefMacro{sketchy-differential-testing-a-single-generator-sketchy_rand-Sk6-fullItrs-fail}{0}
\DefMacro{sketchy-differential-testing-a-single-generator-sketchy_rand-Sk6-fullItrsMaster-fail}{0}
\DefMacro{sketchy-differential-testing-a-single-generator-sketchy_rand-Sk6-fullItrsCo-fail}{0}
\DefMacro{sketchy-differential-testing-a-single-generator-sketchy_rand-Sk6-fullItrsMasterCo-fail}{0}
\DefMacro{sketchy-differential-testing-a-single-generator-sketchy_rand-Sk6-oneItr-fail}{0}
\DefMacro{sketchy-differential-testing-a-single-generator-sketchy_rand-Sk6-oneItrMaster-fail}{0}
\DefMacro{sketchy-differential-testing-a-single-generator-sketchy_rand-Sk6-oneItrCo-fail}{0}
\DefMacro{sketchy-differential-testing-a-single-generator-sketchy_rand-Sk6-oneItrMasterCo-fail}{0}
\DefMacro{sketchy-differential-testing-a-single-generator-sketchy_rand-Sk7-name}{M7}
\DefMacro{sketchy-differential-testing-a-single-generator-sketchy_rand-Sk7-fullItrs-fail}{0}
\DefMacro{sketchy-differential-testing-a-single-generator-sketchy_rand-Sk7-fullItrsMaster-fail}{0}
\DefMacro{sketchy-differential-testing-a-single-generator-sketchy_rand-Sk7-fullItrsCo-fail}{0}
\DefMacro{sketchy-differential-testing-a-single-generator-sketchy_rand-Sk7-fullItrsMasterCo-fail}{0}
\DefMacro{sketchy-differential-testing-a-single-generator-sketchy_rand-Sk7-oneItr-fail}{0}
\DefMacro{sketchy-differential-testing-a-single-generator-sketchy_rand-Sk7-oneItrMaster-fail}{0}
\DefMacro{sketchy-differential-testing-a-single-generator-sketchy_rand-Sk7-oneItrCo-fail}{0}
\DefMacro{sketchy-differential-testing-a-single-generator-sketchy_rand-Sk7-oneItrMasterCo-fail}{0}
\DefMacro{sketchy-differential-testing-a-single-generator-sketchy_rand-Sk8-name}{M8}
\DefMacro{sketchy-differential-testing-a-single-generator-sketchy_rand-Sk8-fullItrs-fail}{0}
\DefMacro{sketchy-differential-testing-a-single-generator-sketchy_rand-Sk8-fullItrsMaster-fail}{0}
\DefMacro{sketchy-differential-testing-a-single-generator-sketchy_rand-Sk8-fullItrsCo-fail}{0}
\DefMacro{sketchy-differential-testing-a-single-generator-sketchy_rand-Sk8-fullItrsMasterCo-fail}{0}
\DefMacro{sketchy-differential-testing-a-single-generator-sketchy_rand-Sk8-oneItr-fail}{0}
\DefMacro{sketchy-differential-testing-a-single-generator-sketchy_rand-Sk8-oneItrMaster-fail}{0}
\DefMacro{sketchy-differential-testing-a-single-generator-sketchy_rand-Sk8-oneItrCo-fail}{0}
\DefMacro{sketchy-differential-testing-a-single-generator-sketchy_rand-Sk8-oneItrMasterCo-fail}{0}
\DefMacro{sketchy-differential-testing-a-single-generator-sketchy_rand-Sk9-name}{M9}
\DefMacro{sketchy-differential-testing-a-single-generator-sketchy_rand-Sk9-fullItrs-fail}{0}
\DefMacro{sketchy-differential-testing-a-single-generator-sketchy_rand-Sk9-fullItrsMaster-fail}{0}
\DefMacro{sketchy-differential-testing-a-single-generator-sketchy_rand-Sk9-fullItrsCo-fail}{0}
\DefMacro{sketchy-differential-testing-a-single-generator-sketchy_rand-Sk9-fullItrsMasterCo-fail}{0}
\DefMacro{sketchy-differential-testing-a-single-generator-sketchy_rand-Sk9-oneItr-fail}{0}
\DefMacro{sketchy-differential-testing-a-single-generator-sketchy_rand-Sk9-oneItrMaster-fail}{0}
\DefMacro{sketchy-differential-testing-a-single-generator-sketchy_rand-Sk9-oneItrCo-fail}{0}
\DefMacro{sketchy-differential-testing-a-single-generator-sketchy_rand-Sk9-oneItrMasterCo-fail}{0}
\DefMacro{sketchy-differential-testing-a-single-generator-sketchy_rand-Sk10-name}{M10}
\DefMacro{sketchy-differential-testing-a-single-generator-sketchy_rand-Sk10-fullItrs-fail}{0}
\DefMacro{sketchy-differential-testing-a-single-generator-sketchy_rand-Sk10-fullItrsMaster-fail}{0}
\DefMacro{sketchy-differential-testing-a-single-generator-sketchy_rand-Sk10-fullItrsCo-fail}{0}
\DefMacro{sketchy-differential-testing-a-single-generator-sketchy_rand-Sk10-fullItrsMasterCo-fail}{0}
\DefMacro{sketchy-differential-testing-a-single-generator-sketchy_rand-Sk10-oneItr-fail}{0}
\DefMacro{sketchy-differential-testing-a-single-generator-sketchy_rand-Sk10-oneItrMaster-fail}{0}
\DefMacro{sketchy-differential-testing-a-single-generator-sketchy_rand-Sk10-oneItrCo-fail}{0}
\DefMacro{sketchy-differential-testing-a-single-generator-sketchy_rand-Sk10-oneItrMasterCo-fail}{0}
\DefMacro{sketchy-differential-testing-a-single-generator-sketchy_rand-Sk11-name}{M11}
\DefMacro{sketchy-differential-testing-a-single-generator-sketchy_rand-Sk11-fullItrs-fail}{0}
\DefMacro{sketchy-differential-testing-a-single-generator-sketchy_rand-Sk11-fullItrsMaster-fail}{0}
\DefMacro{sketchy-differential-testing-a-single-generator-sketchy_rand-Sk11-fullItrsCo-fail}{0}
\DefMacro{sketchy-differential-testing-a-single-generator-sketchy_rand-Sk11-fullItrsMasterCo-fail}{0}
\DefMacro{sketchy-differential-testing-a-single-generator-sketchy_rand-Sk11-oneItr-fail}{0}
\DefMacro{sketchy-differential-testing-a-single-generator-sketchy_rand-Sk11-oneItrMaster-fail}{0}
\DefMacro{sketchy-differential-testing-a-single-generator-sketchy_rand-Sk11-oneItrCo-fail}{0}
\DefMacro{sketchy-differential-testing-a-single-generator-sketchy_rand-Sk11-oneItrMasterCo-fail}{0}
\DefMacro{sketchy-differential-testing-a-single-generator-sketchy_rand-Sk12-name}{M12}
\DefMacro{sketchy-differential-testing-a-single-generator-sketchy_rand-Sk12-fullItrs-fail}{9}
\DefMacro{sketchy-differential-testing-a-single-generator-sketchy_rand-Sk12-fullItrsMaster-fail}{9}
\DefMacro{sketchy-differential-testing-a-single-generator-sketchy_rand-Sk12-fullItrsCo-fail}{9}
\DefMacro{sketchy-differential-testing-a-single-generator-sketchy_rand-Sk12-fullItrsMasterCo-fail}{9}
\DefMacro{sketchy-differential-testing-a-single-generator-sketchy_rand-Sk12-oneItr-fail}{1}
\DefMacro{sketchy-differential-testing-a-single-generator-sketchy_rand-Sk12-oneItrMaster-fail}{1}
\DefMacro{sketchy-differential-testing-a-single-generator-sketchy_rand-Sk12-oneItrCo-fail}{1}
\DefMacro{sketchy-differential-testing-a-single-generator-sketchy_rand-Sk12-oneItrMasterCo-fail}{1}
\DefMacro{sketchy-differential-testing-a-single-generator-sketchy_rand-Sk55-name}{M13}
\DefMacro{sketchy-differential-testing-a-single-generator-sketchy_rand-Sk55-fullItrs-fail}{0}
\DefMacro{sketchy-differential-testing-a-single-generator-sketchy_rand-Sk55-fullItrsMaster-fail}{0}
\DefMacro{sketchy-differential-testing-a-single-generator-sketchy_rand-Sk55-fullItrsCo-fail}{0}
\DefMacro{sketchy-differential-testing-a-single-generator-sketchy_rand-Sk55-fullItrsMasterCo-fail}{0}
\DefMacro{sketchy-differential-testing-a-single-generator-sketchy_rand-Sk55-oneItr-fail}{0}
\DefMacro{sketchy-differential-testing-a-single-generator-sketchy_rand-Sk55-oneItrMaster-fail}{0}
\DefMacro{sketchy-differential-testing-a-single-generator-sketchy_rand-Sk55-oneItrCo-fail}{0}
\DefMacro{sketchy-differential-testing-a-single-generator-sketchy_rand-Sk55-oneItrMasterCo-fail}{0}
\DefMacro{sketchy-differential-testing-a-single-generator-sketchy_rand-SkOptLockCoarsening-name}{M14}
\DefMacro{sketchy-differential-testing-a-single-generator-sketchy_rand-SkOptLockCoarsening-fullItrs-fail}{0}
\DefMacro{sketchy-differential-testing-a-single-generator-sketchy_rand-SkOptLockCoarsening-fullItrsMaster-fail}{0}
\DefMacro{sketchy-differential-testing-a-single-generator-sketchy_rand-SkOptLockCoarsening-fullItrsCo-fail}{0}
\DefMacro{sketchy-differential-testing-a-single-generator-sketchy_rand-SkOptLockCoarsening-fullItrsMasterCo-fail}{0}
\DefMacro{sketchy-differential-testing-a-single-generator-sketchy_rand-SkOptLockCoarsening-oneItr-fail}{0}
\DefMacro{sketchy-differential-testing-a-single-generator-sketchy_rand-SkOptLockCoarsening-oneItrMaster-fail}{0}
\DefMacro{sketchy-differential-testing-a-single-generator-sketchy_rand-SkOptLockCoarsening-oneItrCo-fail}{0}
\DefMacro{sketchy-differential-testing-a-single-generator-sketchy_rand-SkOptLockCoarsening-oneItrMasterCo-fail}{0}
\DefMacro{sketchy-differential-testing-a-single-generator-sketchy_rand-SkOptLoopPeeling-name}{M15}
\DefMacro{sketchy-differential-testing-a-single-generator-sketchy_rand-SkOptLoopPeeling-fullItrs-fail}{0}
\DefMacro{sketchy-differential-testing-a-single-generator-sketchy_rand-SkOptLoopPeeling-fullItrsMaster-fail}{0}
\DefMacro{sketchy-differential-testing-a-single-generator-sketchy_rand-SkOptLoopPeeling-fullItrsCo-fail}{0}
\DefMacro{sketchy-differential-testing-a-single-generator-sketchy_rand-SkOptLoopPeeling-fullItrsMasterCo-fail}{0}
\DefMacro{sketchy-differential-testing-a-single-generator-sketchy_rand-SkOptLoopPeeling-oneItr-fail}{0}
\DefMacro{sketchy-differential-testing-a-single-generator-sketchy_rand-SkOptLoopPeeling-oneItrMaster-fail}{0}
\DefMacro{sketchy-differential-testing-a-single-generator-sketchy_rand-SkOptLoopPeeling-oneItrCo-fail}{0}
\DefMacro{sketchy-differential-testing-a-single-generator-sketchy_rand-SkOptLoopPeeling-oneItrMasterCo-fail}{0}
\DefMacro{sketchy-differential-testing-a-single-generator-sketchy_rand-SkOptLoopUnrolling-name}{M16}
\DefMacro{sketchy-differential-testing-a-single-generator-sketchy_rand-SkOptLoopUnrolling-fullItrs-fail}{0}
\DefMacro{sketchy-differential-testing-a-single-generator-sketchy_rand-SkOptLoopUnrolling-fullItrsMaster-fail}{0}
\DefMacro{sketchy-differential-testing-a-single-generator-sketchy_rand-SkOptLoopUnrolling-fullItrsCo-fail}{0}
\DefMacro{sketchy-differential-testing-a-single-generator-sketchy_rand-SkOptLoopUnrolling-fullItrsMasterCo-fail}{0}
\DefMacro{sketchy-differential-testing-a-single-generator-sketchy_rand-SkOptLoopUnrolling-oneItr-fail}{0}
\DefMacro{sketchy-differential-testing-a-single-generator-sketchy_rand-SkOptLoopUnrolling-oneItrMaster-fail}{0}
\DefMacro{sketchy-differential-testing-a-single-generator-sketchy_rand-SkOptLoopUnrolling-oneItrCo-fail}{0}
\DefMacro{sketchy-differential-testing-a-single-generator-sketchy_rand-SkOptLoopUnrolling-oneItrMasterCo-fail}{0}
\DefMacro{sketchy-differential-testing-a-single-generator-sketchy_rand-SkOptLoopUnswitching-name}{M17}
\DefMacro{sketchy-differential-testing-a-single-generator-sketchy_rand-SkOptLoopUnswitching-fullItrs-fail}{0}
\DefMacro{sketchy-differential-testing-a-single-generator-sketchy_rand-SkOptLoopUnswitching-fullItrsMaster-fail}{0}
\DefMacro{sketchy-differential-testing-a-single-generator-sketchy_rand-SkOptLoopUnswitching-fullItrsCo-fail}{0}
\DefMacro{sketchy-differential-testing-a-single-generator-sketchy_rand-SkOptLoopUnswitching-fullItrsMasterCo-fail}{0}
\DefMacro{sketchy-differential-testing-a-single-generator-sketchy_rand-SkOptLoopUnswitching-oneItr-fail}{0}
\DefMacro{sketchy-differential-testing-a-single-generator-sketchy_rand-SkOptLoopUnswitching-oneItrMaster-fail}{0}
\DefMacro{sketchy-differential-testing-a-single-generator-sketchy_rand-SkOptLoopUnswitching-oneItrCo-fail}{0}
\DefMacro{sketchy-differential-testing-a-single-generator-sketchy_rand-SkOptLoopUnswitching-oneItrMasterCo-fail}{0}
\DefMacro{sketchy-differential-testing-a-single-generator-sketchy_rand-SkOptRangeCheckElimination-name}{M18}
\DefMacro{sketchy-differential-testing-a-single-generator-sketchy_rand-SkOptRangeCheckElimination-fullItrs-fail}{0}
\DefMacro{sketchy-differential-testing-a-single-generator-sketchy_rand-SkOptRangeCheckElimination-fullItrsMaster-fail}{0}
\DefMacro{sketchy-differential-testing-a-single-generator-sketchy_rand-SkOptRangeCheckElimination-fullItrsCo-fail}{0}
\DefMacro{sketchy-differential-testing-a-single-generator-sketchy_rand-SkOptRangeCheckElimination-fullItrsMasterCo-fail}{0}
\DefMacro{sketchy-differential-testing-a-single-generator-sketchy_rand-SkOptRangeCheckElimination-oneItr-fail}{0}
\DefMacro{sketchy-differential-testing-a-single-generator-sketchy_rand-SkOptRangeCheckElimination-oneItrMaster-fail}{0}
\DefMacro{sketchy-differential-testing-a-single-generator-sketchy_rand-SkOptRangeCheckElimination-oneItrCo-fail}{0}
\DefMacro{sketchy-differential-testing-a-single-generator-sketchy_rand-SkOptRangeCheckElimination-oneItrMasterCo-fail}{0}
\DefMacro{sketchy-differential-testing-a-single-generator-sketchy_rand-SkOptScalarReplacement-name}{M19}
\DefMacro{sketchy-differential-testing-a-single-generator-sketchy_rand-SkOptScalarReplacement-fullItrs-fail}{0}
\DefMacro{sketchy-differential-testing-a-single-generator-sketchy_rand-SkOptScalarReplacement-fullItrsMaster-fail}{0}
\DefMacro{sketchy-differential-testing-a-single-generator-sketchy_rand-SkOptScalarReplacement-fullItrsCo-fail}{0}
\DefMacro{sketchy-differential-testing-a-single-generator-sketchy_rand-SkOptScalarReplacement-fullItrsMasterCo-fail}{0}
\DefMacro{sketchy-differential-testing-a-single-generator-sketchy_rand-SkOptScalarReplacement-oneItr-fail}{0}
\DefMacro{sketchy-differential-testing-a-single-generator-sketchy_rand-SkOptScalarReplacement-oneItrMaster-fail}{0}
\DefMacro{sketchy-differential-testing-a-single-generator-sketchy_rand-SkOptScalarReplacement-oneItrCo-fail}{0}
\DefMacro{sketchy-differential-testing-a-single-generator-sketchy_rand-SkOptScalarReplacement-oneItrMasterCo-fail}{0}

\DefMacro{FloatEnvTableSketchyDifferentialTestingRandom}{table*}
\DefMacro{FontsizeTableSketchyDifferentialTestingRandom}{small}
\DefMacro{TableCaptionSketchyDifferentialTestingRandom}{Number of failures
found in \NumGeneratedProgramsPerSketch \generatedprograms in
various configurations per \sketch
of \TotalNumManuallyCreatedSketches manually
created \sketches. \UseMacro{TableCellSketchyDifferentialTestingCrashRandom}
means Master class crashed when run, which indicates some of
the \generatedprograms caused
crashing. \UseMacro{TableHeaderSketchyDifferentialTestingFullItrsRandom}
is \fullitrs; \UseMacro{TableHeaderSketchyDifferentialTestingFullItrsMasterRandom}
is \fullitrsmaster; \UseMacro{TableHeaderSketchyDifferentialTestingOneItrCoRandom}
is \oneitrco; \UseMacro{TableHeaderSketchyDifferentialTestingFullItrsMasterRandom}
is \fullitrsmasterco; \UseMacro{TableHeaderSketchyDifferentialTestingOneItrRandom}
is \oneitr; \UseMacro{TableHeaderSketchyDifferentialTestingOneItrMasterRandom}
is \oneitrmaster; \UseMacro{TableHeaderSketchyDifferentialTestingOneItrMasterCoRandom}
is \oneitrmasterco; \UseMacro{TableHeaderSketchyDifferentialTestingFullItrsCoRandom}
is \fullitrsco.\label{tab:differential-testing-sketchy:a-single-generator:random}}
\DefMacro{TableHeaderSketchyDifferentialTestingSketchesRandom}{\Sketches}
\DefMacro{TableHeaderSketchyDifferentialTestingFullItrsRandom}{F.I.}
\DefMacro{TableHeaderSketchyDifferentialTestingFullItrsMasterRandom}{F.I.M.}
\DefMacro{TableHeaderSketchyDifferentialTestingFullItrsCoRandom}{F.I.C.}
\DefMacro{TableHeaderSketchyDifferentialTestingFullItrsMasterCoRandom}{F.I.M.C.}
\DefMacro{TableHeaderSketchyDifferentialTestingOneItrRandom}{O.I.}
\DefMacro{TableHeaderSketchyDifferentialTestingOneItrMasterRandom}{O.I.M.}
\DefMacro{TableHeaderSketchyDifferentialTestingOneItrCoRandom}{O.I.C.}
\DefMacro{TableHeaderSketchyDifferentialTestingOneItrMasterCoRandom}{O.I.M.C.}
\DefMacro{TableCellSketchyDifferentialTestingCrashRandom}{CRASH}
\DefMacro{TableCellSketchyDifferentialTestingTimeoutRandom}{TIMEOUT}

\DefMacro{sketchy-differential-testing-a-single-generator-sketchy_static-SkJDK8243670-name}{B1}
\DefMacro{sketchy-differential-testing-a-single-generator-sketchy_static-SkJDK8243670-fullItrs-fail}{0}
\DefMacro{sketchy-differential-testing-a-single-generator-sketchy_static-SkJDK8243670-fullItrsMaster-fail}{0}
\DefMacro{sketchy-differential-testing-a-single-generator-sketchy_static-SkJDK8243670-fullItrsCo-fail}{0}
\DefMacro{sketchy-differential-testing-a-single-generator-sketchy_static-SkJDK8243670-fullItrsMasterCo-fail}{0}
\DefMacro{sketchy-differential-testing-a-single-generator-sketchy_static-SkJDK8243670-oneItr-fail}{0}
\DefMacro{sketchy-differential-testing-a-single-generator-sketchy_static-SkJDK8243670-oneItrMaster-fail}{0}
\DefMacro{sketchy-differential-testing-a-single-generator-sketchy_static-SkJDK8243670-oneItrCo-fail}{0}
\DefMacro{sketchy-differential-testing-a-single-generator-sketchy_static-SkJDK8243670-oneItrMasterCo-fail}{0}
\DefMacro{sketchy-differential-testing-a-single-generator-sketchy_static-SkJDK8248552-name}{B2}
\DefMacro{sketchy-differential-testing-a-single-generator-sketchy_static-SkJDK8248552-fullItrs-fail}{3}
\DefMacro{sketchy-differential-testing-a-single-generator-sketchy_static-SkJDK8248552-fullItrsMaster-fail}{\UseMacro{TableCellSketchyDifferentialTestingCrashStatic}}
\DefMacro{sketchy-differential-testing-a-single-generator-sketchy_static-SkJDK8248552-fullItrsCo-fail}{3}
\DefMacro{sketchy-differential-testing-a-single-generator-sketchy_static-SkJDK8248552-fullItrsMasterCo-fail}{\UseMacro{TableCellSketchyDifferentialTestingCrashStatic}}
\DefMacro{sketchy-differential-testing-a-single-generator-sketchy_static-SkJDK8248552-oneItr-fail}{7}
\DefMacro{sketchy-differential-testing-a-single-generator-sketchy_static-SkJDK8248552-oneItrMaster-fail}{\UseMacro{TableCellSketchyDifferentialTestingCrashStatic}}
\DefMacro{sketchy-differential-testing-a-single-generator-sketchy_static-SkJDK8248552-oneItrCo-fail}{7}
\DefMacro{sketchy-differential-testing-a-single-generator-sketchy_static-SkJDK8248552-oneItrMasterCo-fail}{\UseMacro{TableCellSketchyDifferentialTestingCrashStatic}}
\DefMacro{sketchy-differential-testing-a-single-generator-sketchy_static-SkJDK8251535-name}{B3}
\DefMacro{sketchy-differential-testing-a-single-generator-sketchy_static-SkJDK8251535-fullItrs-fail}{0}
\DefMacro{sketchy-differential-testing-a-single-generator-sketchy_static-SkJDK8251535-fullItrsMaster-fail}{0}
\DefMacro{sketchy-differential-testing-a-single-generator-sketchy_static-SkJDK8251535-fullItrsCo-fail}{0}
\DefMacro{sketchy-differential-testing-a-single-generator-sketchy_static-SkJDK8251535-fullItrsMasterCo-fail}{0}
\DefMacro{sketchy-differential-testing-a-single-generator-sketchy_static-SkJDK8251535-oneItr-fail}{0}
\DefMacro{sketchy-differential-testing-a-single-generator-sketchy_static-SkJDK8251535-oneItrMaster-fail}{0}
\DefMacro{sketchy-differential-testing-a-single-generator-sketchy_static-SkJDK8251535-oneItrCo-fail}{0}
\DefMacro{sketchy-differential-testing-a-single-generator-sketchy_static-SkJDK8251535-oneItrMasterCo-fail}{0}
\DefMacro{sketchy-differential-testing-a-single-generator-sketchy_static-SkJDK8255763-name}{B4}
\DefMacro{sketchy-differential-testing-a-single-generator-sketchy_static-SkJDK8255763-fullItrs-fail}{0}
\DefMacro{sketchy-differential-testing-a-single-generator-sketchy_static-SkJDK8255763-fullItrsMaster-fail}{0}
\DefMacro{sketchy-differential-testing-a-single-generator-sketchy_static-SkJDK8255763-fullItrsCo-fail}{0}
\DefMacro{sketchy-differential-testing-a-single-generator-sketchy_static-SkJDK8255763-fullItrsMasterCo-fail}{0}
\DefMacro{sketchy-differential-testing-a-single-generator-sketchy_static-SkJDK8255763-oneItr-fail}{73}
\DefMacro{sketchy-differential-testing-a-single-generator-sketchy_static-SkJDK8255763-oneItrMaster-fail}{73}
\DefMacro{sketchy-differential-testing-a-single-generator-sketchy_static-SkJDK8255763-oneItrCo-fail}{73}
\DefMacro{sketchy-differential-testing-a-single-generator-sketchy_static-SkJDK8255763-oneItrMasterCo-fail}{73}
\DefMacro{sketchy-differential-testing-a-single-generator-sketchy_static-Sk1-name}{M1}
\DefMacro{sketchy-differential-testing-a-single-generator-sketchy_static-Sk1-fullItrs-fail}{0}
\DefMacro{sketchy-differential-testing-a-single-generator-sketchy_static-Sk1-fullItrsMaster-fail}{0}
\DefMacro{sketchy-differential-testing-a-single-generator-sketchy_static-Sk1-fullItrsCo-fail}{0}
\DefMacro{sketchy-differential-testing-a-single-generator-sketchy_static-Sk1-fullItrsMasterCo-fail}{0}
\DefMacro{sketchy-differential-testing-a-single-generator-sketchy_static-Sk1-oneItr-fail}{0}
\DefMacro{sketchy-differential-testing-a-single-generator-sketchy_static-Sk1-oneItrMaster-fail}{0}
\DefMacro{sketchy-differential-testing-a-single-generator-sketchy_static-Sk1-oneItrCo-fail}{0}
\DefMacro{sketchy-differential-testing-a-single-generator-sketchy_static-Sk1-oneItrMasterCo-fail}{0}
\DefMacro{sketchy-differential-testing-a-single-generator-sketchy_static-Sk2-name}{M2}
\DefMacro{sketchy-differential-testing-a-single-generator-sketchy_static-Sk2-fullItrs-fail}{0}
\DefMacro{sketchy-differential-testing-a-single-generator-sketchy_static-Sk2-fullItrsMaster-fail}{0}
\DefMacro{sketchy-differential-testing-a-single-generator-sketchy_static-Sk2-fullItrsCo-fail}{0}
\DefMacro{sketchy-differential-testing-a-single-generator-sketchy_static-Sk2-fullItrsMasterCo-fail}{0}
\DefMacro{sketchy-differential-testing-a-single-generator-sketchy_static-Sk2-oneItr-fail}{0}
\DefMacro{sketchy-differential-testing-a-single-generator-sketchy_static-Sk2-oneItrMaster-fail}{0}
\DefMacro{sketchy-differential-testing-a-single-generator-sketchy_static-Sk2-oneItrCo-fail}{0}
\DefMacro{sketchy-differential-testing-a-single-generator-sketchy_static-Sk2-oneItrMasterCo-fail}{0}
\DefMacro{sketchy-differential-testing-a-single-generator-sketchy_static-Sk3-name}{M3}
\DefMacro{sketchy-differential-testing-a-single-generator-sketchy_static-Sk3-fullItrs-fail}{0}
\DefMacro{sketchy-differential-testing-a-single-generator-sketchy_static-Sk3-fullItrsMaster-fail}{0}
\DefMacro{sketchy-differential-testing-a-single-generator-sketchy_static-Sk3-fullItrsCo-fail}{0}
\DefMacro{sketchy-differential-testing-a-single-generator-sketchy_static-Sk3-fullItrsMasterCo-fail}{0}
\DefMacro{sketchy-differential-testing-a-single-generator-sketchy_static-Sk3-oneItr-fail}{0}
\DefMacro{sketchy-differential-testing-a-single-generator-sketchy_static-Sk3-oneItrMaster-fail}{0}
\DefMacro{sketchy-differential-testing-a-single-generator-sketchy_static-Sk3-oneItrCo-fail}{0}
\DefMacro{sketchy-differential-testing-a-single-generator-sketchy_static-Sk3-oneItrMasterCo-fail}{0}
\DefMacro{sketchy-differential-testing-a-single-generator-sketchy_static-Sk4-name}{M4}
\DefMacro{sketchy-differential-testing-a-single-generator-sketchy_static-Sk4-fullItrs-fail}{0}
\DefMacro{sketchy-differential-testing-a-single-generator-sketchy_static-Sk4-fullItrsMaster-fail}{0}
\DefMacro{sketchy-differential-testing-a-single-generator-sketchy_static-Sk4-fullItrsCo-fail}{0}
\DefMacro{sketchy-differential-testing-a-single-generator-sketchy_static-Sk4-fullItrsMasterCo-fail}{\UseMacro{TableCellSketchyDifferentialTestingTimeoutStatic}}
\DefMacro{sketchy-differential-testing-a-single-generator-sketchy_static-Sk4-oneItr-fail}{0}
\DefMacro{sketchy-differential-testing-a-single-generator-sketchy_static-Sk4-oneItrMaster-fail}{0}
\DefMacro{sketchy-differential-testing-a-single-generator-sketchy_static-Sk4-oneItrCo-fail}{0}
\DefMacro{sketchy-differential-testing-a-single-generator-sketchy_static-Sk4-oneItrMasterCo-fail}{0}
\DefMacro{sketchy-differential-testing-a-single-generator-sketchy_static-Sk5-name}{M5}
\DefMacro{sketchy-differential-testing-a-single-generator-sketchy_static-Sk5-fullItrs-fail}{0}
\DefMacro{sketchy-differential-testing-a-single-generator-sketchy_static-Sk5-fullItrsMaster-fail}{0}
\DefMacro{sketchy-differential-testing-a-single-generator-sketchy_static-Sk5-fullItrsCo-fail}{0}
\DefMacro{sketchy-differential-testing-a-single-generator-sketchy_static-Sk5-fullItrsMasterCo-fail}{0}
\DefMacro{sketchy-differential-testing-a-single-generator-sketchy_static-Sk5-oneItr-fail}{0}
\DefMacro{sketchy-differential-testing-a-single-generator-sketchy_static-Sk5-oneItrMaster-fail}{0}
\DefMacro{sketchy-differential-testing-a-single-generator-sketchy_static-Sk5-oneItrCo-fail}{0}
\DefMacro{sketchy-differential-testing-a-single-generator-sketchy_static-Sk5-oneItrMasterCo-fail}{0}
\DefMacro{sketchy-differential-testing-a-single-generator-sketchy_static-Sk7-name}{M7}
\DefMacro{sketchy-differential-testing-a-single-generator-sketchy_static-Sk7-fullItrs-fail}{0}
\DefMacro{sketchy-differential-testing-a-single-generator-sketchy_static-Sk7-fullItrsMaster-fail}{0}
\DefMacro{sketchy-differential-testing-a-single-generator-sketchy_static-Sk7-fullItrsCo-fail}{0}
\DefMacro{sketchy-differential-testing-a-single-generator-sketchy_static-Sk7-fullItrsMasterCo-fail}{0}
\DefMacro{sketchy-differential-testing-a-single-generator-sketchy_static-Sk7-oneItr-fail}{0}
\DefMacro{sketchy-differential-testing-a-single-generator-sketchy_static-Sk7-oneItrMaster-fail}{0}
\DefMacro{sketchy-differential-testing-a-single-generator-sketchy_static-Sk7-oneItrCo-fail}{0}
\DefMacro{sketchy-differential-testing-a-single-generator-sketchy_static-Sk7-oneItrMasterCo-fail}{0}
\DefMacro{sketchy-differential-testing-a-single-generator-sketchy_static-Sk8-name}{M8}
\DefMacro{sketchy-differential-testing-a-single-generator-sketchy_static-Sk8-fullItrs-fail}{0}
\DefMacro{sketchy-differential-testing-a-single-generator-sketchy_static-Sk8-fullItrsMaster-fail}{0}
\DefMacro{sketchy-differential-testing-a-single-generator-sketchy_static-Sk8-fullItrsCo-fail}{0}
\DefMacro{sketchy-differential-testing-a-single-generator-sketchy_static-Sk8-fullItrsMasterCo-fail}{0}
\DefMacro{sketchy-differential-testing-a-single-generator-sketchy_static-Sk8-oneItr-fail}{0}
\DefMacro{sketchy-differential-testing-a-single-generator-sketchy_static-Sk8-oneItrMaster-fail}{0}
\DefMacro{sketchy-differential-testing-a-single-generator-sketchy_static-Sk8-oneItrCo-fail}{0}
\DefMacro{sketchy-differential-testing-a-single-generator-sketchy_static-Sk8-oneItrMasterCo-fail}{0}
\DefMacro{sketchy-differential-testing-a-single-generator-sketchy_static-Sk9-name}{M9}
\DefMacro{sketchy-differential-testing-a-single-generator-sketchy_static-Sk9-fullItrs-fail}{0}
\DefMacro{sketchy-differential-testing-a-single-generator-sketchy_static-Sk9-fullItrsMaster-fail}{0}
\DefMacro{sketchy-differential-testing-a-single-generator-sketchy_static-Sk9-fullItrsCo-fail}{0}
\DefMacro{sketchy-differential-testing-a-single-generator-sketchy_static-Sk9-fullItrsMasterCo-fail}{0}
\DefMacro{sketchy-differential-testing-a-single-generator-sketchy_static-Sk9-oneItr-fail}{0}
\DefMacro{sketchy-differential-testing-a-single-generator-sketchy_static-Sk9-oneItrMaster-fail}{0}
\DefMacro{sketchy-differential-testing-a-single-generator-sketchy_static-Sk9-oneItrCo-fail}{0}
\DefMacro{sketchy-differential-testing-a-single-generator-sketchy_static-Sk9-oneItrMasterCo-fail}{0}
\DefMacro{sketchy-differential-testing-a-single-generator-sketchy_static-Sk10-name}{M10}
\DefMacro{sketchy-differential-testing-a-single-generator-sketchy_static-Sk10-fullItrs-fail}{0}
\DefMacro{sketchy-differential-testing-a-single-generator-sketchy_static-Sk10-fullItrsMaster-fail}{0}
\DefMacro{sketchy-differential-testing-a-single-generator-sketchy_static-Sk10-fullItrsCo-fail}{0}
\DefMacro{sketchy-differential-testing-a-single-generator-sketchy_static-Sk10-fullItrsMasterCo-fail}{0}
\DefMacro{sketchy-differential-testing-a-single-generator-sketchy_static-Sk10-oneItr-fail}{0}
\DefMacro{sketchy-differential-testing-a-single-generator-sketchy_static-Sk10-oneItrMaster-fail}{0}
\DefMacro{sketchy-differential-testing-a-single-generator-sketchy_static-Sk10-oneItrCo-fail}{0}
\DefMacro{sketchy-differential-testing-a-single-generator-sketchy_static-Sk10-oneItrMasterCo-fail}{0}
\DefMacro{sketchy-differential-testing-a-single-generator-sketchy_static-Sk11-name}{M11}
\DefMacro{sketchy-differential-testing-a-single-generator-sketchy_static-Sk11-fullItrs-fail}{0}
\DefMacro{sketchy-differential-testing-a-single-generator-sketchy_static-Sk11-fullItrsMaster-fail}{0}
\DefMacro{sketchy-differential-testing-a-single-generator-sketchy_static-Sk11-fullItrsCo-fail}{0}
\DefMacro{sketchy-differential-testing-a-single-generator-sketchy_static-Sk11-fullItrsMasterCo-fail}{0}
\DefMacro{sketchy-differential-testing-a-single-generator-sketchy_static-Sk11-oneItr-fail}{0}
\DefMacro{sketchy-differential-testing-a-single-generator-sketchy_static-Sk11-oneItrMaster-fail}{0}
\DefMacro{sketchy-differential-testing-a-single-generator-sketchy_static-Sk11-oneItrCo-fail}{0}
\DefMacro{sketchy-differential-testing-a-single-generator-sketchy_static-Sk11-oneItrMasterCo-fail}{0}
\DefMacro{sketchy-differential-testing-a-single-generator-sketchy_static-Sk12-name}{M12}
\DefMacro{sketchy-differential-testing-a-single-generator-sketchy_static-Sk12-fullItrs-fail}{11}
\DefMacro{sketchy-differential-testing-a-single-generator-sketchy_static-Sk12-fullItrsMaster-fail}{11}
\DefMacro{sketchy-differential-testing-a-single-generator-sketchy_static-Sk12-fullItrsCo-fail}{11}
\DefMacro{sketchy-differential-testing-a-single-generator-sketchy_static-Sk12-fullItrsMasterCo-fail}{11}
\DefMacro{sketchy-differential-testing-a-single-generator-sketchy_static-Sk12-oneItr-fail}{0}
\DefMacro{sketchy-differential-testing-a-single-generator-sketchy_static-Sk12-oneItrMaster-fail}{0}
\DefMacro{sketchy-differential-testing-a-single-generator-sketchy_static-Sk12-oneItrCo-fail}{0}
\DefMacro{sketchy-differential-testing-a-single-generator-sketchy_static-Sk12-oneItrMasterCo-fail}{0}
\DefMacro{sketchy-differential-testing-a-single-generator-sketchy_static-SkOptLockCoarsening-name}{M14}
\DefMacro{sketchy-differential-testing-a-single-generator-sketchy_static-SkOptLockCoarsening-fullItrs-fail}{0}
\DefMacro{sketchy-differential-testing-a-single-generator-sketchy_static-SkOptLockCoarsening-fullItrsMaster-fail}{0}
\DefMacro{sketchy-differential-testing-a-single-generator-sketchy_static-SkOptLockCoarsening-fullItrsCo-fail}{0}
\DefMacro{sketchy-differential-testing-a-single-generator-sketchy_static-SkOptLockCoarsening-fullItrsMasterCo-fail}{0}
\DefMacro{sketchy-differential-testing-a-single-generator-sketchy_static-SkOptLockCoarsening-oneItr-fail}{0}
\DefMacro{sketchy-differential-testing-a-single-generator-sketchy_static-SkOptLockCoarsening-oneItrMaster-fail}{0}
\DefMacro{sketchy-differential-testing-a-single-generator-sketchy_static-SkOptLockCoarsening-oneItrCo-fail}{0}
\DefMacro{sketchy-differential-testing-a-single-generator-sketchy_static-SkOptLockCoarsening-oneItrMasterCo-fail}{0}
\DefMacro{sketchy-differential-testing-a-single-generator-sketchy_static-SkOptLoopPeeling-name}{M15}
\DefMacro{sketchy-differential-testing-a-single-generator-sketchy_static-SkOptLoopPeeling-fullItrs-fail}{0}
\DefMacro{sketchy-differential-testing-a-single-generator-sketchy_static-SkOptLoopPeeling-fullItrsMaster-fail}{0}
\DefMacro{sketchy-differential-testing-a-single-generator-sketchy_static-SkOptLoopPeeling-fullItrsCo-fail}{0}
\DefMacro{sketchy-differential-testing-a-single-generator-sketchy_static-SkOptLoopPeeling-fullItrsMasterCo-fail}{0}
\DefMacro{sketchy-differential-testing-a-single-generator-sketchy_static-SkOptLoopPeeling-oneItr-fail}{0}
\DefMacro{sketchy-differential-testing-a-single-generator-sketchy_static-SkOptLoopPeeling-oneItrMaster-fail}{0}
\DefMacro{sketchy-differential-testing-a-single-generator-sketchy_static-SkOptLoopPeeling-oneItrCo-fail}{0}
\DefMacro{sketchy-differential-testing-a-single-generator-sketchy_static-SkOptLoopPeeling-oneItrMasterCo-fail}{0}
\DefMacro{sketchy-differential-testing-a-single-generator-sketchy_static-SkOptLoopUnrolling-name}{M16}
\DefMacro{sketchy-differential-testing-a-single-generator-sketchy_static-SkOptLoopUnrolling-fullItrs-fail}{0}
\DefMacro{sketchy-differential-testing-a-single-generator-sketchy_static-SkOptLoopUnrolling-fullItrsMaster-fail}{0}
\DefMacro{sketchy-differential-testing-a-single-generator-sketchy_static-SkOptLoopUnrolling-fullItrsCo-fail}{0}
\DefMacro{sketchy-differential-testing-a-single-generator-sketchy_static-SkOptLoopUnrolling-fullItrsMasterCo-fail}{0}
\DefMacro{sketchy-differential-testing-a-single-generator-sketchy_static-SkOptLoopUnrolling-oneItr-fail}{0}
\DefMacro{sketchy-differential-testing-a-single-generator-sketchy_static-SkOptLoopUnrolling-oneItrMaster-fail}{0}
\DefMacro{sketchy-differential-testing-a-single-generator-sketchy_static-SkOptLoopUnrolling-oneItrCo-fail}{0}
\DefMacro{sketchy-differential-testing-a-single-generator-sketchy_static-SkOptLoopUnrolling-oneItrMasterCo-fail}{0}
\DefMacro{sketchy-differential-testing-a-single-generator-sketchy_static-SkOptLoopUnswitching-name}{M17}
\DefMacro{sketchy-differential-testing-a-single-generator-sketchy_static-SkOptLoopUnswitching-fullItrs-fail}{0}
\DefMacro{sketchy-differential-testing-a-single-generator-sketchy_static-SkOptLoopUnswitching-fullItrsMaster-fail}{0}
\DefMacro{sketchy-differential-testing-a-single-generator-sketchy_static-SkOptLoopUnswitching-fullItrsCo-fail}{0}
\DefMacro{sketchy-differential-testing-a-single-generator-sketchy_static-SkOptLoopUnswitching-fullItrsMasterCo-fail}{0}
\DefMacro{sketchy-differential-testing-a-single-generator-sketchy_static-SkOptLoopUnswitching-oneItr-fail}{0}
\DefMacro{sketchy-differential-testing-a-single-generator-sketchy_static-SkOptLoopUnswitching-oneItrMaster-fail}{0}
\DefMacro{sketchy-differential-testing-a-single-generator-sketchy_static-SkOptLoopUnswitching-oneItrCo-fail}{0}
\DefMacro{sketchy-differential-testing-a-single-generator-sketchy_static-SkOptLoopUnswitching-oneItrMasterCo-fail}{0}
\DefMacro{sketchy-differential-testing-a-single-generator-sketchy_static-SkOptScalarReplacement-name}{M19}
\DefMacro{sketchy-differential-testing-a-single-generator-sketchy_static-SkOptScalarReplacement-fullItrs-fail}{0}
\DefMacro{sketchy-differential-testing-a-single-generator-sketchy_static-SkOptScalarReplacement-fullItrsMaster-fail}{0}
\DefMacro{sketchy-differential-testing-a-single-generator-sketchy_static-SkOptScalarReplacement-fullItrsCo-fail}{0}
\DefMacro{sketchy-differential-testing-a-single-generator-sketchy_static-SkOptScalarReplacement-fullItrsMasterCo-fail}{0}
\DefMacro{sketchy-differential-testing-a-single-generator-sketchy_static-SkOptScalarReplacement-oneItr-fail}{0}
\DefMacro{sketchy-differential-testing-a-single-generator-sketchy_static-SkOptScalarReplacement-oneItrMaster-fail}{0}
\DefMacro{sketchy-differential-testing-a-single-generator-sketchy_static-SkOptScalarReplacement-oneItrCo-fail}{0}
\DefMacro{sketchy-differential-testing-a-single-generator-sketchy_static-SkOptScalarReplacement-oneItrMasterCo-fail}{0}

\DefMacro{FloatEnvTableSketchyDifferentialTestingStatic}{table*}
\DefMacro{FontsizeTableSketchyDifferentialTestingStatic}{small}
\DefMacro{TableCaptionSketchyDifferentialTestingStatic}{Number of failures
found in \NumGeneratedProgramsPerSketch \generatedprograms in
static approach in various configurations per \sketch
of \TotalNumManuallyCreatedSketches manually
created \sketches. \UseMacro{TableCellSketchyDifferentialTestingCrashStatic}
means Master class crashed when run, which indicates some of
the \generatedprograms caused
crashing. \UseMacro{TableHeaderSketchyDifferentialTestingFullItrsStatic}
is \fullitrs; \UseMacro{TableHeaderSketchyDifferentialTestingFullItrsMasterStatic}
is \fullitrsmaster; \UseMacro{TableHeaderSketchyDifferentialTestingOneItrCoStatic}
is \oneitrco; \UseMacro{TableHeaderSketchyDifferentialTestingFullItrsMasterStatic}
is \fullitrsmasterco; \UseMacro{TableHeaderSketchyDifferentialTestingOneItrStatic}
is \oneitr; \UseMacro{TableHeaderSketchyDifferentialTestingOneItrMasterStatic}
is \oneitrmaster; \UseMacro{TableHeaderSketchyDifferentialTestingOneItrMasterCoStatic}
is \oneitrmasterco; \UseMacro{TableHeaderSketchyDifferentialTestingFullItrsCoStatic}
is \fullitrsco.\label{tab:differential-testing-sketchy:a-single-generator:static}}
\DefMacro{TableHeaderSketchyDifferentialTestingSketchesStatic}{\Sketches}
\DefMacro{TableHeaderSketchyDifferentialTestingFullItrsStatic}{F.I.}
\DefMacro{TableHeaderSketchyDifferentialTestingFullItrsMasterStatic}{F.I.M.}
\DefMacro{TableHeaderSketchyDifferentialTestingFullItrsCoStatic}{F.I.C.}
\DefMacro{TableHeaderSketchyDifferentialTestingFullItrsMasterCoStatic}{F.I.M.C.}
\DefMacro{TableHeaderSketchyDifferentialTestingOneItrStatic}{O.I.}
\DefMacro{TableHeaderSketchyDifferentialTestingOneItrMasterStatic}{O.I.M.}
\DefMacro{TableHeaderSketchyDifferentialTestingOneItrCoStatic}{O.I.C.}
\DefMacro{TableHeaderSketchyDifferentialTestingOneItrMasterCoStatic}{O.I.M.C.}
\DefMacro{TableCellSketchyDifferentialTestingCrashStatic}{CRASH}
\DefMacro{TableCellSketchyDifferentialTestingTimeoutStatic}{TIMEOUT}

\DefMacro{FloatEnvTableSketchyIterationsRandom}{table*}
\DefMacro{FontsizeTableSketchyIterationsRandom}{scriptsize}
\DefMacro{TableCaptionSketchyIterationsRandom}{Numbers of iterations, hot transformation, and condition holes during generation of \NumGeneratedProgramsPerSketch programs from each manually created \sketch.
\label{tab:iterations-sketchy:a-single-generator:random}}
\DefMacro{TableHeaderSketchyIterationsSketchesRandom}{\Sketches}
\DefMacro{TableHeaderSketchyIterationsNumIterationsRandom}{\# Iterations}
\DefMacro{TableHeaderSketchyIterationsNumHotTransformingRandom}{\# Hot Transformation}
\DefMacro{TableHeaderSketchyIterationsNumConditionHolesEvaluatedRandom}{\# Condition Holes}
\DefMacro{TableHeaderSketchyIterationsNonOptRandom}{\nonopt/\hotfillingopt}
\DefMacro{TableHeaderSketchyIterationsStopEarlyOptRandom}{\stopearlyopt}
\DefMacro{TableHeaderSketchyIterationsSolverAidOptRandom}{\solveraidopt}
\DefMacro{TableHeaderSketchyIterationsHotFillingOptRandom}{\hotfillingopt}
\DefMacro{TableHeaderSketchyIterationsFullOptRandom}{\fullopt}
\DefMacro{TableHeaderSketchyIterationsAlwaysTrueRandom}{Always True}
\DefMacro{TableHeaderSketchyIterationsAlwaysFalseRandom}{Always False}

\DefMacro{FontsizeTableSketchCharacteristics}{small}
\DefMacro{FloatEnvTableSketchCharacteristics}{table}
\DefMacro{TableCaptionSketchCharacteristics}{Characteristics of
\sketches.\label{tab:sketch-chracteristics}}
\DefMacro{TableHeaderSketchCharacteristicsNumNonDetChoices}{\# Placeholders}

\DefMacro{FloatEnvTableSketchyExecTimeFullVersionFullItrsRandom}{table*}
\DefMacro{FloatEnvTableSketchyExecTimeFullVersionFullItrsMasterRandom}{table*}
\DefMacro{FloatEnvTableSketchyExecTimeFullVersionFullItrsCoRandom}{table*}
\DefMacro{FloatEnvTableSketchyExecTimeFullVersionFullItrsMasterCoRandom}{table*}
\DefMacro{FloatEnvTableSketchyExecTimeFullVersionOneItrRandom}{table*}
\DefMacro{FloatEnvTableSketchyExecTimeFullVersionOneItrMasterRandom}{table*}
\DefMacro{FloatEnvTableSketchyExecTimeFullVersionOneItrCoRandom}{table*}
\DefMacro{FloatEnvTableSketchyExecTimeFullVersionOneItrMasterCoRandom}{table*}
\DefMacro{FontsizeTableSketchyExecTimeFullVersionFullItrsRandom}{small}
\DefMacro{FontsizeTableSketchyExecTimeFullVersionFullItrsMasterRandom}{small}
\DefMacro{FontsizeTableSketchyExecTimeFullVersionFullItrsCoRandom}{small}
\DefMacro{FontsizeTableSketchyExecTimeFullVersionFullItrsMasterCoRandom}{small}
\DefMacro{FontsizeTableSketchyExecTimeFullVersionOneItrRandom}{small}
\DefMacro{FontsizeTableSketchyExecTimeFullVersionOneItrMasterRandom}{small}
\DefMacro{FontsizeTableSketchyExecTimeFullVersionOneItrCoRandom}{small}
\DefMacro{FontsizeTableSketchyExecTimeFullVersionOneItrMasterCoRandom}{small}
\DefMacro{TableCaptionSketchyExecTimeFullVersionFullItrsRandom}{Time ((d:hh:)mm:ss.S)
to execute \NumGeneratedProgramsPerSketch programs in \fullitrs
configuration. \label{tab:exec-time-sketchy-full-version:full-itrs:random}}
\DefMacro{TableCaptionSketchyExecTimeFullVersionFullItrsMasterRandom}{Time ((d:hh:)mm:ss.S)
to execute \NumGeneratedProgramsPerSketch programs in \fullitrsmaster
configuration. \label{tab:exec-time-sketchy-full-version:full-itrs-master:random}}
\DefMacro{TableCaptionSketchyExecTimeFullVersionFullItrsCoRandom}{Time ((d:hh:)mm:ss.S)
to execute \NumGeneratedProgramsPerSketch programs in \fullitrsco
configuration. \label{tab:exec-time-sketchy-full-version:full-itrs-co:random}}
\DefMacro{TableCaptionSketchyExecTimeFullVersionFullItrsMasterCoRandom}{Time ((d:hh:)mm:ss.S)
to execute \NumGeneratedProgramsPerSketch programs in \fullitrsmasterco
configuration. \label{tab:exec-time-sketchy-full-version:full-itrs-master-co:random}}
\DefMacro{TableCaptionSketchyExecTimeFullVersionOneItrRandom}{Time ((d:hh:)mm:ss.S)
to execute \NumGeneratedProgramsPerSketch programs in \oneitr
configuration. \label{tab:exec-time-sketchy-full-version:one-itr:random}}
\DefMacro{TableCaptionSketchyExecTimeFullVersionOneItrMasterRandom}{Time ((d:hh:)mm:ss.S)
to execute \NumGeneratedProgramsPerSketch programs in \oneitrmaster
configuration. \label{tab:exec-time-sketchy-full-version:one-itr-master:random}}
\DefMacro{TableCaptionSketchyExecTimeFullVersionOneItrCoRandom}{Time ((d:hh:)mm:ss.S)
to execute \NumGeneratedProgramsPerSketch programs in \oneitrco
configuration. \label{tab:exec-time-sketchy-full-version:one-itr-co:random}}
\DefMacro{TableCaptionSketchyExecTimeFullVersionOneItrMasterCoRandom}{Time ((d:hh:)mm:ss.S)
to execute \NumGeneratedProgramsPerSketch programs in \oneitrmasterco
configuration.\label{tab:exec-time-sketchy-full-version:one-itr-master-co:random}}

\DefMacro{FloatEnvTableSketchyExecTimeFullVersionFullItrsStatic}{table*}
\DefMacro{FloatEnvTableSketchyExecTimeFullVersionFullItrsMasterStatic}{table*}
\DefMacro{FloatEnvTableSketchyExecTimeFullVersionFullItrsCoStatic}{table*}
\DefMacro{FloatEnvTableSketchyExecTimeFullVersionFullItrsMasterCoStatic}{table*}
\DefMacro{FloatEnvTableSketchyExecTimeFullVersionOneItrStatic}{table*}
\DefMacro{FloatEnvTableSketchyExecTimeFullVersionOneItrMasterStatic}{table*}
\DefMacro{FloatEnvTableSketchyExecTimeFullVersionOneItrCoStatic}{table*}
\DefMacro{FloatEnvTableSketchyExecTimeFullVersionOneItrMasterCoStatic}{table*}
\DefMacro{FontsizeTableSketchyExecTimeFullVersionFullItrsStatic}{small}
\DefMacro{FontsizeTableSketchyExecTimeFullVersionFullItrsMasterStatic}{small}
\DefMacro{FontsizeTableSketchyExecTimeFullVersionFullItrsCoStatic}{small}
\DefMacro{FontsizeTableSketchyExecTimeFullVersionFullItrsMasterCoStatic}{small}
\DefMacro{FontsizeTableSketchyExecTimeFullVersionOneItrStatic}{small}
\DefMacro{FontsizeTableSketchyExecTimeFullVersionOneItrMasterStatic}{small}
\DefMacro{FontsizeTableSketchyExecTimeFullVersionOneItrCoStatic}{small}
\DefMacro{FontsizeTableSketchyExecTimeFullVersionOneItrMasterCoStatic}{small}
\DefMacro{TableCaptionSketchyExecTimeFullVersionFullItrsStatic}{Time ((d:hh:)mm:ss.S)
to execute \NumGeneratedProgramsPerSketch programs generated in static approach in \fullitrs
configuration. \label{tab:exec-time-sketchy-full-version:full-itrs:static}}
\DefMacro{TableCaptionSketchyExecTimeFullVersionFullItrsMasterStatic}{Time ((d:hh:)mm:ss.S)
to execute \NumGeneratedProgramsPerSketch programs generated in static approach in \fullitrsmaster
configuration. \label{tab:exec-time-sketchy-full-version:full-itrs-master:static}}
\DefMacro{TableCaptionSketchyExecTimeFullVersionFullItrsCoStatic}{Time ((d:hh:)mm:ss.S)
to execute \NumGeneratedProgramsPerSketch programs generated in static approach in \fullitrsco
configuration. \label{tab:exec-time-sketchy-full-version:full-itrs-co:static}}
\DefMacro{TableCaptionSketchyExecTimeFullVersionFullItrsMasterCoStatic}{Time ((d:hh:)mm:ss.S)
to execute \NumGeneratedProgramsPerSketch programs generated in static approach in \fullitrsmasterco
configuration. \label{tab:exec-time-sketchy-full-version:full-itrs-master-co:static}}
\DefMacro{TableCaptionSketchyExecTimeFullVersionOneItrStatic}{Time ((d:hh:)mm:ss.S)
to execute \NumGeneratedProgramsPerSketch programs generated in static approach in \oneitr
configuration. \label{tab:exec-time-sketchy-full-version:one-itr:static}}
\DefMacro{TableCaptionSketchyExecTimeFullVersionOneItrMasterStatic}{Time ((d:hh:)mm:ss.S)
to execute \NumGeneratedProgramsPerSketch programs generated in static approach in \oneitrmaster
configuration. \label{tab:exec-time-sketchy-full-version:one-itr-master:static}}
\DefMacro{TableCaptionSketchyExecTimeFullVersionOneItrCoStatic}{Time ((d:hh:)mm:ss.S)
to execute \NumGeneratedProgramsPerSketch programs generated in static approach in \oneitrco
configuration. \label{tab:exec-time-sketchy-full-version:one-itr-co:static}}
\DefMacro{TableCaptionSketchyExecTimeFullVersionOneItrMasterCoStatic}{Time ((d:hh:)mm:ss.S)
to execute \NumGeneratedProgramsPerSketch programs generated in static approach in \oneitrmasterco
configuration.\label{tab:exec-time-sketchy-full-version:one-itr-master-co:static}}

\DefMacro{CaptionFigureExample}{An example of a \sketch and one of the \generatedprograms
from the \sketch.\label{fig:example}}
\DefMacro{CaptionFigureExampleSketch}{An example of a \sketch.\label{fig:example:sketch}}
\DefMacro{CaptionFigureExampleGenerated}{An example of a \generatedprogram.\label{fig:example:generated}}
\DefMacro{CaptionFigureSyntax}{Syntax for an imperative language with
\holes.\label{fig:syntax}}
\DefMacro{CaptionFigureSemantics}{Semantics for the simple language from
Figure~\ref{fig:syntax}.\label{fig:semantics}}
\DefMacro{CaptionFigureAPI}{API for writing \holes; a call to any of the methods in the
API instantiates an \eAST node.\label{fig:technique:api-list}}
\DefMacro{CaptionFigureAST}{\eAST corresponding to \CodeIn{logic} \hole from
Figure~\ref{fig:example:sketch}.\label{fig:technique:ast}}
\DefMacro{CaptionAlgorithmGeneration}{Generation algorithm.\label{algo:driver}}
\DefMacro{CaptionAlgorithmTesting}{\JIT-testing algorithm.\label{algo:testing}}
\DefMacro{FigureCaptionJITLevelTransition}{Common transitions
among optimization levels (L0-L4) of the tiered compilation
in \OracleJDK 7+.\label{fig:jit-level-transition}}
\DefMacro{CaptionFigureBugc67}{Bug \UseMacro{bugc67}.\label{fig:bugs:c67}}
\DefMacro{CaptionFigureBugsk4gen619}{Bug \UseMacro{bugsk4gen619}.\label{fig:bugs:sk4gen619}}
\DefMacro{CaptionFigureBugmath182}{Bug \UseMacro{bugmath182}.\label{fig:bugs:math182}}
\DefMacro{CaptionFigureBugcheckstyle106}{Bug \UseMacro{bugcheckstyle106}.\label{fig:bugs:checkstyle106}}
\DefMacro{CaptionFigureBugcodec292}{Bug \UseMacro{bugcodec292}.\label{fig:bugs:codec292}}
\DefMacro{CaptionFigureBugcompress208}{Bug \UseMacro{bugcompress208}.\label{fig:bugs:compress208}}
\DefMacro{CaptionFigureBugs}{Minimized detected \JIT-bugs. %
\label{fig:bugs}}

\DefMacro{FigureCaptionSketchyGenTime}{Percent of \nonopt to
generate \NumGeneratedProgramsPerSketch programs in various
configurations.\label{fig:sketchy:gen-time}}
\DefMacro{FigureCaptionTrackingHolesDetails}{Reachability of filled \holes in \generatedprograms from execution-based and static approaches respectively. \label{fig:tracking-holes-details}}

\newcommand{\RQGenerationTime}{RQ1\xspace}
\newcommand{\RQExtraction}{RQ2\xspace}

\newcommand{\RQBugsFound}{RQ3\xspace}

\newcommand{\NumHandCraftedSketches}{13\xspace}
\newcommand{\NumBugReportSketches}{four\xspace}
\newcommand{\NumOptimizationSketches}{six\xspace}
\newcommand{\TotalNumManuallyCreatedSketches}{23\xspace}
\newcommand{\TotalNumExtractedSketches}{\UseMacro{table-extraction-short-version-total-numTemplates}\xspace}
\newcommand{\TotalNumJavaProjects}{\UseMacro{table-extraction-short-version-total-numProjects}\xspace}
\newcommand{\TotalNumClassesUsedForExtraction}{\UseMacro{table-extraction-short-version-total-numFiles}\xspace}
\newcommand{\TotalNumMethodsUsedForExtraction}{\UseMacro{table-extraction-short-version-total-numMethods}\xspace}

\newcommand{\NumOfBugsFromManualWriting}{two\xspace}
\newcommand{\NumOfBugsFromExtraction}{four\xspace}
\newcommand{\NumPreviouslyUnknownBugs}{four\xspace}
\newcommand{\NumOfCVEsFromManualWriting}{one\xspace}
\newcommand{\NumOfCVEsFromExtraction}{one\xspace}
\newcommand{\TotalNumOfCVEs}{two\xspace}
\newcommand{\TotalNumOfBugs}{six\xspace}

\newcommand{\NumGeneratedProgramsPerSketch}{1{,}000\xspace}
\newcommand{\NumGeneratedProgramsPerExtractedSketch}{10\xspace}
\newcommand{\NumIterationsPerGeneratedProgram}{100{,}000\xspace}
\newcommand{\NumTotalGeneratedPrograms}{23{,}000\xspace}
\newcommand{\NumTotalGeneratedProgramsFromExtractedSketches}{\UseMacro{table-extraction-total-numGen}\xspace}
\newcommand{\NumTotalGeneratedProgramsFromExtractedSketchesThereotically}{54{,}190\xspace} %

\newcommand{\TotalGenTimeWithStatic}{around \Num{three} minutes\xspace}
\newcommand{\TotalGenTimeWithFullOpt}{around \Num{20} minutes\xspace}
\newcommand{\TotalGenTimeWithNonOpt}{over \Num{two} days\xspace}
\newcommand{\TotalExecTimeOnLFourAndLOneFullItrs}{almost \Num{four} and a half hours\xspace}
\newcommand{\TotalExecTimeOnLFourFullItrs}{around \Num{two} hours\xspace}

\newcommand{\TotalExecTimeOnLOneFullItrs}{around \Num{two} and a half hours\xspace}

\newcommand{\NumOfOptimizations}{three\xspace}

\DefMacro{bugc67}{m12gen61\xspace} %
\DefMacro{ReportOfBugc67}{\url{https://bugs.openjdk.java.net/browse/JDK-8239244} (Login required)} %
\DefMacro{CVEOfBugc67}{\url{https://cve.mitre.org/cgi-bin/cvename.cgi?name=CVE-2020-14792}}
\DefMacro{OracleCPUOfBugc67}{\url{https://www.oracle.com/security-alerts/cpuoct2020.html}}

\DefMacro{bugsk4gen619}{m4gen152\xspace}
\DefMacro{ReportOfBugsk4gen619}{\url{https://bugs.openjdk.java.net/browse/JDK-8258981}}
\DefMacro{PriorityOfBugsk4gen619}{P3}

\DefMacro{bugmath182}{math182\xspace} %
\DefMacro{ReportOfBugmath182}{\url{https://bugs.openjdk.java.net/browse/JDK-8271130} (Login required)} %
\DefMacro{CVEOfBugmath182}{\url{https://cve.mitre.org/cgi-bin/cvename.cgi?name=CVE-2022-21305}}
\DefMacro{OracleCPUOfBugmath182}{\url{https://www.oracle.com/security-alerts/cpujan2022.html}}

\DefMacro{bugcheckstyle106}{checkstyle106\xspace}
\DefMacro{ReportOfBugcheckstyle106}{\url{https://bugs.openjdk.java.net/browse/JDK-8271276}}
\DefMacro{PriorityOfBugcheckstyle106}{P2}

\DefMacro{bugcodec292}{codec292\xspace}
\DefMacro{ReportOfBugcodec292}{\url{https://bugs.openjdk.java.net/browse/JDK-8271459}}
\DefMacro{PriorityOfBugcodec292}{P2}

\DefMacro{bugcompress208}{compress208\xspace}
\DefMacro{ReportOfBugcompress208}{\url{https://bugs.openjdk.java.net/browse/JDK-8271926}}
\DefMacro{PriorityOfBugcompress208}{P3}

\newcommand{\ToolURL}{\url{https://github.com/EngineeringSoftware/jattack}}

\begin{document}

\title{\Title}

\author{Zhiqiang Zang}
\affiliation{%
\institution{The University of Texas at Austin}
\city{Austin}
\state{Texas}
\country{USA}
}
\email{zhiqiang.zang@utexas.edu}

\author{Nathan Wiatrek}
\affiliation{%
\institution{The University of Texas at Austin}
\city{Austin}
\state{Texas}
\country{USA}
}
\email{nwiatrek@utexas.edu}

\author{Milos Gligoric}
\affiliation{%
\institution{The University of Texas at Austin}
\city{Austin}
\state{Texas}
\country{USA}
}
\email{gligoric@utexas.edu}

\author{August Shi}
\affiliation{%
\institution{The University of Texas at Austin}
\city{Austin}
\state{Texas}
\country{USA}
}
\email{august@utexas.edu}

\begin{abstract}
We present \JTool, a framework that enables
template-based testing for compilers.
Using \JTool, a developer writes a \sketchprogram that describes a set
of programs to be generated and given as test inputs to a
compiler.
Such a framework enables developers to incorporate their domain
knowledge on testing compilers, giving a basic program structure that
allows for exploring complex programs that can trigger sophisticated
compiler optimizations.
A developer writes a \sketchprogram in the host language (Java) that
contains \holes to be filled by \JTool.
Each \hole, written using a domain-specific language, constructs a \node
within an extended abstract syntax tree (\eAST).  An \eAST \node
defines the search space for the \hole, i.e., a set of expressions and
values.
\JTool generates programs by executing \sketches and filling each \hole
by randomly choosing expressions and values (available within the
search space defined by the \hole).
Additionally, we introduce several optimizations to reduce \JTool's
generation cost.
While \JTool could be used to test various compiler features, we
demonstrate its capabilities in helping test just-in-time (\JIT) Java
compilers, whose optimizations occur at runtime after a sufficient
number of executions.
Using \JTool, we have found \TotalNumOfBugs critical bugs that were
confirmed
by Oracle developers.
\Capitalize{\NumPreviouslyUnknownBugs} of them were previously
unknown, including \TotalNumOfCVEs unknown CVEs (Common
Vulnerabilities and Exposures).
\JTool shows the power of combining developers' domain knowledge (via
templates) with random testing to detect bugs in \JIT compilers.
\end{abstract}



\maketitle

\vspace{-3pt}
\section{Introduction}
\label{sec:introduction}

Compilers are among the most critical components in the software
development toolchain, and
their correctness is of utmost importance.
A bug in a compiler might lead to a crash during the translation, an
incorrect output (native code does not match the semantics
of the program written by developers~\cite{BugJDK8251535}), or
even expose security vulnerabilities in the generated code.

Compiler developers have manually written thousands of tests, i.e.,
programs in the compiler's target programming language, as to
check for correctness~\cite{jtregWebpage, GCCTestSuitesWebpage,
LLVMTestingInfraGuide}.  Although manually-written tests nicely
capture \emph{developers' intuition} of what programs are expected to
trigger corner cases, it is
time-consuming to write a large number of such tests. As a result,
researchers and practitioners have developed a number of automated
techniques for testing compilers~\cite{YangPLDI11CSmith,
Groce12SwarmTesting, Livinskii20YARPGen,
ChenPLDI16DifferentialTestingOfJVM,
ChenICSE19DeepDifferentialTestingOfJVMImplementations,
LePLDI14CompilerValidationViaEMI,
YoshikawaETAL03RandomGeneratorForJIT, PadhyeISSTA19JQFZest,
BrownPLDI20VerifiedRangeAnalysisForJavaScriptJITs,
Holler12FuzzingWithCodeFragments}, namely by generating a large number
of programs on which the compiler can run.
Existing compiler-testing techniques mainly fall in two categories:
grammar-based~\cite{YoshikawaETAL03RandomGeneratorForJIT,
YangPLDI11CSmith, Livinskii20YARPGen,
Holler12FuzzingWithCodeFragments} and
mutation-based~\cite{ChenPLDI16DifferentialTestingOfJVM,
ChenICSE19DeepDifferentialTestingOfJVMImplementations,
LePLDI14CompilerValidationViaEMI, PadhyeISSTA19JQFZest}.
The former group generates programs from scratch following the
production rules available in the language grammar.  The latter group
usually starts with some seed programs and then genetically mutates
the seeds.

Although existing approaches are valuable, they have shortcomings.
First, they provide limited ways for compiler developers to fully
embed their knowledge into the testing process.
Second, they frequently support only a subset of the language grammar
(e.g., only integer values~\cite{YangPLDI11CSmith}).

We present \JTool, a framework that enables \emph{compiler testing
using \sketches}.
Using \JTool, a developer writes a \emph{\sketchprogram} (\sketch for
short) that describes a \emph{set of concrete programs} to be used as
inputs
to a compiler.
Unlike prior work, \emph{our framework enables developers to express richer manual
tests for compilers}.
Our design of a \sketch
captures the developers' intuition in very much the same way as
manually-written tests but provides an opportunity to express variants
of those tests that can be obtained by \emph{testing} the \sketches.
The goal is similar to parameterized unit
testing~\cite{Tillmann05ParameterizedUnitTests}, where developers
manually write unit tests that encapsulate some features they want to
test in their code but have parameters that a backend framework
explores as to obtain deeper testing around the insights the
developers initially provide. Unlike with mutation-based fuzzers,
compiler developers can use
\sketches to specify
exactly how to generate program variants.
(Figure~\ref{fig:example} shows an example \sketch, which is discussed
in detail in Section~\ref{sec:example}.)
\JTool complements existing automated compiler-testing techniques that
can provide a structure of a program on which \JTool can further build
\sketches.

In \JTool, a developer writes a \sketch in the host language (Java),
which contains \emph{\holes} to be filled by \JTool.
Each \hole is written in a domain-specific language (\DSL) embedded in
the host language, i.e., we do not change the syntax, compiler, nor runtime
environment of the host language.
We define the \DSL as a set of \APIs that allow developers to specify
characteristics of the \hole they want explored in a \sketch, where each
API call produces an instance of an \emph{extended abstract syntax
tree} (\eAST) \node; an \eAST \node bounds the search space for the
\hole, i.e., defines a set of possible statements, expressions, and
values.  As an example, consider the following API call that defines a
\hole: \CodeIn{relation(intVal(), intVal(), GT, LT).eval()},
\noindent
which represents a logical relation between two integer literals (each
can take any value between \CodeIn{Integer.MIN\_VALUE} and
\CodeIn{Integer.MAX\_VALUE}) using either $>$ (\CodeIn{GT}) or $<$
(\CodeIn{LT}) relational operators; this hole evaluates to a boolean.
Using the \JTool's API, each \hole can then be type-checked by the
host compiler.

\JTool is useful for augmenting testing for many complex compiler
features as it leverages the developer insights from the provided
\sketches. For our evaluation, we focus specifically on testing
\emph{just-in-time} (\JIT) compilers. Unlike traditional ahead-of-time
compilers that translate a program into native code prior to
deployment~\cite{Aho07compilers}, \JIT compilers translate the program
during execution~\cite{Aycok03ABriefHistoryOfJustinTime}. Certain
optimizations only occur after executing specific program structures a
sufficient number of times.

\JTool takes two inputs: (1)~a \sketch, and (2)~an iteration count
($N$), i.e., the number of times each \generatedprogram will be
executed in a loop to ensure that \JIT compilation is triggered.
\JTool generates a program by repeatedly \emph{executing}
the \sketch (up to $N$ times) and filling each \hole, when the \hole
is reached the first time, by \emph{randomly} choosing expressions and
values available within the search space defined by the \hole. (In
theory, a \sketch can be exhaustively explored, but it is generally
not feasible.) Next, to detect any \JIT-related bugs,
each \generatedprogram is executed $N$ times using different \JIT
compilers, potentially detecting bugs via differential
testing~\cite{Mckeeman98DifferentialTesting}.

We also introduce \NumOfOptimizations optimizations into \JTool to
reduce the generation cost. The first optimization, \emph{\stopearly},
involves stopping after detecting that further generation
would not fill any more \holes.
The second optimization, \emph{\hotfilling}, dynamically transforms
the \sketch when a \hole is reached the very first time; the \API call
is transformed into the concrete expression that the call would
produce.
The final optimization, \emph{\solveraid}, uses a modern constraint
solver (Z3~\cite{DeMoura08Z3})
to detect \holes for conditional statements (e.g., \CodeIn{if}) that
always evaluate to a constant value.

To demonstrate \JTool's capabilities in testing \JIT compilers, we
wrote \TotalNumManuallyCreatedSketches \sketches. We focused on
interesting Java language features and took inspiration from existing
tests for the Java compiler. We report the cost of generation and
execution, as well as benefits of our optimizations; our optimizations
reduce the \emph{generation time} by
\UseMacro{table-timing-gen-reduction-full_opt-sum}\%. We used the
\generatedprograms as inputs to multiple commercial \JIT compilers,
including the \OracleJDK \JIT compiler.
Using just our own \sketches, we were able to discover
\NumOfBugsFromManualWriting bugs in the \OracleJDK \JIT compiler.
These bugs were confirmed and fixed by Oracle developers, and
\NumOfCVEsFromManualWriting of the bugs was previously unknown and
acknowledged on Oracle's list of CVEs.

We also evaluated how well \JTool can be used for automated compiler
testing by \emph{extracting \sketches from existing Java
projects}.
This evaluation is inspired by mutation
testing~\cite{DeMillo78MutationTesting, JiaAndHarman2011TSE,
JustETAL2014FSE, AndrewsETAL2006TSE,
ChenPLDI16DifferentialTestingOfJVM, LePLDI14CompilerValidationViaEMI,
Le15FindingDeepCompilerBugsViaGuidedStochasticProgramMutation}, where
we essentially ``mutate'' existing code to construct different tests
for compilers. Note that, unlike in traditional mutation
testing, \holes in our case are filled by randomly generating values
and expressions. Moreover, mutation testing uses a predefined set of
mutation operators while \JTool provides a way for a user to write
holes anywhere in their code following their intuition (e.g., the way
we wrote templates for \JIT compilers). Each hole has its own set of
values (and the set is determined by the developer, not by a tool).

Using \TotalNumJavaProjects open-source Java projects that span a wide
variety of domains and therefore use of different Java language
features, we automatically extracted \TotalNumExtractedSketches
\sketches. By running these \sketches through \JTool, we
found \NumOfBugsFromExtraction more bugs in the \OracleJDK \JIT
compiler (out of which only one was previously known) including
\NumOfCVEsFromExtraction previously unknown CVE recently
acknowledged by Oracle.

\vspace{5pt}
\noindent
The key contributions of this paper include:

\begin{itemize}[topsep=2pt,itemsep=0pt,partopsep=0ex,parsep=0ex,leftmargin=*]
\item \textbf{Framework}. We introduce \JTool, the framework for \sketching tests for compilers.
\JTool is designed to complement
manually-written tests and
blend developer's intuition (via templates) and random testing to
increase likelihood to detect bugs in Java \JIT compilers.
\item \textbf{Programming and execution models}.  We introduce a
programming and an execution model to integrate
\sketches entirely in the host language (Java), without changing the syntax
or the runtime environment.  \Sketches are like
manually-written programs with \holes; each \hole, expressed using a \DSL,
builds an \eAST \node
that specifies values that the \hole can take
(i.e., defines a search space).
We introduced \NumOfOptimizations optimizations that are applied
when generating programs from \sketches.
\item \textbf{Use case}. We implemented \JTool for the Java
programming language and applied it to testing Java \JIT compilers.
We evaluated \JTool by writing \TotalNumManuallyCreatedSketches \sketchprograms.
Our results show that the optimizations substantially reduce test
generation time, making \JTool practical. Furthermore, we discovered
\NumOfBugsFromManualWriting bugs in the \OracleJDK \JIT compiler.
\item \textbf{\Sketch extraction}. We evaluated \JTool as
an automated framework for \JIT compiler testing by automatically extracting
\sketches from existing Java projects.  Using
\TotalNumExtractedSketches \sketches from
\TotalNumJavaProjects open-source Java projects, \JTool discovered
four more bugs in the Oracle JDK \JIT compiler.
\end{itemize}

\noindent
\JTool is available at \ToolURL.

\section{Example}
\label{sec:example}

Figure~\ref{fig:example:sketch} shows a \sketchprogram that we wrote
while developing \JTool for Java.  Our motivation for this \sketch was
to exercise Java \JIT optimizations for programs that use local arrays
and static variables.
It is important to note that \emph{every \sketch for \JTool} \emph{is
a valid Java program}.
This \sketch uses static methods (e.g., \CodeIn{logic}) that are
defined in the \CodeIn{\aTool.\aAttack} class.
As such, the
Java compiler can also type-check the \sketch.

\begin{figure}[t]
\begin{subfigure}{\linewidth}
\begin{lstlisting}[language=sketchy]
import static (*@\aTool@*).(*@\aAttack@*).*;
public class C {
  static int s1;
  static int s2;
  @Entry
  public static int m() {
    int[] arr1 = { s1++, s2, (*@\colorbox{colorhole1}{\highlightApi{intVal}().\highlightEval{eval}()}@*)(*@\annotation{1}{annot:hole1:unfilled}@*),(*@ \label{line:staticupdate} @*) (*@ \label{line:holes1:start} @*)
                     (*@\colorbox{colorhole2}{\highlightApi{intVal}().\highlightEval{eval}()}@*)(*@\annotation{2}{annot:hole2:unfilled}@*), (*@\colorbox{colorhole3}{\highlightApi{intVal}().\highlightEval{eval}()}@*)(*@\annotation{3}{annot:hole3:unfilled}@*) };(*@ \label{line:holes1:end} @*)
    for (int i = 0; i < arr1.length; ++i)
      if ((*@\colorbox{colorhole4}{\highlightApi{logic}(\highlightApi{relation}(\highlightApi{intId}(), \highlightApi{intId}(), \highlightApi{LE}),}@*) (*@ \label{line:holes2:start} @*)
                 (*@\colorbox{colorhole4}{\highlightApi{relation}(\highlightApi{intId}(), \highlightApi{intId}(), \highlightApi{LE}),}@*)
                 (*@\colorbox{colorhole4}{\highlightApi{AND}, \highlightApi{OR}).\highlightEval{eval}()}@*)(*@\annotation{4}{annot:hole4:unfilled}@*)) (*@ \label{line:holes2:end} @*)
        arr1[i] &=
            (*@\colorbox{colorhole5}{\highlightApi{arithmetic}(\highlightApi{intId}(), \highlightApi{intId}(), \highlightApi{ADD}, \highlightApi{MUL}).\highlightEval{eval}()}@*)(*@\annotation{5}{annot:hole5:unfilled}@*); (*@ \label{line:holes3} @*)
    return 0; } }
\end{lstlisting}
\caption{\UseMacro{CaptionFigureExampleSketch}}
\end{subfigure}
\begin{subfigure}{\linewidth}
\begin{lstlisting}[language=sketchy]
import static (*@\aTool@*).(*@\aAttack@*).*;
public class C {
  static int s1;
  static int s2;
  @Entry
  public static int m() {
    int[] arr1 = { s1++, s2, (*@\colorbox{colorhole1}{45350238}@*)(*@\annotation{1}{annot:hole1:filled}@*),
                     (*@\colorbox{colorhole2}{681339300}@*)(*@\annotation{2}{annot:hole2:filled}@*), (*@\colorbox{colorhole3}{125652422}@*)(*@\annotation{3}{annot:hole3:filled}@*) };
    for (int i = 0; i < arr1.length; ++i)
      if ((*@\colorbox{colorhole4}{arr1[3] <= s2 || s2 <= arr1[2]}@*)(*@\annotation{4}{annot:hole4:filled}@*))
        arr1[i] &= (*@\colorbox{colorhole5}{arr1[1] * s1}@*)(*@\annotation{5}{annot:hole5:filled}@*);
    return 0; } }
\end{lstlisting}
\caption{\UseMacro{CaptionFigureExampleGenerated}}
\end{subfigure}
\caption{\UseMacro{CaptionFigureExample}}
\end{figure}

The \sketch contains five \emph{\holes} representing places where
\JTool should generate expressions, filling them in to create a
concrete \emph{\generatedprogram}. Three \holes are between
lines~\ref{line:holes1:start} and \ref{line:holes1:end}, one between
lines~\ref{line:holes2:start} and \ref{line:holes2:end}, and one on
line~\ref{line:holes3}. The number of holes is equal to the number of
\CodeIn{eval} invocations. The \CodeIn{eval} invocation as well as the
type information of the expression calling the \CodeIn{eval} allows
JAttack to tell a \hole from actual code.

The first three \holes are defined by the \CodeIn{intVal} method
calls; each call to \CodeIn{intVal} represents a \hole that will be
filled by an integer literal; note that without any arguments,
\CodeIn{intVal} produces an integer between
\CodeIn{Integer.MIN\_VALUE} and \CodeIn{Integer.MAX\_VALUE}. The next
\hole (lines~\ref{line:holes2:start}-\ref{line:holes2:end}) defines a
logical ``and'' or ``or'' expression (\CodeIn{logic} with the
\CodeIn{AND} and \CodeIn{OR} arguments) between two relational
expressions. Each relational expression (\CodeIn{relation}) connects
two free integer variables \CodeIn{intId}, which can be \CodeIn{s1},
\CodeIn{s2}, \CodeIn{i}, or any element of \CodeIn{arr1} (the array
index is randomly picked between 0 and the size of the array) at this point, using the \CodeIn{<=}
operator (\CodeIn{LE}). The final \hole (line~\ref{line:holes3}) is an
arithmetic expression (\CodeIn{arithmetic}) of two free integer
variables, which are combined using either a $+$ or a $*$ operator
(\CodeIn{ADD} or \CodeIn{MUL}). We describe more details on what type
of expressions we support for \holes along with our API in
Section~\ref{sec:technique}.

\JTool \emph{generates} programs through an \emph{execution-based}
model. In other words, \JTool fills the \holes in a \sketch after
executing the \sketch.
(Unlike static generation, an execution-based model prunes the search
space by only filling \holes reached during execution. See
Section~\ref{sec:execution-based-vs-static-generation} for details.)
A \sketch must have a \sketchmethod,
annotated with \entryannot as shown in the example (method \CodeIn{m}).
When the execution reaches any unfilled \hole, \JTool
generates a valid expression for that \hole based on the used API
calls.
When all reachable \holes are filled (see
Section~\ref{sec:generation:procedure} for how this is determined),
\JTool outputs the corresponding \generatedprogram. \JTool then calls
the \sketchmethod again to generate the next program up to the
specified maximum number of programs.
An example of a \generatedprogram that \JTool outputs for the \sketch in
Figure~\ref{fig:example:sketch} is shown in
Figure~\ref{fig:example:generated}. The numbered circles in the
\generatedprogram correspond to the same ones next to \holes in
Figure~\ref{fig:example:sketch}.

Finally, to detect any \JIT-related bugs, \JTool \emph{executes} each
\generatedprogram starting from the same entry method a large number
of times
using different \JIT compilers,
potentially detecting bugs via differential
testing~\cite{Mckeeman98DifferentialTesting}.
The large number of re-executions is necessary as to trigger \JIT
optimizations. In Java, a \JVM starts executing a program with an
interpreter and monitors the execution for ``hot'' code sections,
i.e., code that is frequently executed. The \JIT compiler
then optimizes ``hot'' sections.
Through repeated executions of the generated method \CodeIn{m}, the \generatedprogram shown
in Figure~\ref{fig:example:generated} revealed a bug in the \OracleJDK \JIT
compiler, crashing the \JVM.

\section{\JTool Framework}
\label{sec:technique}

\JTool introduces test \sketching, a way to define a \emph{set of
programs} used for testing compilers. \Sketches written for \JTool
could be useful for testing other program analysis tools, but we leave
such studies for future work.
We designed \JTool guided by the following requirements:
(1)~developers
decide where the \holes should be placed and bound the search space of
each \hole, and (2)~the domain-specific language (\DSL) for writing
\holes is non-intrusive, i.e.,
it requires no changes to the
host compiler.

In this section, we
describe our programming and execution models
(Section~\ref{sec:programming:model}), implementation for Java
(Section~\ref{sec:execution:model}), the generation procedure
(Section~\ref{sec:generation:procedure}), the optimizations for
generation (\ref{sec:optimization}) and the overall \JIT-testing
procedure (Section~\ref{sec:jit-testing}).

\subsection{Programming and Execution Models}
\label{sec:programming:model}

We define the syntax and operational semantics of a simple imperative
language with an extension to support \sketches. Note the language
shown here includes only integer type for ease of presentation; we
greatly extend the scope in our implementation for Java.  The simple
imperative language and extensions represent the foundations for
supporting \sketches for general imperative languages, and our
implementation in Java, described in later sections, is based
on these extensions.

\MyPara{Syntax}
Figure~\ref{fig:syntax} defines the syntax of the simple language.  A program
is a sequence of zero or more statements.  Each statement is either an
assignment, conditional, goto, or halt.  An expression in a
program can combine relational, arithmetic, and logical operators.  On
top of these basic imperative features, the language also introduces
the concept of a \hole, denoted with \aHole{}.
These \holes can be used around a sequence of comma-separated
expressions, or they can be around individual operands, where
\aHole{\CodeIn{c}} represents a \hole for a literal/constant and
\aHole{\CodeIn{v}} represents a \hole for a variable.

\begin{figure}[t]
\begin{lstlisting}[language=syntax]
Stmts := (LABEL:)? Stmt ";" Stmts | (*@$\epsilon$@*)
Stmt := AssignStmt | IfStmt | GotoStmt | halt
AssignStmt := ID "=" Exp
IfStmt := "if" "(" Exp ")" LABEL
GotoStmt := goto LABEL

Exp := ExpBasic | "(*@$[\![$@*)" ExpAlt "(*@$]\!]$@*)"
ExpBasic := ExpBasic Op Operand | Operand
ExpAlt := ExpAlt "," ExpBasic | ExpBasic

Operand := ID | NUM | "(*@$[\![$@*)c(*@$]\!]$@*)" | "(*@$[\![$@*)v(*@$]\!]$@*)"
Op := "+" | "-" | "||" | "&&" | "==" | "!=" | "<"
\end{lstlisting}
\caption{\UseMacro{CaptionFigureSyntax}}
\end{figure}

\begin{figure}[t]
\begin{footnotesize}

\inference[ASSIGN\_EXP: ]{ \aConfig{pc+1, I, M, L} exp => \aConfig{pc', I', M, L} v}
{\aConfig{pc, I, M, L} id = exp => \aConfig{pc', I', M, L} id = v}
\inference[ASSIGN\_VAL: ]{ M' = M[v/id]}
{\aConfig{pc, I, M, L} id = v => \aConfig{pc+1, I, M', L} I(pc+1)}
\inference[GOTO: ]{ }
{\aConfig{pc, I, M, L} \mathrm{goto}\ l => \aConfig{L(l), I, M, L} I(L(l))}
\inference[IF\_EXP: ]{ \aConfig{pc+1, I, M, L} exp => \aConfig{pc', I', M,  L} v}
{\aConfig{pc, I, M, L} \mathrm{if}(exp)\ N => \aConfig{pc', I', M, L} \mathrm{if}(v)\ N}
\inference[IF\_VAL: ]{ pc' = (v == 0)\ ?\ pc + 1 : L(l) }
{\aConfig{pc, I, M, L} \mathrm{if}(v)\ l => \aConfig{pc', I', M,  L} I(pc')}
\inference[C\_HOLE: ]{ val = alt(range(MIN, MAX)) & I' = I[val/pc] }
{\aConfig{pc, I, M, L} \aHole{c} => \aConfig{pc, I', M, L} val}
\inference[V\_HOLE: ]{ id = alt(identifiers(M)) & I' = I[id/pc] }
{\aConfig{pc, I, M, L} \aHole{v} => \aConfig{pc, I', M, L} id}
\inference[E\_HOLE: ]{ e = alt([e_1, e_2, ...., e_n]) & I' = I[e/pc] }
{\aConfig{pc, I, M, L} \aHole{e_1, e_2, ..., e_n} => \aConfig{pc, I', M, L} e}

\end{footnotesize}
\caption{\UseMacro{CaptionFigureSemantics}}
\end{figure}

\MyPara{Semantics (core language)}
Valid programs can only use integer literals.  We define the state of
a program with the following configuration: $\langle pc, I, M, L
\rangle$, where $pc$ is the program counter (initially $0$), $I$ is the
instruction memory (i.e., mapping from program counter to statements or
expressions),
$M$ is the main memory (i.e., mapping from identifiers to integer
values), and $L$ is a map from labels to indices in $I$.
Prior to the execution, statements and expression indices are placed
into $I$ by performing a pre-order traversal of the program's abstract syntax tree (first
statement is at index 0).  Also, $L$ is initialized to map each label
to the appropriate index in $I$.
We also use the following operations: (1)~map lookup $\_(val)$, and
(2)~map update $\_[val/\_]$.

Figure~\ref{fig:semantics} shows the operational semantics of the
language.  For simplicity, the rules do not include error handling.
The assignment statement simply updates the value of a variable in
memory.  The conditional statement evaluates the expression and then
jumps if the expression evaluates to true ($val \ne 0$).  The goto
statement unconditionally jumps to the statement with the specified
label.
We do not show the rules for computing basic expressions, as we assume
the same semantics as in the C programming language.
The halt statement terminates the execution.

\MyPara{Semantics (\sketch language)}
We define several utility functions: (1)~$\mathtt{identifiers}(M)$ -
returns a list of available variable names in $M$ at the point of an
invocation, (2)~$\mathtt{range}(x, y)$ - returns a list of integers
between $x$ and $y$, and (3)~$\mathtt{alt}([...])$ - takes a
sequence as input and outputs one of its elements.

A \hole for an integer literal (C\_HOLE) evaluates to an integer
literal and rewrites itself to that literal.
A \hole for a variable (V\_HOLE) evaluates to an available
identifier and rewrites itself to that identifier.
Finally, a \hole for an expression (E\_HOLE) evaluates to one of the given
expressions (and rewrites itself to that expression).
Note that the rewrite rules are such that the entire \hole is replaced
with the choice of a concrete expression upon execution, so the \hole no
longer exists after the first evaluation, ensuring
that each hole evaluates only to a single expression per execution.
The program in $I$ at the end of execution is the \generatedprogram
in the same language as the original \sketch.

\MyPara{Filling a \hole}
Given a list of \candidates for a \hole, we need to explore different
\candidates every time we execute the program, which would in turn
rewrite the \sketch into new concrete programs that we can later use
for testing compilers.
While one can try and systematically explore all the possible
\candidates, the search space can be incredibly large (e.g., for
\aHole{\CodeIn{c}} the range of possible integers go from \CodeIn{MIN}
to \CodeIn{MAX}), especially when considering
combinations of \candidates chosen across all \holes in the program.

For this work, we choose \candidates for a \hole randomly.  Random
exploration has been found effective in prior
work~\cite{YoshikawaETAL03RandomGeneratorForJIT, YangPLDI11CSmith,
PadhyeISSTA19JQFZest, Livinskii20YARPGen}. We keep re-executing the
\sketch to rewrite into concrete programs up until we reach a
specified limit for number of \generatedprograms.
Note that each execution of a \sketch is independent of other
executions, i.e., any modifications to the \sketch during one run is
\emph{not} observable in another run.

\MyPara{Example}
Consider the following example in our language: \CodeIn{s1 =
\aHole{c}; s2 = \aHole{c}; if (\aHole{v} < \aHole{v}) l9; l9: halt;}.
Executing this \sketch once might generate: \CodeIn{s1 = 45350238; s2
= 681339300; if (s1 < s2) l9; l9: halt;}.  Another execution can lead to a different \generatedprogram: \CodeIn{s1
= 125652422; s2 = 23297; if (s2 < s2) l9; l9: halt;}.

\subsection{\JTool Implementation for Java}
\label{sec:execution:model}

We implement the semantics of \JTool for the host Java programming
language. To support the concept of \holes while integrating it into
Java,
we introduce a set of API methods that construct \holes.

\begin{figure}[t]
\begin{lstlisting}[language=api-list]
BoolVal boolVal();
IntVal intVal(int min, int max);
IntVal intVal();
BoolId boolId(String... names);
IntId intId(String... names);
...
<T> RefId<T> refId(Class<T> type, String... names);
<T extends Number> BAriExp<T> arithmetic(
    Exp<T> left, Exp<T> right, Op... ops);
<T extends Number> RelExp<T> relation(
    Exp<T> left, Exp<T> right, Op... ops);
LogExp logic(
    Exp<Boolean> left, Exp<Boolean> right, Op... ops);
...
<T> ExprStmt exprStmt(Exp<T> exp);
IfStmt ifStmt(
    Exp<Boolean> cond, Stmt thenStmt, Stmt elseStmt);
WhileStmt whileStmt(Exp<Boolean> cond, Stmt body);
...
<T> Exp<T> alt(Exp<T>... exps);
Stmt alt(Stmt... stmts);
\end{lstlisting}
\caption{\UseMacro{CaptionFigureAPI}}
\end{figure}

Figure~\ref{fig:technique:api-list} shows a subset of the full \API we
provide. This \API represents a \DSL that maps to our simple
imperative language. All methods in the \API return an instance of a
\node rooting an \emph{extended abstract syntax tree} (\eAST).
An \CodeIn{Exp<Integer>} \node corresponds to an expression
(evaluating to integer due to Java typing, so \CodeIn{Exp<Boolean>} is
the same for boolean type).
An \CodeIn{IntVal} (extends \CodeIn{Exp<Integer>}) \node and an
\CodeIn{IntId} (extends \CodeIn{Exp<Integer>}) \node correspond to an
integer literal and variable, respectively. We define \CodeIn{BoolVal}
and \CodeIn{BoolId} to correspond to the boolean type for Java.
\CodeIn{BAriExp<Integer>} (extends \CodeIn{Exp<Integer>}) is for
binary arithmetic expressions, while \CodeIn{RelExp} (extends
\CodeIn{Exp<Boolean>}) and \CodeIn{LogExp} (extends \CodeIn{Exp<Boolean>})
are for relational and logical expressions.
These \nodes are therefore placeholders for the actual, concrete expressions to
be generated at runtime, so they represent \holes to be filled. For
example, an \CodeIn{IntVal} \node created using the method
\CodeIn{intVal} represents \aHole{\CodeIn{c}}, a \hole that can
evaluate to any integer from \CodeIn{Integer.MIN\_VALUE} to
\CodeIn{Integer.MAX\_VALUE}. We also provide an \CodeIn{intVal} API
that can specify the range of integer values,
and an \CodeIn{intId} API that can enumerate available variables (at
any point) by analyzing bytecode, when no variable is specified.

While specific \API methods can create corresponding
\nodes, e.g., \CodeIn{arithmetic} for \CodeIn{BAriExp}, we also
provide method \CodeIn{alt} that can choose from a provided list of
\CodeIn{Exp<Integer>} or \CodeIn{Exp<Boolean>} \nodes, like the
semantics for an expression \hole (E\_HOLE) in our simple imperative
language.

Although we illustrate the implementation of \JTool using \CodeIn{int}
and \CodeIn{boolean}, we support any other primitive type, e.g.,
\CodeIn{long} and \CodeIn{double}, or any reference type. For
instance, \CodeIn{refId} can enumerate available variables (at any
point) with type \CodeIn{T} that can be specified using argument
\CodeIn{Class<T>~type}, e.g., \CodeIn{refId(String.class)} returns a
\CodeIn{RefId<String>} node corresponding to any available
\CodeIn{String} variables at this execution point.
We provide API methods to create statements as well, such as
\CodeIn{exprStmt}, \CodeIn{ifStmt} and \CodeIn{whileStmt}.
Furthermore, one can extend our implementation to include more
language constructs in Java, as along as they can be represented in an
\eAST.

As an alternative, we originally designed our \API to use a list of
concrete Java expressions to choose from, e.g., \CodeIn{alt(i++, j++)}.
However, these expressions would get executed and result in
side-effects, and the final execution would not match executing the
corresponding \generatedprogram with the concrete expressions
substituting for the \hole, so we abandoned that direction.

Instead, when using \eAST \nodes, we do not actually generate an
expression to fill a \hole until the \CodeIn{eval} method is invoked
on the \node, e.g., \CodeIn{intVal().eval()}. Only after calling
\CodeIn{eval} does a concrete \node get generated for that \hole. Once
generated, the node is interpreted to compute the result of the
expression.
Furthermore, all subsequent calls to the same API method (from the
same location) will always return the same \node. For our Java
implementation, we define a \hole to be where the developer calls
\CodeIn{eval} for a built \eAST.
\noindent The \eAST constructed for an API call represents a range of
\candidates to fill the \hole.
As an example, consider the \hole specified by the \CodeIn{logic} call
(lines~\ref{line:holes2:start}-\ref{line:holes2:end} in
Figure~\ref{fig:example:sketch}).
Executing the \CodeIn{logic} method returns a root \node of an \eAST,
illustrated in Figure~\ref{fig:technique:ast}.
\Candidates for the \hole are obtained by recursively obtaining
\candidates for \nodes in subtrees and combining them together.
\begin{wrapfigure}{l}{4.6cm}
\centering

\begin{forest}
for tree={ellipse, draw=none, fill=red!20, thick, text centered, text depth=.25ex, inner sep=0, outer sep=0, minimum width=4em, minimum height=1.5em, font=\tt\footnotesize,
for leaves={rectangle, fill=blue!20, minimum width=3em},
edge={semithick},
annot/.style={label={[align=left, font=\tt\footnotesize]above:{#1}}},
s sep=0.6em,
l=2em,
l sep=0em
}
[LogExp, annot={\&\&, ||}
[RelExp, annot={<=}
[IntId]
[IntId]
]
[RelExp, annot={<=}
[IntId]
[IntId]
]
]
\end{forest}

\caption{\UseMacro{CaptionFigureAST}}
\end{wrapfigure}
In this example, the \CodeIn{RelExp} \nodes would result in
\candidates that combine choice of integer variables combined with the
specified operators (just \CodeIn{LE} in the example); the top-level
\CodeIn{LogExp} \node would use the returned \candidates and combine
with the specified \CodeIn{AND} or \CodeIn{OR} to create the final
\candidates. This \eAST structure corresponds to an expression \hole
in our imperative language, namely the following:

\noindent \CodeIn{\aHole{\aHole{v} <= \aHole{v} \&\& \aHole{v} <= \aHole{v}, \aHole{v} <= \aHole{v} || \aHole{v} <= \aHole{v}}}.

\noindent
Unlike our simple imperative language, our \API provides syntactic
sugars to describe a large set of similar \candidates without having
to enumerate all of them by specifying multiple operators at once (see
\CodeIn{Op... ops} in Figure~\ref{fig:technique:api-list}).

\subsection{Generation Procedure}
\label{sec:generation:procedure}

Figure~\ref{algo:driver} shows the overall algorithm for \JTool's
\CodeIn{Generate} function that executes a \sketch repeatedly
to generate concrete program instances. The input to \CodeIn{Generate} is a
\sketch $T$ and the number of programs to generate $N$.  The
output is a set of \generatedprograms $G$.

Function \CodeIn{Generate} starts by initializing the empty set of
\generatedprograms $G$ and then capturing the initial global state of
the \sketch $T$ into variable $S$
(line~\ref{algo:driver:line:capture}). We currently support capturing
static fields with primitive and array types as the global state;
future work could also capture reference types~\cite{bell14vmvm,
bell18crochet}. We capture the global state to be used later when
generating programs, ensuring the generation of each (out of $N$)
program is done from the clean state. (We use the Java reflection
mechanism to capture the state.) Additionally, \CodeIn{Generate} finds
the \sketchmethod (line~\ref{algo:driver:line:find}), which is the
entry point for executing $T$ (in our Java implementation, this is the
method annotated with \CodeIn{@Entry}),
and also counts the total number of \holes that should be filled in
the \sketch (line~\ref{algo:driver:line:count}).

Next, \CodeIn{Generate} repeatedly calls \CodeIn{RunTemplate}, which
executes the \sketch, resulting in a \generatedprogram that is added
to $G$. Assume that a \sketch always terminates, which can be
guaranteed through carefully specifying the search space for the
\holes in conditions,
the overall loop in \CodeIn{Generate} ends when the number of uniquely
\generatedprograms has reached the maximum necessary number $N$.
We set a timeout, for \CodeIn{RunTemplate}, as it
might not be feasible to generate the specified number of unique
programs.

Before calling \CodeIn{RunTemplate} in each iteration,
\CodeIn{Generate} sets the global state to be the same as the initial
global state $S$ (line~\ref{algo:driver:line:resetstate}). Setting the
initial state to $S$ ensures that
subsequent runs of \CodeIn{RunTemplate} always start the generation
process, which executes the same \sketchmethod, in the clean state.
\newline
\MyPara{Example} Consider the \sketch from
Figure~\ref{fig:example:sketch}. The \sketch has a static variable
\CodeIn{s1} that is modified (line~\ref{line:staticupdate}).
Subsequent executions should make sure \CodeIn{s1} starts at $0$
again, otherwise they would not be starting at the same state and
would not generate programs that are
even possible.

Function \CodeIn{RunTemplate} is responsible for generating a single
concrete program from the given \sketch $T$. First, it initializes $H$
as an empty mapping from \holes to their filled expressions
(line~\ref{algo:driver:line:init-map}).
\CodeIn{RunTemplate} then
sets an intermediate program $Q$ to be the input \sketchprogram $T$ to start
with (line~\ref{algo:driver:line:init-intermediate-program}), and then it repeatedly executes the entry method $entryMeth$ on
$Q$ (line~\ref{algo:driver:line:exec}).
The \CodeIn{ExecEntryMethod} returns a mapping $H'$ of \holes it
filled to the actual expressions.
\newline
\MyPara{Example} In
Figure~\ref{fig:example:sketch}, executing the \hole on
line~\ref{line:holes1:start} would result in a mapping of that \hole
to concrete value $45350238$ (line~\ref{annot:hole1:filled} in
Figure~\ref{fig:example:generated}).

The overall mapping $H$ gets updated with $H'$.
If all \holes have been filled, then the loop terminates
(line~\ref{algo:driver:line:stopping}). The reason for executing the
\sketchmethod $entryMeth$ many times is to ensure all \holes that can
be reached get filled. Eventually, our goal is to execute a
corresponding \generatedmethod up to \CodeIn{MAX\_NUM\_ITERATIONS}
times as to trigger \JIT optimizations
(Section~\ref{sec:jit-testing}). Some \holes may only be reachable
after multiple iterations, so executing just once would not fill those
\holes.
\newline
\MyPara{Example} Consider the \sketch from Figure \ref{fig:example:sketch}. The
last \hole (line~\ref{line:holes3}) could be skipped in the first run
because the condition
(line~\ref{line:holes2:start}-\ref{line:holes2:end}) is evaluated to
\CodeIn{false}. However, the \hole{} could be filled later when static
variable \CodeIn{s1} gets updated (line~\ref{line:staticupdate}), making the
condition \CodeIn{true}.

Some choice of \candidate for a \hole may possibly make another, later
\hole unreachable, putting it in dead code. \JTool may fill a \hole in
a condition, such as for an \CodeIn{if} statement,
that always evaluates to \CodeIn{false}, and therefore any \holes
within the block of these conditional statements cannot be
reached.
To prevent the execution from \CodeIn{RunTemplate} from continuously
executing while being unable to fill those unreachable \holes,
\CodeIn{RunTemplate} stops after the \CodeIn{MAX\_NUM\_ITERATIONS}
maximum number of iterations.
Having unfilled dead-code \holes in a \generatedprogram is fine
because such code should never even be executed within the maximum
number of iterations later (and if it is
executed, that would indicate a bug in the \JIT
compiler). \CodeIn{RunTemplate} does stop earlier when all \holes are
filled (line~\ref{algo:driver:line:stopping}).

Three optimizations are introduced to reduce generation cost
(line~\ref{algo:driver:line:optimizations:start}-\ref{algo:driver:line:optimizations:end})
and we describe the optimizations in detail in
Section~\ref{sec:optimization}. Note that $Q$ is an intermediate
program, and we do not directly return $Q$. As such, we can optimize
and make extra changes in $Q$
to speed up generation, and these changes do \emph{not} belong in a
final generated program $P$.

The final returned program $P$ is then the original \sketch $T$ with
all its \holes filled using the mapping $H$ (computed using function
\CodeIn{ApplyFilledHoles} in line
\ref{algo:driver:line:applyfilledHoles}, which is not shown).
Essentially, each node corresponding to a filled \hole can output the
concrete code snippet for the expression it currently holds, and the
\hole expression in the \sketch $T$ gets replaced with this concrete
code snippet. \CodeIn{Generate} then takes the returned program $P$
and adds it to the running set of \generatedprograms $G$. Note that
\CodeIn{Generate} will keep calling \CodeIn{RunTemplate} until
obtaining a sufficient number of programs; each time,
\CodeIn{Generate} will use the fresh \sketchprogram $T$, which has no
filled \holes, as to create a brand new \generatedprogram.

\begin{figure}[t]
\begin{algorithmic}[1]
\Input \sketchprogram $T$
\Input number of programs to generate $N$
\Output set of \generatedprograms $G$
\Function{Generate}{$T$, $N$}
\State $G \gets \emptyset$
\State $S \gets \Call{CaptureGlobalState}{T}$ \label{algo:driver:line:capture}
\State $\mathit{entryMeth} \gets \Call{FindEntryMethod}{T}$ \label{algo:driver:line:find}
\State $num \gets \Call{CountHoles}{T}$ \label{algo:driver:line:count}
\Repeat
\State \Call{resetGlobalState}{$S$} \label{algo:driver:line:resetstate}
\State $P \gets \Call{RunTemplate}{T, \mathit{entryMeth}, num}$ \label{algo:driver:line:callRunSketch}
\State $G \gets G \cup \{P\}$
\Until{$|G| =$ $N$}
\State \Return $G$
\EndFunction
\\
\Input \sketchprogram $T$
\Input entry method $\mathit{entryMeth}$
\Input number of \holes $num$
\Output \generatedprogram $P$
\Function{RunTemplate}{$T, \mathit{entryMeth}, num$}
\State $H \gets \{\}$ \label{algo:driver:line:init-map}
\State $seenStates \gets \emptyset$
\State $Q \gets T$ \label{algo:driver:line:init-intermediate-program}
\For{$i$ $\gets$ $1$ to \textit{MAX\_NUM\_ITERATIONS} } \label{algo:driver:line:loop}
\State $H' \gets \Call{ExecEntryMethod}{\mathit{entryMeth}, Q}$ \label{algo:driver:line:exec}
\State $H \gets H \cup H'$
\If{$|H| = num$} \Break \label{algo:driver:line:stopping}
\EndIf
\State $R \gets \Call{CaptureGlobalState}{Q}$ \label{algo:driver:line:optimizations:start}
\If{$R \in seenStates$} \label{algo:driver:line:earlystopping}
\Break
\EndIf
\State $seenStates \gets seenStates \cup \{R\}$
\State $Q \gets \Call{HotFill}{Q, H}$ \label{algo:driver:line:hotfill}
\State $Q \gets \Call{RemoveDeadCode}{Q, H}$ \label{algo:driver:line:deadcode} \label{algo:driver:line:optimizations:end}
\EndFor
\State \Return \Call{ApplyFilledHoles}{$T, H$} \label{algo:driver:line:applyfilledHoles}
\EndFunction
\end{algorithmic}
\caption{\UseMacro{CaptionAlgorithmGeneration}}
\end{figure}

\subsection{Optimizations for Generation}
\label{sec:optimization}

We develop \NumOfOptimizations optimizations to speed up the
generation process. Note that these optimizations all apply only for a
single run of \CodeIn{RunTemplate}, to just a single
\generatedprogram at a time.  Also, these optimizations do \emph{not} impact
the generated programs, they only speed up the generation process.

\MyPara{\Stopearly}
We can have an even earlier stopping condition based on the insight
that if the global state after execution is the same as an already
seen state, then any future run would lead to the same behavior (as a
previous execution). Starting execution in the same global state
cannot lead to new executions that fill new \holes.  In
\CodeIn{RunTemplate}, we keep track of the seen global states in
\CodeIn{seenStates} and check the global state after each execution
(line~\ref{algo:driver:line:earlystopping} in
Figure~\ref{algo:driver}).  This type of program state hashing has
been extensively used in software model
checking~\cite{Iosif02SymmetryReductionCriteriaForSoftwareModelChecking}.

\MyPara{\Hotfilling}
In our preliminary experiments, we found that executing a
\sketchmethod many times is time-consuming, especially compared to
executing the \generatedmethod as part of our evaluation.
The extra overhead comes from repeated executions of our Java \API
methods that build and evaluate \eAST \nodes. Recall during generation
the filled \hole is not rewritten into the concrete expression, but
just evaluated to produce the same value as the concrete expression.
The filled \holes get replaced with the actual code only when the
entire \sketch gets translated into a new \generatedprogram
(line~\ref{algo:driver:line:applyfilledHoles} in
Figure~\ref{algo:driver}). Thus, while our implementation ensures that
repeated execution of the same API method returns the same \eAST node,
invoking the \CodeIn{eval} method to evaluate the node is still
expensive compared to evaluating the concrete code that replaces the \hole in the \generatedprogram.

The \hotfilling optimization replaces the \hole at runtime
(during generation) with the concrete expression when the \hole is
evaluated for the first time, so that execution in the following
iterations can use concrete code rather than invoking our Java API
methods, including the \CodeIn{eval} method that evaluates the filled \hole.
In \CodeIn{RunTemplate}, we invoke the method \CodeIn{HotFill}
(line~\ref{algo:driver:line:hotfill} in Figure~\ref{algo:driver}) on
the resulting $Q$ after execution that finds all API calls with set
\nodes and, using the mapping of \holes to expressions $H$, replaces
those calls with the concrete expressions.
Then, using interfaces provided from package \CodeIn{javax.tools},
e.g., \CodeIn{javax.tools.JavaCompiler}, we implement an in-memory
Java compiler, file manager, and associated class loader to
dynamically compile $Q$ and then reload this modified \sketch's class,
resulting in a new $Q$. The next iteration starts from the new $Q$ as
the \sketch (line~\ref{algo:driver:line:exec} in
Figure~\ref{algo:driver}). This technique is conceptually similar to
``quickening'' optimization implemented in self-optimizing
interpreters~\cite{Brunthaler10Quickening, Humer14Truffle}.

\MyPara{\Solveraid}
In our preliminary experiments, we also noticed a significant number
of \generatedprograms with conditional expressions that are trivially
false, e.g., \CodeIn{(var1 > var1)}. The body of such conditional
statements would never be executed, so it is unnecessary to execute
any further to fill \holes within statements guarded by that
condition. After executing the \sketchmethod and obtaining filled
\holes in $H$, we invoke function \CodeIn{RemoveDeadCode} to eliminate
any such dead code in the program $Q$
(line~\ref{algo:driver:line:deadcode} in Figure~\ref{algo:driver}),
completely rewriting the body into an empty statement. This technique
is conceptually similar to partial
evaluation~\cite{Jones93PartialEvaluation}.
We leverage a modern SMT solver (Z3~\cite{DeMoura08Z3}) in our
implementation to determine whether any conditional expression is
satisfiable or not, eliminating code in case the expression is
unsatisfiable. Note that we only temporarily remove
the code as a means to speed up generating a single program. The
returned \generatedprogram does not have any unreachable code removed.
Later calls to \CodeIn{RunTemplate} always start with the same
\sketch $T$ that has all the code still there.

\subsection{\JIT-Testing Procedure}
\label{sec:jit-testing}

Revealing \JIT-related bugs requires not just programs but also
executing those programs many times. For each \generatedprogram, we
iterate though different \JIT compilers. For each \JIT compiler, we
repeatedly execute the \generatedmethod, hashing the output of each
execution (a \generatedmethod's return value is always encoded into an
integer) into a running total. After executing
\CodeIn{MAX\_NUM\_ITERATIONS} times (the same limit in
Figure~\ref{algo:driver}), we capture the global state of
\generatedprogram (values of all static fields) and encode it within a
checksum value, adding this to the running total. The final total
representing the combination of all the executions.

For a given \generatedprogram, we use differential
testing~\cite{Mckeeman98DifferentialTesting} to
check if the running totals computed from all \JIT compilers are all
the same. Any difference should indicate that the \generatedprogram
detected a bug within some \JIT compiler.
However, the program may itself be non-deterministic, i.e., having
different outputs when run multiple times on the same \JIT compiler.
Note that non-determinism was only observed from the \sketches
extracted from existing Java projects, e.g., involving randomness,
while our manually created \sketches are guaranteed to be
deterministic. To avoid being misled by non-determinism, when there
are differences in output across different \JIT compilers, we choose a
\JIT compiler as a reference point and run the program twice using
that same compiler. If the outputs from running on the same \JIT
compiler differ, then output differences between \JIT compilers do not
indicate a bug. While this step may potentially miss
detecting some bugs, it gives higher guarantees that reported bugs are
true bugs.

Besides checking for differences in final running totals, we also report a
bug if the execution crashes on some \JIT compiler. Executing any
unfilled \hole (left as the API method call in the \generatedprogram)
would also trigger a crash, because an unfilled \hole should not be
reachable. The ultimate output of the entire \JIT-testing procedure is
a subset of \generatedprograms that expose a bug in one of the input
\JIT compilers.

\section{Extracting \Sketches}
\label{sec:extraction}

We also evaluate \JTool for automated end-to-end compiler testing.
Namely, we provide an approach to automatically extract
\sketches from existing Java projects.  The code written
for these projects are naturally representative of Java
language features and can be used as the foundation for
\sketches that can find bugs in Java \JIT compilers. We can then
also easily and automatically scale up the number of \sketches
to run through \JTool.

Given a Java class, we first parse all the available methods in the
class to detect potential \holes. For each statement, we recursively
convert each subexpression into the corresponding \hole, starting from
the leaves of the expression tree. For example, the expression
\CodeIn{a + b} would be converted into \CodeIn{arithmetic(intId(),
intId()).eval()} (specifying no operator argument means using all
valid operators), which matches the expression structure. Note that
the final call to \CodeIn{eval} is on the outermost API call,
allowing for the greatest space of combination of
values that \JTool can explore.

After inserting \holes into the Java class, we then scan the class for
available static methods, which are the candidate \sketchmethods. If
the static method takes any parameters, we insert additional \emph{parameter
methods}, one for each parameter; a parameter method returns a
concrete value for the corresponding parameter type upon
execution. For primitive values, we leverage \JTool to provide a
possible value, e.g., if the parameter is an \CodeIn{int} type we
simply use \CodeIn{intVal} to represent an integer value. For
non-primitive types, i.e., classes, we search if such classes have
default constructors or constructors with primitive arguments that we can
simply use to create an instance of that class. If there are no such
constructors, we search from other classes for a public static method
that returns an instance of the class. If none of the above cases
applies, we then use \CodeIn{null} as the concrete value.

Note that our approach for extracting \sketches and then using
them to generate concrete programs is similar in nature to concepts in
mutation testing~\cite{DeMillo78MutationTesting}, where existing
programs are mutated into other similar programs through syntactic
mutation operators~\cite{ChenPLDI16DifferentialTestingOfJVM,
LePLDI14CompilerValidationViaEMI,
Le15FindingDeepCompilerBugsViaGuidedStochasticProgramMutation}.
Conceptually, converting a program into a \sketchprogram and then
generating additional programs through \JTool is like mutating the
original program. However, our way of generating programs leverages
the capabilities of \JTool and its DSL to allow more expressive
transformations that are beyond traditional mutation operators, and
more similar to what can be generated using higher-order mutation
operators~\cite{Jia08HigherOrderMutationTesting}.

\section{Experimental Setup}
\label{sec:setup}

We briefly describe our evaluation setup.

\subsection{Evaluation Subjects}
We wrote \NumHandCraftedSketches \sketches that exercise Java language
features. We also studied the available optimizations used in the
\OracleJDK \JIT compiler, creating \NumOptimizationSketches \sketches
whose basic structure would trigger those optimizations while
including \holes for \JTool to explore. Finally, we studied existing
bug reports for \JIT-related bugs, creating \NumBugReportSketches
\sketches by modifying the programs attached to bug reports to include
\holes. In our evaluation, we refer to the \sketches based on our own
understanding of Java and the compiler developers' intuition of
optimizations using prefix ``M''. We refer to the \sketches based on
bug reports using prefix ``B''. Overall, we created
\TotalNumManuallyCreatedSketches \sketches, with the goal to evaluate
the effectiveness of our optimizations.

To evaluate \JTool in the context of automated test generation, we
collect
\sketches automatically from existing Java code. We use
\TotalNumJavaProjects open-source Java Maven projects from GitHub to
extract \sketches from their classes.
Given a project or a module of a multi-module Maven project, we find
classes defined in all ``.java'' files.
We extract \sketches from these classes following the procedures
described in Section~\ref{sec:extraction}.

\subsection{Configuring \JTool}
For each \sketch we created ourselves, we configure \JTool to generate
\NumGeneratedProgramsPerSketch concrete programs (\CodeIn{N} in
Figure~\ref{algo:driver}). While generation is fastest when we turn on
all \NumOfOptimizations generation optimizations
(Section~\ref{sec:optimization}), we also evaluate running generation
without any optimization and with each optimization separately,
measuring each one's effectiveness.
For each of the \generatedprograms, we execute it
\NumIterationsPerGeneratedProgram times (\CodeIn{MAX\_NUM\_ITERATIONS}
in Figure~\ref{algo:driver}) on different \JIT compilers. The \JIT compilers we evaluate on
are \OracleJDK, \OpenJDK, and \OpenJNine, all based on JDK
11.0.8.

For \sketches extracted from existing Java projects, we follow the
same approach, except we configure \JTool to generate only
\NumGeneratedProgramsPerExtractedSketch concrete programs from each
\sketchprogram, because of the large number of \sketches, and we test
only on the latest \OracleJDK, which was 16.0.2 at that time.

In our evaluation, we configure \OracleJDK to restrict the specific
tiers, L1 and L4, using the option \CodeIn{-XX:TieredStopAtLevel}, in
order to test \COne and \CTwo
compilers~\cite{JavaHotSpotVMWhitePaper}, respectively.
We treat each restricted tier configuration for \OracleJDK as
conceptually a new \JIT compiler for use in our \JIT-testing
procedure (Section~\ref{sec:jit-testing}).

We run all experiments on a 64-bit Ubuntu 18.04.1 desktop with an
Intel(R) Core(TM) i7-8700 CPU @3.20GHz and 64GB RAM. For all time
measurements, we run the evaluation five times and report the average
of those times.

\section{Evaluation}
\label{sec:evaluation}

We evaluate \JTool by asking:

\begin{enumerate}[label={\textbf{RQ\arabic*}:}, leftmargin=*]
\item How efficient is \JTool at generating programs and executing
those \generatedprograms with different \JIT compilers?
\item How well can \JTool be used for \emph{automated} compiler
testing via extracted \sketches from a large number of
existing Java programs, and how does it compare with
state-of-the-art automated \JIT compiler testing?
\item What critical bugs does \JTool detect in \JIT compilers?
\end{enumerate}

\noindent
We address \RQGenerationTime as to better understand how efficient
\JTool is at generating programs from \sketches, as well as the impact
of our optimizations for generation, and to understand the efficiency
of \JTool's testing procedure. We address \RQExtraction to understand
how well \JTool can be used for automated compiler testing and compare
the effectiveness with tools used in industry. We address
\RQBugsFound to understand the bugs that we exposed.
\JTool and all \JIT-bugs we detected, including associated \sketches
and \generatedprograms, are available at \ToolURL.

\subsection{Performance and Optimizations}
\label{sec:evaluation:generation}

Table~\ref{tab:timing-gen} shows the
time for \JTool to generate \NumGeneratedProgramsPerSketch programs
for each of our manually created \TotalNumManuallyCreatedSketches
\sketches. The different columns show the total time to generate all
\NumGeneratedProgramsPerSketch programs when using different
generation optimizations (Section~\ref{sec:optimization}). Namely,
``\nonopt'' means no optimizations, ``\stopearlyopt'' means using only
\stopearly, ``\hotfillingopt'' means using only \hotfilling, and
``\solveraidopt'' means using only \solveraid.
The final column for ``\fullopt'' is the time when using all
optimizations. In addition to time, for each optimization column, we
also show the percentage of time reduced relative to ``\nonopt'' time
(the higher the reduction the better). The final row shows the sum of
generation time across all \sketches and the overall reduction
over this total time.

\bgroup
\def\arraystretch{1.2}

\begin{\UseMacro{FloatEnvTableTimingGen}}[t]
\centering
\caption{\UseMacro{CaptionTableTimingGen}}
\begin{\UseMacro{FontsizeTableTimingGen}}
\begin{tabular}{lrrrrrrrrr}
\toprule
\multirow{2}{*}{\textbf{\UseMacro{TableTimingGenHeadSketch}}}& \textbf{\UseMacro{TableTimingGenHeadNonOpt}}
& \multicolumn{2}{c}{\textbf{\UseMacro{TableTimingGenHeadStopEarlyOpt}}}
& \multicolumn{2}{c}{\textbf{\UseMacro{TableTimingGenHeadHotFillingOpt}}}
& \multicolumn{2}{c}{\textbf{\UseMacro{TableTimingGenHeadSolverAidOpt}}}
& \multicolumn{2}{c}{\textbf{\UseMacro{TableTimingGenHeadFullOpt}}}
\\
\cmidrule(lr){2-2}\cmidrule(lr){3-4}\cmidrule(lr){5-6}\cmidrule(lr){7-8}\cmidrule(lr){9-10}
& \textbf{\UseMacro{TableTimingGenHeadTime}}
& \textbf{\UseMacro{TableTimingGenHeadTime}}
& \textbf{\UseMacro{TableTimingGenHeadReduction}}
& \textbf{\UseMacro{TableTimingGenHeadTime}}
& \textbf{\UseMacro{TableTimingGenHeadReduction}}
& \textbf{\UseMacro{TableTimingGenHeadTime}}
& \textbf{\UseMacro{TableTimingGenHeadReduction}}
& \textbf{\UseMacro{TableTimingGenHeadTime}}
& \textbf{\UseMacro{TableTimingGenHeadReduction}}
\\
\midrule
\UseMacro{table-timing-gen-SkJDK8243670-name}
& \UseMacro{table-timing-gen-SkJDK8243670-genTime-non_opt}
& \UseMacro{table-timing-gen-SkJDK8243670-genTime-stop_early_opt}
& \UseMacro{table-timing-gen-SkJDK8243670-reduction-stop_early_opt}
& \UseMacro{table-timing-gen-SkJDK8243670-genTime-hot_filling_opt}
& \UseMacro{table-timing-gen-SkJDK8243670-reduction-hot_filling_opt}
& \UseMacro{table-timing-gen-SkJDK8243670-genTime-solver_aid_opt}
& \UseMacro{table-timing-gen-SkJDK8243670-reduction-solver_aid_opt}
& \UseMacro{table-timing-gen-SkJDK8243670-genTime-full_opt}
& \UseMacro{table-timing-gen-SkJDK8243670-reduction-full_opt}
\\
\midrule
\UseMacro{table-timing-gen-SkJDK8248552-name}
& \UseMacro{table-timing-gen-SkJDK8248552-genTime-non_opt}
& \UseMacro{table-timing-gen-SkJDK8248552-genTime-stop_early_opt}
& \UseMacro{table-timing-gen-SkJDK8248552-reduction-stop_early_opt}
& \UseMacro{table-timing-gen-SkJDK8248552-genTime-hot_filling_opt}
& \UseMacro{table-timing-gen-SkJDK8248552-reduction-hot_filling_opt}
& \UseMacro{table-timing-gen-SkJDK8248552-genTime-solver_aid_opt}
& \UseMacro{table-timing-gen-SkJDK8248552-reduction-solver_aid_opt}
& \UseMacro{table-timing-gen-SkJDK8248552-genTime-full_opt}
& \UseMacro{table-timing-gen-SkJDK8248552-reduction-full_opt}
\\
\midrule
\UseMacro{table-timing-gen-SkJDK8251535-name}
& \UseMacro{table-timing-gen-SkJDK8251535-genTime-non_opt}
& \UseMacro{table-timing-gen-SkJDK8251535-genTime-stop_early_opt}
& \UseMacro{table-timing-gen-SkJDK8251535-reduction-stop_early_opt}
& \UseMacro{table-timing-gen-SkJDK8251535-genTime-hot_filling_opt}
& \UseMacro{table-timing-gen-SkJDK8251535-reduction-hot_filling_opt}
& \UseMacro{table-timing-gen-SkJDK8251535-genTime-solver_aid_opt}
& \UseMacro{table-timing-gen-SkJDK8251535-reduction-solver_aid_opt}
& \UseMacro{table-timing-gen-SkJDK8251535-genTime-full_opt}
& \UseMacro{table-timing-gen-SkJDK8251535-reduction-full_opt}
\\
\midrule
\UseMacro{table-timing-gen-SkJDK8255763-name}
& \UseMacro{table-timing-gen-SkJDK8255763-genTime-non_opt}
& \UseMacro{table-timing-gen-SkJDK8255763-genTime-stop_early_opt}
& \UseMacro{table-timing-gen-SkJDK8255763-reduction-stop_early_opt}
& \UseMacro{table-timing-gen-SkJDK8255763-genTime-hot_filling_opt}
& \UseMacro{table-timing-gen-SkJDK8255763-reduction-hot_filling_opt}
& \UseMacro{table-timing-gen-SkJDK8255763-genTime-solver_aid_opt}
& \UseMacro{table-timing-gen-SkJDK8255763-reduction-solver_aid_opt}
& \UseMacro{table-timing-gen-SkJDK8255763-genTime-full_opt}
& \UseMacro{table-timing-gen-SkJDK8255763-reduction-full_opt}
\\
\midrule
\UseMacro{table-timing-gen-Sk1-name}
& \UseMacro{table-timing-gen-Sk1-genTime-non_opt}
& \UseMacro{table-timing-gen-Sk1-genTime-stop_early_opt}
& \UseMacro{table-timing-gen-Sk1-reduction-stop_early_opt}
& \UseMacro{table-timing-gen-Sk1-genTime-hot_filling_opt}
& \UseMacro{table-timing-gen-Sk1-reduction-hot_filling_opt}
& \UseMacro{table-timing-gen-Sk1-genTime-solver_aid_opt}
& \UseMacro{table-timing-gen-Sk1-reduction-solver_aid_opt}
& \UseMacro{table-timing-gen-Sk1-genTime-full_opt}
& \UseMacro{table-timing-gen-Sk1-reduction-full_opt}
\\
\midrule
\UseMacro{table-timing-gen-Sk2-name}
& \UseMacro{table-timing-gen-Sk2-genTime-non_opt}
& \UseMacro{table-timing-gen-Sk2-genTime-stop_early_opt}
& \UseMacro{table-timing-gen-Sk2-reduction-stop_early_opt}
& \UseMacro{table-timing-gen-Sk2-genTime-hot_filling_opt}
& \UseMacro{table-timing-gen-Sk2-reduction-hot_filling_opt}
& \UseMacro{table-timing-gen-Sk2-genTime-solver_aid_opt}
& \UseMacro{table-timing-gen-Sk2-reduction-solver_aid_opt}
& \UseMacro{table-timing-gen-Sk2-genTime-full_opt}
& \UseMacro{table-timing-gen-Sk2-reduction-full_opt}
\\
\midrule
\UseMacro{table-timing-gen-Sk3-name}
& \UseMacro{table-timing-gen-Sk3-genTime-non_opt}
& \UseMacro{table-timing-gen-Sk3-genTime-stop_early_opt}
& \UseMacro{table-timing-gen-Sk3-reduction-stop_early_opt}
& \UseMacro{table-timing-gen-Sk3-genTime-hot_filling_opt}
& \UseMacro{table-timing-gen-Sk3-reduction-hot_filling_opt}
& \UseMacro{table-timing-gen-Sk3-genTime-solver_aid_opt}
& \UseMacro{table-timing-gen-Sk3-reduction-solver_aid_opt}
& \UseMacro{table-timing-gen-Sk3-genTime-full_opt}
& \UseMacro{table-timing-gen-Sk3-reduction-full_opt}
\\
\midrule
\UseMacro{table-timing-gen-Sk4-name}
& \UseMacro{table-timing-gen-Sk4-genTime-non_opt}
& \UseMacro{table-timing-gen-Sk4-genTime-stop_early_opt}
& \UseMacro{table-timing-gen-Sk4-reduction-stop_early_opt}
& \UseMacro{table-timing-gen-Sk4-genTime-hot_filling_opt}
& \UseMacro{table-timing-gen-Sk4-reduction-hot_filling_opt}
& \UseMacro{table-timing-gen-Sk4-genTime-solver_aid_opt}
& \UseMacro{table-timing-gen-Sk4-reduction-solver_aid_opt}
& \UseMacro{table-timing-gen-Sk4-genTime-full_opt}
& \UseMacro{table-timing-gen-Sk4-reduction-full_opt}
\\
\midrule
\UseMacro{table-timing-gen-Sk5-name}
& \UseMacro{table-timing-gen-Sk5-genTime-non_opt}
& \UseMacro{table-timing-gen-Sk5-genTime-stop_early_opt}
& \UseMacro{table-timing-gen-Sk5-reduction-stop_early_opt}
& \UseMacro{table-timing-gen-Sk5-genTime-hot_filling_opt}
& \UseMacro{table-timing-gen-Sk5-reduction-hot_filling_opt}
& \UseMacro{table-timing-gen-Sk5-genTime-solver_aid_opt}
& \UseMacro{table-timing-gen-Sk5-reduction-solver_aid_opt}
& \UseMacro{table-timing-gen-Sk5-genTime-full_opt}
& \UseMacro{table-timing-gen-Sk5-reduction-full_opt}
\\
\midrule
\UseMacro{table-timing-gen-Sk6-name}
& \UseMacro{table-timing-gen-Sk6-genTime-non_opt}
& \UseMacro{table-timing-gen-Sk6-genTime-stop_early_opt}
& \UseMacro{table-timing-gen-Sk6-reduction-stop_early_opt}
& \UseMacro{table-timing-gen-Sk6-genTime-hot_filling_opt}
& \UseMacro{table-timing-gen-Sk6-reduction-hot_filling_opt}
& \UseMacro{table-timing-gen-Sk6-genTime-solver_aid_opt}
& \UseMacro{table-timing-gen-Sk6-reduction-solver_aid_opt}
& \UseMacro{table-timing-gen-Sk6-genTime-full_opt}
& \UseMacro{table-timing-gen-Sk6-reduction-full_opt}
\\
\midrule
\UseMacro{table-timing-gen-Sk7-name}
& \UseMacro{table-timing-gen-Sk7-genTime-non_opt}
& \UseMacro{table-timing-gen-Sk7-genTime-stop_early_opt}
& \UseMacro{table-timing-gen-Sk7-reduction-stop_early_opt}
& \UseMacro{table-timing-gen-Sk7-genTime-hot_filling_opt}
& \UseMacro{table-timing-gen-Sk7-reduction-hot_filling_opt}
& \UseMacro{table-timing-gen-Sk7-genTime-solver_aid_opt}
& \UseMacro{table-timing-gen-Sk7-reduction-solver_aid_opt}
& \UseMacro{table-timing-gen-Sk7-genTime-full_opt}
& \UseMacro{table-timing-gen-Sk7-reduction-full_opt}
\\
\midrule
\UseMacro{table-timing-gen-Sk8-name}
& \UseMacro{table-timing-gen-Sk8-genTime-non_opt}
& \UseMacro{table-timing-gen-Sk8-genTime-stop_early_opt}
& \UseMacro{table-timing-gen-Sk8-reduction-stop_early_opt}
& \UseMacro{table-timing-gen-Sk8-genTime-hot_filling_opt}
& \UseMacro{table-timing-gen-Sk8-reduction-hot_filling_opt}
& \UseMacro{table-timing-gen-Sk8-genTime-solver_aid_opt}
& \UseMacro{table-timing-gen-Sk8-reduction-solver_aid_opt}
& \UseMacro{table-timing-gen-Sk8-genTime-full_opt}
& \UseMacro{table-timing-gen-Sk8-reduction-full_opt}
\\
\midrule
\UseMacro{table-timing-gen-Sk9-name}
& \UseMacro{table-timing-gen-Sk9-genTime-non_opt}
& \UseMacro{table-timing-gen-Sk9-genTime-stop_early_opt}
& \UseMacro{table-timing-gen-Sk9-reduction-stop_early_opt}
& \UseMacro{table-timing-gen-Sk9-genTime-hot_filling_opt}
& \UseMacro{table-timing-gen-Sk9-reduction-hot_filling_opt}
& \UseMacro{table-timing-gen-Sk9-genTime-solver_aid_opt}
& \UseMacro{table-timing-gen-Sk9-reduction-solver_aid_opt}
& \UseMacro{table-timing-gen-Sk9-genTime-full_opt}
& \UseMacro{table-timing-gen-Sk9-reduction-full_opt}
\\
\midrule
\UseMacro{table-timing-gen-Sk10-name}
& \UseMacro{table-timing-gen-Sk10-genTime-non_opt}
& \UseMacro{table-timing-gen-Sk10-genTime-stop_early_opt}
& \UseMacro{table-timing-gen-Sk10-reduction-stop_early_opt}
& \UseMacro{table-timing-gen-Sk10-genTime-hot_filling_opt}
& \UseMacro{table-timing-gen-Sk10-reduction-hot_filling_opt}
& \UseMacro{table-timing-gen-Sk10-genTime-solver_aid_opt}
& \UseMacro{table-timing-gen-Sk10-reduction-solver_aid_opt}
& \UseMacro{table-timing-gen-Sk10-genTime-full_opt}
& \UseMacro{table-timing-gen-Sk10-reduction-full_opt}
\\
\midrule
\UseMacro{table-timing-gen-Sk11-name}
& \UseMacro{table-timing-gen-Sk11-genTime-non_opt}
& \UseMacro{table-timing-gen-Sk11-genTime-stop_early_opt}
& \UseMacro{table-timing-gen-Sk11-reduction-stop_early_opt}
& \UseMacro{table-timing-gen-Sk11-genTime-hot_filling_opt}
& \UseMacro{table-timing-gen-Sk11-reduction-hot_filling_opt}
& \UseMacro{table-timing-gen-Sk11-genTime-solver_aid_opt}
& \UseMacro{table-timing-gen-Sk11-reduction-solver_aid_opt}
& \UseMacro{table-timing-gen-Sk11-genTime-full_opt}
& \UseMacro{table-timing-gen-Sk11-reduction-full_opt}
\\
\midrule
\UseMacro{table-timing-gen-Sk12-name}
& \UseMacro{table-timing-gen-Sk12-genTime-non_opt}
& \UseMacro{table-timing-gen-Sk12-genTime-stop_early_opt}
& \UseMacro{table-timing-gen-Sk12-reduction-stop_early_opt}
& \UseMacro{table-timing-gen-Sk12-genTime-hot_filling_opt}
& \UseMacro{table-timing-gen-Sk12-reduction-hot_filling_opt}
& \UseMacro{table-timing-gen-Sk12-genTime-solver_aid_opt}
& \UseMacro{table-timing-gen-Sk12-reduction-solver_aid_opt}
& \UseMacro{table-timing-gen-Sk12-genTime-full_opt}
& \UseMacro{table-timing-gen-Sk12-reduction-full_opt}
\\
\midrule
\UseMacro{table-timing-gen-Sk55-name}
& \UseMacro{table-timing-gen-Sk55-genTime-non_opt}
& \UseMacro{table-timing-gen-Sk55-genTime-stop_early_opt}
& \UseMacro{table-timing-gen-Sk55-reduction-stop_early_opt}
& \UseMacro{table-timing-gen-Sk55-genTime-hot_filling_opt}
& \UseMacro{table-timing-gen-Sk55-reduction-hot_filling_opt}
& \UseMacro{table-timing-gen-Sk55-genTime-solver_aid_opt}
& \UseMacro{table-timing-gen-Sk55-reduction-solver_aid_opt}
& \UseMacro{table-timing-gen-Sk55-genTime-full_opt}
& \UseMacro{table-timing-gen-Sk55-reduction-full_opt}
\\
\midrule
\UseMacro{table-timing-gen-SkOptLockCoarsening-name}
& \UseMacro{table-timing-gen-SkOptLockCoarsening-genTime-non_opt}
& \UseMacro{table-timing-gen-SkOptLockCoarsening-genTime-stop_early_opt}
& \UseMacro{table-timing-gen-SkOptLockCoarsening-reduction-stop_early_opt}
& \UseMacro{table-timing-gen-SkOptLockCoarsening-genTime-hot_filling_opt}
& \UseMacro{table-timing-gen-SkOptLockCoarsening-reduction-hot_filling_opt}
& \UseMacro{table-timing-gen-SkOptLockCoarsening-genTime-solver_aid_opt}
& \UseMacro{table-timing-gen-SkOptLockCoarsening-reduction-solver_aid_opt}
& \UseMacro{table-timing-gen-SkOptLockCoarsening-genTime-full_opt}
& \UseMacro{table-timing-gen-SkOptLockCoarsening-reduction-full_opt}
\\
\midrule
\UseMacro{table-timing-gen-SkOptLoopPeeling-name}
& \UseMacro{table-timing-gen-SkOptLoopPeeling-genTime-non_opt}
& \UseMacro{table-timing-gen-SkOptLoopPeeling-genTime-stop_early_opt}
& \UseMacro{table-timing-gen-SkOptLoopPeeling-reduction-stop_early_opt}
& \UseMacro{table-timing-gen-SkOptLoopPeeling-genTime-hot_filling_opt}
& \UseMacro{table-timing-gen-SkOptLoopPeeling-reduction-hot_filling_opt}
& \UseMacro{table-timing-gen-SkOptLoopPeeling-genTime-solver_aid_opt}
& \UseMacro{table-timing-gen-SkOptLoopPeeling-reduction-solver_aid_opt}
& \UseMacro{table-timing-gen-SkOptLoopPeeling-genTime-full_opt}
& \UseMacro{table-timing-gen-SkOptLoopPeeling-reduction-full_opt}
\\
\midrule
\UseMacro{table-timing-gen-SkOptLoopUnrolling-name}
& \UseMacro{table-timing-gen-SkOptLoopUnrolling-genTime-non_opt}
& \UseMacro{table-timing-gen-SkOptLoopUnrolling-genTime-stop_early_opt}
& \UseMacro{table-timing-gen-SkOptLoopUnrolling-reduction-stop_early_opt}
& \UseMacro{table-timing-gen-SkOptLoopUnrolling-genTime-hot_filling_opt}
& \UseMacro{table-timing-gen-SkOptLoopUnrolling-reduction-hot_filling_opt}
& \UseMacro{table-timing-gen-SkOptLoopUnrolling-genTime-solver_aid_opt}
& \UseMacro{table-timing-gen-SkOptLoopUnrolling-reduction-solver_aid_opt}
& \UseMacro{table-timing-gen-SkOptLoopUnrolling-genTime-full_opt}
& \UseMacro{table-timing-gen-SkOptLoopUnrolling-reduction-full_opt}
\\
\midrule
\UseMacro{table-timing-gen-SkOptLoopUnswitching-name}
& \UseMacro{table-timing-gen-SkOptLoopUnswitching-genTime-non_opt}
& \UseMacro{table-timing-gen-SkOptLoopUnswitching-genTime-stop_early_opt}
& \UseMacro{table-timing-gen-SkOptLoopUnswitching-reduction-stop_early_opt}
& \UseMacro{table-timing-gen-SkOptLoopUnswitching-genTime-hot_filling_opt}
& \UseMacro{table-timing-gen-SkOptLoopUnswitching-reduction-hot_filling_opt}
& \UseMacro{table-timing-gen-SkOptLoopUnswitching-genTime-solver_aid_opt}
& \UseMacro{table-timing-gen-SkOptLoopUnswitching-reduction-solver_aid_opt}
& \UseMacro{table-timing-gen-SkOptLoopUnswitching-genTime-full_opt}
& \UseMacro{table-timing-gen-SkOptLoopUnswitching-reduction-full_opt}
\\
\midrule
\UseMacro{table-timing-gen-SkOptRangeCheckElimination-name}
& \UseMacro{table-timing-gen-SkOptRangeCheckElimination-genTime-non_opt}
& \UseMacro{table-timing-gen-SkOptRangeCheckElimination-genTime-stop_early_opt}
& \UseMacro{table-timing-gen-SkOptRangeCheckElimination-reduction-stop_early_opt}
& \UseMacro{table-timing-gen-SkOptRangeCheckElimination-genTime-hot_filling_opt}
& \UseMacro{table-timing-gen-SkOptRangeCheckElimination-reduction-hot_filling_opt}
& \UseMacro{table-timing-gen-SkOptRangeCheckElimination-genTime-solver_aid_opt}
& \UseMacro{table-timing-gen-SkOptRangeCheckElimination-reduction-solver_aid_opt}
& \UseMacro{table-timing-gen-SkOptRangeCheckElimination-genTime-full_opt}
& \UseMacro{table-timing-gen-SkOptRangeCheckElimination-reduction-full_opt}
\\
\midrule
\UseMacro{table-timing-gen-SkOptScalarReplacement-name}
& \UseMacro{table-timing-gen-SkOptScalarReplacement-genTime-non_opt}
& \UseMacro{table-timing-gen-SkOptScalarReplacement-genTime-stop_early_opt}
& \UseMacro{table-timing-gen-SkOptScalarReplacement-reduction-stop_early_opt}
& \UseMacro{table-timing-gen-SkOptScalarReplacement-genTime-hot_filling_opt}
& \UseMacro{table-timing-gen-SkOptScalarReplacement-reduction-hot_filling_opt}
& \UseMacro{table-timing-gen-SkOptScalarReplacement-genTime-solver_aid_opt}
& \UseMacro{table-timing-gen-SkOptScalarReplacement-reduction-solver_aid_opt}
& \UseMacro{table-timing-gen-SkOptScalarReplacement-genTime-full_opt}
& \UseMacro{table-timing-gen-SkOptScalarReplacement-reduction-full_opt}
\\
\midrule
\textbf{\UseMacro{TableTimingGenCellSum}}
& \UseMacro{table-timing-gen-genTime-non_opt-sum}
& \UseMacro{table-timing-gen-genTime-stop_early_opt-sum}
& \UseMacro{table-timing-gen-reduction-stop_early_opt-sum}
& \UseMacro{table-timing-gen-genTime-hot_filling_opt-sum}
& \UseMacro{table-timing-gen-reduction-hot_filling_opt-sum}
& \UseMacro{table-timing-gen-genTime-solver_aid_opt-sum}
& \UseMacro{table-timing-gen-reduction-solver_aid_opt-sum}
& \UseMacro{table-timing-gen-genTime-full_opt-sum}
& \UseMacro{table-timing-gen-reduction-full_opt-sum}
\\
\bottomrule
\end{tabular}
\end{\UseMacro{FontsizeTableTimingGen}}
\end{\UseMacro{FloatEnvTableTimingGen}}

\egroup

Without any optimizations, the total time for generation across all
\sketches (essentially $\NumGeneratedProgramsPerSketch * \TotalNumManuallyCreatedSketches =
\NumTotalGeneratedPrograms$ programs total) is
\TotalGenTimeWithNonOpt.
We find that the overall time drops tremendously after the
optimizations are in place. When all optimizations are enabled
(``\fullopt''), the overall time to generate all programs for all
\sketches is \TotalGenTimeWithFullOpt, which is a
\UseMacro{table-timing-gen-reduction-full_opt-sum}\% reduction over
the time it takes to generate all programs without any optimization.

Breaking down the effectiveness of our optimizations even further, we
find that the \hotfilling optimization is in general the most
effective, with \hotfilling reducing the generation time by
\UseMacro{table-timing-gen-reduction-hot_filling_opt-sum}\% versus
\UseMacro{table-timing-gen-reduction-stop_early_opt-sum}\% for
\stopearly and
\UseMacro{table-timing-gen-reduction-solver_aid_opt-sum}\% for
\solveraid. Furthermore, we also see that \stopearly and \solveraid
have cases where they result in taking more time to generate programs
than without any optimization (the negative percentages in the table),
which suggests the extra checks required by \stopearly and the time to
invoke Z3 to solve constraints
end up introducing more overhead than actually helping. (We did not
set a timeout for Z3, because we did not observe Z3 getting stuck;
however, setting a timeout could impact the performance of the
\solveraid optimization.)
We see just one case for \hotfilling (and ultimately for when all
optimizations are on) where there is little reduction in time.
However, this one case (\UseMacro{table-timing-gen-SkJDK8255763-name})
takes very little time even without any optimizations, and the
difference in time is seemingly just noise.
Ultimately, all optimizations do help overall, with the reduction in
time when using all optimizations still higher than each individually.

We also measure the time to execute the \generatedprograms
from each of the \TotalNumManuallyCreatedSketches manually created
\sketches. The total time across all \generatedprograms is
\TotalExecTimeOnLFourFullItrs on L4 and \TotalExecTimeOnLOneFullItrs
on L1.

\subsection{\Sketch Extraction}
\label{sec:evaluation:extraction}
We extract \TotalNumExtractedSketches \sketches from
\TotalNumMethodsUsedForExtraction methods in
\TotalNumClassesUsedForExtraction classes, resulting in
\NumTotalGeneratedProgramsFromExtractedSketches \generatedprograms.
Recall that we let \JTool generate
\NumGeneratedProgramsPerExtractedSketch programs from every \sketch
(Section~\ref{sec:setup}), but not every \sketch includes sufficient
number of \holes from which
\NumGeneratedProgramsPerExtractedSketch programs can be generated
(\JTool only
explores the reachable \holes), which is why the total number of
\generatedprograms is less than
$\NumGeneratedProgramsPerExtractedSketch * \TotalNumExtractedSketches
= \NumTotalGeneratedProgramsFromExtractedSketchesThereotically$.
We found \UseMacro{table-extraction-total-numFail} out of
\NumTotalGeneratedProgramsFromExtractedSketches \generatedprograms
failed during our \JIT-testing procedure. We inspected all these
\UseMacro{table-extraction-total-numFail} programs and discovered
\NumOfBugsFromExtraction unique bugs
(Section~\ref{sec:evaluation:bugs}).

\label{sec:evaluation:javafuzzer}
In addition, we compare \JTool against an existing automated compiler
testing tool, \javafuzzer~\cite{JavaFuzzerIntelGitHubPage}, which is a
fuzzer tool Oracle has been using daily for years and
has been successful at detecting bugs in the \OracleJDK (\JIT)
compiler. Guided by grammar rules and pre-defined heuristics on
program structures, \javafuzzer generates hundreds of thousands of
small, random Java programs as tests, and it then performs
differential testing between a JVM under test and a reference JVM. In
contrast, \JTool is primarily developed for developers to embed their
knowledge into program generation by specifying \holes in \sketches
with automated \sketch extraction from existing Java programs.
Although \JTool and \javafuzzer have similar intentions,
they work quite differently, which is why the comparison results should be
taken with a grain of salt. We run \javafuzzer using the same
resources (CPU/RAM) for the same amount of time (which matches the
total execution time for \JTool in Section
\ref{sec:evaluation:extraction}). We perform differential testing by
comparing outputs from executions across different \JIT tiers, same as
for \JTool. We also collect code coverage of both the \COne
(\CodeIn{src/hotspot/share/c1/}) and \CTwo
(\CodeIn{src/hotspot/share/opto/}) compilers from executing the
programs generated by both tools separately.
Table~\ref{tab:comparing-results-of-jitattack-and-javafuzzer} compares
the results of \JTool and \javafuzzer. \javafuzzer did not generate
any program that would expose a bug in the \OracleJDK \JIT compiler in
this time frame. The code coverage achieved using both tools are close
to each other, though \JTool
achieves slightly higher code coverage on both \COne and \CTwo.

\bgroup
\def\arraystretch{1.2}

\begin{\UseMacro{table-jittack-javafuzzer-results-comparison-float-env}}[t]
\centering
\caption{\UseMacro{table-jittack-javafuzzer-results-comparison-caption}}
\begin{\UseMacro{table-jittack-javafuzzer-results-comparison-fontsize}}
\begin{tabular}{lrrrrr}
\toprule
\multirow{2}{*}{\textbf{\UseMacro{table-jittack-javafuzzer-results-comparison-col-tool}}} & \multirow{2}{*}{\textbf{\UseMacro{table-jittack-javafuzzer-results-comparison-col-gen}}} & \multirow{2}{*}{\textbf{\UseMacro{table-jittack-javafuzzer-results-comparison-col-timeout}}} & \multirow{2}{*}{\textbf{\UseMacro{table-jittack-javafuzzer-results-comparison-col-fail}}} & \multicolumn{2}{c}{\textbf{\UseMacro{table-jittack-javafuzzer-results-comparison-col-coverage}}}\\
\cmidrule(lr){5-6}
&  &  &  & \textbf{\UseMacro{table-jittack-javafuzzer-results-comparison-col-coverage-c1}} & \textbf{\UseMacro{table-jittack-javafuzzer-results-comparison-col-coverage-c2}}\\
\midrule
\UseMacro{table-jittack-javafuzzer-results-comparison-row0-tool} & \UseMacro{table-jittack-javafuzzer-results-comparison-row0-gen} & \UseMacro{table-jittack-javafuzzer-results-comparison-row0-timeout} & \UseMacro{table-jittack-javafuzzer-results-comparison-row0-fail} & \UseMacro{table-jittack-javafuzzer-results-comparison-row0-coverage-c1} & \UseMacro{table-jittack-javafuzzer-results-comparison-row0-coverage-c2}\\
\UseMacro{table-jittack-javafuzzer-results-comparison-row1-tool} & \UseMacro{table-jittack-javafuzzer-results-comparison-row1-gen} & \UseMacro{table-jittack-javafuzzer-results-comparison-row1-timeout} & \UseMacro{table-jittack-javafuzzer-results-comparison-row1-fail} & \UseMacro{table-jittack-javafuzzer-results-comparison-row1-coverage-c1} & \UseMacro{table-jittack-javafuzzer-results-comparison-row1-coverage-c2}\\
\bottomrule
\end{tabular}
\end{\UseMacro{table-jittack-javafuzzer-results-comparison-fontsize}}
\end{\UseMacro{table-jittack-javafuzzer-results-comparison-float-env}}

\egroup

\subsection{Detected \JIT-Bugs}
\label{sec:evaluation:bugs}

Bug \UseMacro{bugc67}\footnote{\UseMacro{ReportOfBugc67}},
from \sketch \UseMacro{table-timing-gen-Sk12-name}, showed mismatching
outputs on different tiers because
\CTwo's range-check elimination leads to incorrect loop executions.
The Oracle JDK developers labeled the bug we reported as a CVE (Common
Vulnerabilities and Exposures)\footnote{\UseMacro{CVEOfBugc67}}, and
they fixed the bug in a recent Oracle Critical Patch
Update\footnote{\UseMacro{OracleCPUOfBugc67}}.
The JDK developers also confirmed Bug
\UseMacro{bugsk4gen619}\footnote{\UseMacro{ReportOfBugsk4gen619}},
where a crash occurred from \CTwo, as a
\UseMacro{PriorityOfBugsk4gen619}\footnote{P3: Major loss of
function.} bug; this bug was discovered in parallel by others and
was fixed in JDK 16.
Our \sketch that exposed this bug is shown in
Figure~\ref{fig:example:sketch}.

Additionally, we discovered four bugs using extracted \sketches from
existing Java projects.
Bug \UseMacro{bugmath182}\footnote{\UseMacro{ReportOfBugmath182}}
crashed on tiers L1 and L4 because an array store in \COne compiled
code writes to an arbitrary location due to index overflow. The JDK
developers labeled the bug as a
CVE\footnote{\UseMacro{CVEOfBugmath182}}, and they fixed the bug in
another recent Oracle Critical Patch
Update\footnote{\UseMacro{OracleCPUOfBugmath182}}.
Bug
\UseMacro{bugcheckstyle106}\footnote{\UseMacro{ReportOfBugcheckstyle106}}
was confirmed as a crash bug, with priority
\UseMacro{PriorityOfBugcheckstyle106}\footnote{P2: Crashes, loss of
data, severe memory leak.}, related to wrong JVM state used for a
receiver null check, and it was fixed in JDK 17.
Bug \UseMacro{bugcodec292}\footnote{\UseMacro{ReportOfBugcodec292}}
missed throwing some \CodeIn{NegativeArraySizeException} on tier L4
caused by \CTwo optimizations; the JDK developers labeled it as a
\UseMacro{PriorityOfBugcodec292} bug and fixed it in JDK 18.
Bug
\UseMacro{bugcompress208}\footnote{\UseMacro{ReportOfBugcompress208}}
crashed due to incorrect \CTwo loop optimizations before calling
\CodeIn{Arrays.copyOf} with a negative parameter; this bug was
confirmed with priority \UseMacro{PriorityOfBugcompress208} and was
also discovered in parallel by others. The bug was fixed in JDK 18.

\section{Discussion}
\label{sec:discussion}

In this section, we contrast \JTool's execution-based generation to
static generation, describe limitations of \JTool, and provide
directions for future work.

\subsection{Execution-Based vs. Static Generation}
\label{sec:execution-based-vs-static-generation}
Recall \JTool generates programs through an execution-based model
(Section~\ref{sec:generation:procedure}) but we could have generated
programs statically by processing an entire \sketch and replacing all
\holes with concrete expressions. Static generation would process the
\sketch repeatedly, putting in different concrete expressions per
\hole to output a new \generatedprogram, up to some maximum number.
Generating programs statically could be faster, because it would not
be executing the program at any point.

However, the execution-based generation provides a number of
advantages over static generation. Execution-based generation
(1)~knows what exactly would be executed in a \generatedprogram after
being compiled, i.e., which parts are dead code or which parts are
executable (such information can be leveraged to guide the exploration
of \holes instead of relying on randomness, which we leave as future
work)
and (2)~makes it possible to use values available
at runtime to construct \holes; consider:
\begin{lstlisting}[language=sketchy-display]
int m(int[] a) {
return a[intVal(0, a.length).eval()];
}
\end{lstlisting}
where the \hole would be a random integer between \CodeIn{0}
and the length of the array \CodeIn{a},
which depends on the value of \CodeIn{a}, known only when
actually executing the \sketch. An execution-based model allows
for expressing more complex programs that static generation cannot
generate, as it does not have such runtime information.

To compare execution-based and static generation, we
create a variant of \JTool that generates programs statically.
This variant relies on the same syntax and semantics, but it
statically processes the
\sketch once to replace all the \holes with concrete expressions.
Similarly to execution-based generation, we construct
\eASTs for all the \holes. Each \eAST per \hole contains all the
choices for the \hole, i.e., concrete expressions that can
be filled in the \hole. For each \hole written in the \sketch, we
randomly choose one of the concrete expressions to replace the
\hole, resulting in a \generatedprogram. The \generatedprogram
has every \hole filled, unlike for execution-based generation where
some \holes may remain unfilled if they are not reached during
execution.

For this evaluation, we use the same configuration
(\NumGeneratedProgramsPerSketch programs for each \sketch) for our
static variant of \JTool as to allow for proper comparison against the
execution-based model. The total generation time with static
generation is \TotalGenTimeWithStatic, which is shorter
than execution-based generation (\TotalGenTimeWithFullOpt),
but they both generate a large number of programs, and executing all
\generatedprograms for both approaches still takes
\TotalExecTimeOnLFourAndLOneFullItrs. As such, generation time is
practically negligible compared against the differential testing part
of \JTool. Furthermore,
since static generation fills every \hole
in the \sketch, some \generatedprograms could be syntactically
different from each other, but their differences are only for
expressions in the unreachable \holes, so essentially the same code
would be executed. Execution-based generation would skip unreachable
\holes, ensuring every \generatedprogram is not only syntactically
different but also executed differently. We collected reachability of
the filled \holes when executing the \generatedprograms.
\UseMacro{table-tracking-holes-static-percentage-sum}\% of filled
\holes are reached during execution of \generatedprograms from static
generation while execution-based generation guarantees
\UseMacro{table-tracking-holes-dynamic-percentage-sum}\% reachability.
In terms of detected bugs, compared against execution-based
generation, statically \generatedprograms detected only Bug
\UseMacro{bugc67} and missed detecting Bug \UseMacro{bugsk4gen619}.

\subsection{Limitations \& Future Work}

There are two main reasons why a relatively small number
(\TotalNumExtractedSketches) of \sketches are extracted from a
relatively large number (\TotalNumClassesUsedForExtraction) of classes
in existing projects. First, \JTool currently supports only static
Java methods as \sketchmethods.
We leave support of instance methods as \sketchmethods for future
work, e.g., using EvoSuite~\cite{Fraser11EvoSuite} to create
receiver objects and inputs for instance methods. Second, we use a different name for the
extracted \sketch class from the original class, which sometimes made
the \sketch not pass Java type-checking due to circular dependencies
between the \sketch class and other classes.
We will explore editing the original class in place instead of
creating a renamed \sketch class as future work.

\JTool requires re-executing programs many times just to trigger \JIT
optimizations for testing.
We considered other options such as \CodeIn{-XX:CompileThreshold} that
controls the number of interpreted method invocations before optimization.
We also considered the option \CodeIn{-XX:Tier4InvocationThreshold} that controls the minimum number
of method invocations before transitioning to L4. However, we found
these other options also have a big effect on when \JIT
optimizations occur, so
just using these options would not truly reflect actual \JIT
usage, similar to just enabling C2 from the
beginning~\cite{DisadvantagesOfForcingEarlyCompilation}.

Not all bugs we detected are reproducible each time due to the
non-deterministic nature of executions and \JIT profiling (which is
different from non-determinism within programs under execution,
discussed in Section~\ref{sec:jit-testing}). For example, in one of
the \generatedprograms for
\UseMacro{table-timing-gen-Sk4-name},
we could not always observe failure (crashing
the JVM) when run on the same \JIT compiler multiple times. We
plan to investigate such flakiness in the future.

Although we designed and implemented \JTool for Java and \JIT
compilers, the simple imperative language and extensions, as shown in
Section~\ref{sec:programming:model}, represent the foundations for
supporting \sketches for general imperative languages. Thus the ideas
of template-based testing and execution-based generation, can be also
applied to other languages and compilers, e.g., Scala, C\#, etc., or
even software systems in general. We leave this as future work.

\section{Related Work}
\label{sec:related}

\MyPara{Compiler testing} There is a large body of work on compiler
testing, systematically reviewed in recent
surveys~\cite{Chen20SurveyofCompilerTesting, Tang20CompilerTesting}.
For example, \csmith~\cite{YangPLDI11CSmith, Regehr12CReduce} is a
well-known tool for testing C compilers by randomly generating C
programs. It found bugs in mainstream
compilers~\cite{GCCBugListFoundByCsmith, LLVMBugListFoundByCsmith} and
led to significant attention for compiler
testing~\cite{Morisset13CompilerTestingViaATheoryOfSoundOptimisationsInTheC11C++11MemoryModel,
Holler12FuzzingWithCodeFragments,
Alipour16GeneratingFocusedRandomTestsUsingDirectedSwarmTesting,
Chen19HistoryGuidedConfigurationDiversificationForCompilerTestProgramGeneration,
Livinskii20YARPGen}. Mutation-based
fuzzing~\cite{Manes21FuzzingSurvery, Wang10TaintScope,
Lee21ConstraintGuidedDirectedGreyboxFuzzing, Padhye19FuzzFactory,
You19ProFuzzer, Peng18TFuzz, Park20Aspect} is another approach
to testing compilers by mutating
existing programs, with several techniques and tools specifically for
Java~\cite{ChenPLDI16DifferentialTestingOfJVM,
ChenICSE19DeepDifferentialTestingOfJVMImplementations,
PadhyeISSTA19JQFZest, PadhyeISSTA19JQFTool, Jayaraman09jFuzzAC,
Kersten17Kelinci, Blair20HotFuzz, Vikram21BonsaiFuzzing,
Chaliasos22FindingTypingCompilerBugs,
Zhao22HistoryDrivenTestProgramSynthesisJVMTesting}.
Concerning \JIT compilers, Yoshikawa et
al.~\cite{YoshikawaETAL03RandomGeneratorForJIT} presented a generation
approach
that produces random Java bytecode
. \javafuzzer~\cite{JavaFuzzerIntelGitHubPage,
JavaFuzzerAzulGitHubPage, JavaFuzzerRedHatGitHubPage} generates
random Java programs to test Java \JIT compilers. There is also work
on testing the C\# \JIT compilers~\cite{FuzzlynGitHubRepoPage,
FuzzlynPost} and Smalltalk \JIT compilers~\cite{Poli22Interpreter}.
Unlike all these techniques, \JTool was primarily developed to
complement manually-written tests. Developers can embed
their knowledge into program generation by specifying \holes for
exploration, enabling better testing of \JIT compilers that require
complex structures and execution to reveal bugs.
We do compare against \javafuzzer as part of our evaluation.

While \JTool relies on differential
testing~\cite{Mckeeman98DifferentialTesting} to determine whether a
test fails, other means to construct a test oracle include metamorphic
testing~\cite{Chen98MetamorphicTesting} and specification-based
testing~\cite{Schumi21SpecTest, Ye2021AutomatedCT}. Equivalence Modulo
Inputs (EMI)~\cite{LePLDI14CompilerValidationViaEMI,
Le15RandomizedStressTestingOfLinkTimeOptimizers,
Le15FindingDeepCompilerBugsViaGuidedStochasticProgramMutation,
Sun16FindingCompilerBugsViaLiveCodeMutation,
Lidbury15ManyCoreCompilerFuzzing, Chowdhury20SLEMI,
Donaldson21TestCaseReductionandDeduplicationAlmostforFreewithTransformationBasedCompilerTesting,
Nakamura16EMIForCCompilers} is a representative of metamorphic
testing technique. EMI produces equivalent but different test programs
and compares behaviors across these programs on a single compiler.
Chen et al.~\cite{ChenICSE16ComparisonOfCompilerTesting} compared
differential testing and EMI techniques.

\MyPara{\Sketch-based program synthesis} There has been work on
synthesizing programs given initial \sketches. These techniques can
either synthesize programs using SAT/SMT
solvers~\cite{Feser15SynthesizingDataStructureTransformationsFromInputOutputExamples,
Jha10OracleGuidedComponentBasedProgramSynthesis,
Singh11SynthesizingDataStructureManipulationsFromStoryboards,
Gvero11InteractiveSynthesisOfCodeSnippets,
Solar-LezamaETAL06CombinatorialSketchingForFinatePrograms,
Solar-LezamaSTTT13ProgramSketching, Jeon15JSketch}, using
combinatorial techniques focusing on just
variables~\cite{ZhangPLDI17SkeletalProgramEnumeration}, or define
\holes using domain-specific
languages~\cite{SirerETAL00UsingProductionGrammarsInSoftwareTesting,Bruno05FlexibleDatabaseGenerators}.
Ching and
Katz~\cite{ChingKatz93TestingAPLCompiler} proposed to generate tests
for APL-to-C compiler COMPC through a template that denotes functions
and data types, which leverages the dynamic nature of APL to
execute programs as soon as \holes get filled.
CodeHint~\cite{GalensonICSE14CodeHint} synthesizes sequences of API
method invocations by running code with \holes to be filled.
\edsketch~\cite{HuaSPIN17EdSketchExecution-DrivenSketchingfForJava}
synthesizes implementations for \holes through test executions, aiming
to pass the tests it runs on, focusing on exploring field references.
In contrast to all of these, \JTool generates concrete programs for
testing a \JIT compiler by executing \sketches and allows richer
expressions to be generated in \holes.

\MyPara{Test input generation} Randoop~\cite{Pacheco07Randoop} and
EvoSuite~\cite{Fraser11EvoSuite} automatically generate JUnit tests by
incrementally extending sequences of method invocations. Neither is
able to effectively explore inside a method, e.g., at the expression
level, as \JTool. \astgen~\cite{DanielFSE07ASTGen} and
\udita~\cite{GligoricICSE10UDITA} bounded-exhaustively generate
complex test inputs, including programs, but both require developers
to manually write extra predicates or generators to encode their
intuition of guiding exploration. Similarly,
QuickCheck~\cite{Claessen00QuickCheck,juniquickcheck}, as a
property-based testing tool, is also capable of randomly generating
test inputs but requires developers to provide generators for complex
data types. In contrast to all of these, \JTool provides a different
way for developers to write \sketches. Developers simply write
programs with \holes to make \sketches or even automatically extract
\sketches from existing code. \Sketches are written in the host
language the developers are already using.

\MyPara{Other research on \JIT and \JVM} There has
been work on formal verification of \JIT
compilers~\cite{MyreenPOPL10VerifiedJITCompilerOnX86,
GuoPOPL11TheEssenceOfCompilingWithTraces,
ShingarovVMIL19FormalVerificationOfJITBySymbolicExecution,
BarriereCoqPL20TowardsFormallyVerifiedJITCompilation,
BrownPLDI20VerifiedRangeAnalysisForJavaScriptJITs,
WangOSDI14JitkATrustworthyInKernelInterpreterInfrastructure,
GeffenCAD20SynthesizingJITCompilersForInKernelDSLs,
Hawblitzel13WillYouStillCompileMeTomorrow},
\JIT-induced side-channel detection~\cite{Brennan2020JITLeaks,
BrennanICSE20JVMFuzzingforJITInducedSideChannelDetection}, and on
identifying hard-to-optimize code~\cite{GongFSE15JITProf} and
unspecified JNI behaviors of a
\JVM~\cite{Hwang21JUSTGenEffectiveTestGenerationForUnspecifiedJNIBehaviorsOnJVMs}.

\section{Conclusion}

We presented \JTool, a framework that enables \sketch-based testing for
compilers.
Using \JTool, compiler developers can write \sketches in the same
language as the compiler they are testing (Java), enabling them to
leverage their domain knowledge to set up a code structure likely to
lead to compiler optimizations while leaving \holes representing
expressions they want explored. \JTool executes \sketches, exploring
possible expressions for \holes and filling them in, generating
programs to later be run on compilers. To speed up the generation
process, we introduced \NumOfOptimizations optimizations that reduced
overall generation time by
\UseMacro{table-timing-gen-reduction-full_opt-sum}\% in our
experiments.
Using \TotalNumManuallyCreatedSketches \sketches created on our own and
\TotalNumExtractedSketches \sketches extracted from existing Java
projects,
\JTool found \TotalNumOfBugs critical (P3 or higher) bugs in
\OracleJDK, all of which were confirmed and fixed by Oracle
developers. \Capitalize{\NumPreviouslyUnknownBugs} of them were
previously unknown, including \TotalNumOfCVEs unknown CVEs.
\JTool blends the power of developers insights, who are providing
templates, and random testing to detect critical bugs.

\begin{acks}
We thank Nader Al Awar, Kush Jain, Sandeep Konchady, Owolabi Legunsen,
Yu Liu, Pengyu Nie, Aditya Thimmaiah, Jiyang Zhang, and the anonymous
reviewers for their comments and feedback. This work is partially
supported by a Google Faculty Research Award and the US National
Science Foundation under Grant Nos. CCF-1652517, CCF-2107291, and
CCF-2217696.
\end{acks}

\balance
\bibliography{bib}
\bibliographystyle{ACM-Reference-Format}

\end{document}